\documentclass[aps,prc,twocolumn,preprintnumbers,amsmath,amssymb,amsfonts,nofootinbib,superscriptaddress,altaffilletter]{revtex4-2}

\usepackage{graphicx}
\usepackage[unicode]{hyperref}
\hypersetup{
  bookmarksopen=true
}
\usepackage{amssymb}
\usepackage{amsmath}
\usepackage{braket}
\usepackage{mathtools, lipsum}
\usepackage{color}
\usepackage{supertabular}
\usepackage{dcolumn}
\usepackage{multirow}
\usepackage{upgreek}
\usepackage{xcolor}
\usepackage{booktabs}
\usepackage{natbib}
\usepackage{float}
\usepackage{orcidlink}

\newif\ifshowfigs
\showfigstrue

\usepackage{soul}

\newcommand{\useMSU}[1]{}

\begin{document}

\preprint{INT-PUB-26-023}
\title[Implications on astrophysical properties]{On the effect of higher order symmetry energy corrections in Skyrme models for neutron star matter}

\author{Md. Emanuel Hoque, \orcidlink{0009-0002-8488-8758}}
\email{emanuel.hoque@saha.ac.in}

\author{Arunava Mukherjee, \orcidlink{0000-0003-1274-5846}}
\email{Corresponding author: arunava.mukherjee@saha.ac.in}

\affiliation{Saha Institute of Nuclear Physics, 1/AF Bidhannagar, Kolkata-700064, India}
\affiliation{Homi Bhabha National Institute, Anushakti Nagar, Mumbai 400094, India}

\date[\relax]{compiled on \today}

\begin{abstract}
   
Neutron stars consist of cold, dense, neutron-rich nuclear matter under charge neutrality and $\beta$-equilibrium. In most nuclear equation of state (EOS) studies, the isospin dependence of asymmetric nuclear matter is described using the conventional quadratic/parabolic approximation to the nuclear symmetry energy. However, its validity in the highly neutron-rich inner core of neutron stars remains uncertain. In this work, we systematically investigate the role of higher-order isospin corrections to the symmetry energy within the framework of Skyrme-like effective nuclear interactions. We first analyze the standard SLy4 parametrization and quantify deviations arising from successive higher-order terms in the expansion of the energy per nucleon with respect to the isospin asymmetry parameter. We then extend the analysis to a large population of physically viable Skyrme EOSs sampled over a broad parameter space constrained by conventional nuclear saturation density bounds, thermodynamic stability, and causality, as well as the requirement to support astrophysical neutron-star mass observations exceeding $2\text{M}_{\odot}$. We find that higher-order isospin corrections become increasingly important at supra-nuclear densities and can significantly modify composition-sensitive quantities under $\beta$-equilibrium, including the neutron-proton chemical potential difference, proton fraction, leptonic sector properties, and the direct-Urca process. In contrast, the $\beta$-equilibrated EOS, energy density, pressure, and sound speed remain comparatively insensitive to these corrections for most viable EOSs. Our results demonstrate that while the quadratic approximation captures bulk thermodynamic behavior reasonably well, higher-order isospin contributions play a non-negligible role in determining the detailed composition and microscopic properties of dense matter in neutron-star interiors.

\end{abstract}

\maketitle

\section{Introduction}
\label{sec:introduction}

Ultra-dense matter within the interior of a neutron star exists under cold, degenerate, and highly neutron-rich conditions. Within this old astrophysical environment, both the leptonic and hadronic sectors of the isospin-asymmetric matter are maintained in $\beta$-equilibrium~\cite{LattimerPrakash_2004Sci}. At sub-saturation densities (less than the nuclear saturation density, $n_0$), the properties of this hadronic matter closely resemble those of terrestrial heavy nuclei. Conversely, at intermediate densities ($\sim n_0 - 2.5n_0$), the matter exhibits characteristics analogous to those probed in terrestrial 
Heavy-Ion Collisions (HICs)~\cite{Busza_etal_HIC_2018ARNPS}, subject to minor thermal corrections in pressure arising from finite-temperature effects~\cite{LattimerPrakash_2007PR}. 

Formulating the equation of state (EOS) for cold, $\beta$-equilibrated stellar matter requires solving the quantum many-body problem within a strongly coupled framework. To address this, various sophisticated many-body techniques have been implemented, including Relativistic Mean-Field Theory (RMFT)~\cite{Walecka_1975PLB, WaleckaSerot_1979PLB}, Dirac-Brueckner-Hartree-Fock (DBHF) theory~\cite{Bethe_1971AnnuRev, Lee_etal_1998PRC, KatayamaSaito_2013PRC, Sammarruca_2014article}, variational methods~\cite{PandharipandeWiringa_RMP1979, APR_1998PRC}, Skyrme Hartree-Fock (SHF) functionals~\cite{Skyrme_1958, VautherinBrink_1972PRC, Chabanat_etal_1997NuPhA}, and the Gogny interaction~\cite{GonzalezBoquera_etal_2017PRC}. While each approach possesses distinct conceptual advantages and inherent limitations~\cite{Forbes_etal_prd2019}, certain frameworks remain more widely adopted in the literature~\cite{diff_sym_en_LattimerGroup_prc2024}. 

Skyrme-like interactions provide a tractable parameterization of the short-range nuclear force through density-dependent zero-range effective interactions~\cite{VautherinBrink_1972PRC}. In the Skyrme-Hartree-Fock framework, the complicated many-body nucleon-nucleon interaction is replaced by an effective interaction expanded in powers of the relative momenta and local densities, allowing the nuclear energy-density functional to be expressed in a simple analytic form. This formulation enables self-consistent calculations of bulk nuclear matter properties, finite nuclei, and neutron-rich systems over a wide range of densities and isospin asymmetries. Owing to their computational simplicity and flexibility, Skyrme parameterizations have become one of the standard approaches for constructing nuclear equations of state (EOSs) relevant to both nuclear physics and astrophysical applications. 

In particular, the density-dependent terms effectively incorporate many-body correlations and saturation properties of nuclear matter, making these interactions especially suitable for studies of neutron-star matter and highly isospin asymmetric systems. Within Skyrme energy-density functionals, the isovector sector of the interaction directly governs the density dependence of the nuclear symmetry energy and therefore plays a central role in determining the composition and thermodynamic properties of neutron-star matter.

The isospin dependence of asymmetric nuclear matter is commonly characterized by the nuclear symmetry energy, which quantifies the energy cost of making nuclear matter increasingly neutron-rich relative to symmetric matter. In most practical applications, the energy per nucleon is expanded in powers of the isospin asymmetry parameter around symmetric nuclear matter while retaining only the leading quadratic order contribution. This so-called parabolic approximation has been extensively employed in studies of neutron-rich matter and neutron-star EOSs. However, matter inside neutron stars remains highly neutron-rich across the entire density range, raising concerns regarding the validity of the conventional quadratic approximation and the role of higher-order isospin corrections.

Higher-order corrections to the symmetry energy can potentially influence several important properties of neutron-star matter, including the proton fraction, chemical composition, leptonic content, number density, pressure, and transport properties under $\beta$-equilibrium conditions. In this work, we thoroughly explore the properties of Skyrme-like effective nuclear interactions over a wide parameter space under the minimal essential physical and observational constraints, and systematically investigate the effects of higher-order isospin corrections in the symmetry energy and its density dependence.

\section{Skyrme effective interaction for neutron star matter}\label{sec:skyrme_model} 

In this section, we briefly introduce the concepts and notation used explicitly throughout this paper. The effective potential for Skyrme-like non-relativistic interactions~\cite{VautherinBrink_1972PRC} has been discussed in detail~\cite{Chabanat_etal_1997NuPhA, DouchinHaensel_AA2001} for high-density nuclear matter. This type of interaction is given in the standard form:
\begin{equation}\label{skyrme-potential}
\begin{split}
    V(\vec{r}_1,\vec{r}_2) =& \; t_{0}(1 + x_{0} P_{\sigma})\delta(\vec{r}) \\
    +& \frac{1}{2}t_1(1+x_1 P_{\sigma})[\vec{p'}^{2} \delta(\vec{r})+\delta(\vec{r}) \vec{p}^2] \\
    +& t_2(1+x_2 P_{\sigma}) \vec{p'}\cdot\delta(\vec{r}) \vec{p} \\
    +& \frac{1}{6} t_3 (1+ x_3 P_{\sigma})[\rho(\vec{R})]^{\iota} \delta(\vec{r}) \\
    +& i W_0 \vec{\sigma} \cdot [\vec{p'} \times \delta(\vec{r}) \vec{p}]
\end{split}
\end{equation}
where $\vec{R} = \frac{1}{2}(\vec{r}_1+\vec{r}_2)$. The first three terms in this equation represent a Gaussian central potential, with the third term being related to non-local effects. The last term originates from the two-body spin-orbit coupling, while the fourth term introduces an additional density dependence. 

Using this standard form, the total binding energy of a nucleus can be expressed as the integral of an energy density functional:
\begin{equation}\label{total-binding-energy-0}
\langle \psi|H|\psi \rangle = \int \mathcal{H}(\vec{r}) d^3 r
\end{equation}
where
\begin{equation}\label{total-binding-energy-1}
    \mathcal{H} = \mathcal{K} + \mathcal{H}_0 + \mathcal{H}_3 + \mathcal{H}_{\text{eff}} + \mathcal{H}_{\text{fin}} + \mathcal{H}_{\text{so}} + \mathcal{H}_{\text{sg}} + \mathcal{H}_{\text{Coul}}
\end{equation}

Here, $\mathcal{K}$ is the kinetic energy term, $\mathcal{H}_0$ is a zero-range term, $\mathcal{H}_3$ is the density-dependent term, $\mathcal{H}_{\text{eff}}$ is an effective-mass term, $\mathcal{H}_{\text{fin}}$ is a finite-range term, $\mathcal{H}_{\text{so}}$ is a spin-orbit term, and $\mathcal{H}_{\text{sg}}$ is a term due to the tensor coupling with spin and gradient: 
\begin{equation}\label{eq_K}
    \mathcal{K}  = \frac{\hbar^2}{2m}\tau
\end{equation}

\begin{align}\label{eq_H}
    \mathcal{H}_0 &= \frac{1}{4} t_0 \left[(2+x_0)n_b^2 - (2x_0+1)(n_p^2+n_n^2)\right] \\
    \mathcal{H}_3 &= \frac{1}{24} t_3 n_b^{\iota} \left[(2+x_3)n_b^2 - (2x_3+1)(n_p^2+n_n^2)\right] \\
    \mathcal{H}_{\text{eff}} &= \frac{1}{8} [t_1(2+x_1) + t_2(2+x_2)]\tau n_b \nonumber \\
    &+ \frac{1}{8} [t_2(2x_2+1) - t_1(2x_1+1)](\tau_p n_p+\tau_n n_n) \\
    \mathcal{H}_{\text{fin}} &= \frac{1}{32} [3t_1(2+x_1) - t_2(2+x_2)](\vec{\nabla}n_b)^2 \nonumber \\
    &- \frac{1}{32} [3t_1(2x_1+1) + t_2(2x_2+1)][(\vec{\nabla}n_p)^2 + (\vec{\nabla}n_n)^2] \\
    \mathcal{H}_{\text{so}} &= \frac{1}{2} W_0 [\vec{J}\cdot\vec{\nabla}n_b + \vec{J}_p\cdot\vec{\nabla}n_p + \vec{J}_n\cdot\vec{\nabla}n_n] \\
    \mathcal{H}_{\text{sg}} &= -\frac{1}{16} (t_1x_1 + t_2x_2)\vec{J}^2 + \frac{1}{16} (t_1 - t_2)[\vec{J}_p^2 + \vec{J}_n^2]
\end{align}

and $\mathcal{H}_{\text{Coul}}$ is the Coulomb interaction term. The various total densities are defined as $n_b = n_p + n_n$, $\tau = \tau_p + \tau_n$, and $\vec{J} = \vec{J}_p + \vec{J}_n$. The local matter densities for neutrons ($n$) and protons ($p$) are represented by
\begin{align}\label{local-matter-densities}
    n_q(\vec{r}) = \sum_{l,s} |\phi_l^q(\vec{r},s)|^2 g_l^q  
\end{align}
with $q \in \{n, p\}$, where $l$ and $s$ denote the orbital and spin quantum numbers, respectively. Similarly, the kinetic density and spin density read
\begin{align}\label{kinetic-densities}
    \tau_q(\vec{r}) = \sum_{l,s} |\vec{\nabla} \phi_l^q(\vec{r},s)|^2 g_l^q 
\end{align}
and 
\begin{align}\label{spin-densities}
    \vec{J}_q(\vec{r}) = \sum_{l,s,s'} [\phi_l^{*q}(\vec{r},s')\vec{\nabla} \phi_l^q(\vec{r},s)] \times \langle s'|\vec{\sigma}|s \rangle g_l^q  
\end{align}
respectively, where $\phi_l^q(\vec{r},s)$ is the single-particle wave function with orbital, spin, and isospin quantum numbers $l$, $s$, and $q$, respectively, and $g_l^q$ is the occupation number of the corresponding state. 
\par
In this paper, we initially use the Skyrme interaction parameters corresponding to the SLy4 equation of state (EOS) as the \emph{standard reference case to benchmark our analysis and validate several results} reported by~\citet{DouchinHaensel_AA2001}. Subsequently, we investigate the higher-order corrections of isospin dependence on various quantities beyond the standard parabolic approximations of symmetry energy for infinite nuclear matter and $\beta$-equilibrated matter under the cold, degenerate conditions suitable for the interior of a neutron star.

\subsection{Infinite Nuclear Matter}\label{inf_nuc_mat} 

In the case of a uniform and infinite nuclear matter relevant for neutron stars, the $\mathcal{H}_{\text{fin}}$, $\mathcal{H}_{\text{so}}$, and $\mathcal{H}_{\text{sg}}$ terms do not contribute significantly; therefore, they can be dropped from further considerations regarding the stellar matter and the properties of its equation of state (EOS). Here, we also restrict our study to nucleonic matter EOS containing only neutrons ($n$), protons ($p$), electrons ($e$), and muons ($\mu$). 

For a single species of non-relativistic fermions of mass $m$ and number density $n$, the kinetic energy density is given by
\begin{equation}\label{kinetic-densities-fermion}
\begin{split}
    \mathcal{K} =& \frac{\hbar^2}{2m}\tau = \bigg(\frac{\hbar^2}{2m}\bigg) g \int \frac{k^2 d^3k}{(2\pi)^3} = \bigg(\frac{\hbar^2}{2m}\bigg) g \frac{1}{2\pi^2}\frac{k_f^5}{5} \\
    =& g \frac{\hbar^2}{20\pi^2 m}(3\pi^2 n)^{\frac{5}{3}} 
\end{split}
\end{equation}
with 
\begin{align}\label{densities-fermion}
    n = g \int \frac{d^3p}{(2\pi\hbar)^3} = g \int \frac{d^3k}{(2\pi)^3} = g \frac{1}{2\pi^2}\int k^2 dk = \frac{gk_f^3}{6\pi^2}
\end{align}
where $g$ is the occupation number. Hence, the kinetic-energy term (Eq.~\ref{eq_K}) simplifies to 
\begin{align}\label{K-1}
    \mathcal{K}  = \frac{\hbar^2}{2m}(\tau_n + \tau_p)
\end{align}
with
\begin{align}\label{kin-n}
    \tau_n = \frac{1}{5\pi^2}(3\pi^2n_n)^{\frac{5}{3}} = \frac{1}{5\pi^2}(3\pi^2)^{\frac{5}{3}}\left(\frac{1+I}{2}\right)^{\frac{5}{3}}n_b^{\frac{5}{3}} 
\end{align}
\begin{align}\label{kin-p}
    \tau_p = \frac{1}{5\pi^2}(3\pi^2n_p)^{\frac{5}{3}} = \frac{1}{5\pi^2}(3\pi^2)^{\frac{5}{3}}\left(\frac{1-I}{2}\right)^{\frac{5}{3}}n_b^{\frac{5}{3}} 
\end{align}
which can be expressed in the combined form 
\begin{align}\label{K-2}
    \mathcal{K}  = \frac{3}{5}\frac{\hbar^2}{2m}\left( \frac{3\pi^2}{2} \right)^{\frac{2}{3}}n_b^{\frac{5}{3}}F_{5/3}(I)
\end{align}
where $m (=m_n=m_p)$ is the baryon mass, and 
\begin{equation}\label{isospin-n-p}
\begin{split}
    & I = \frac{(n_n - n_p)}{(n_n + n_p)} ,\\
    & \implies \;\; n_n = \frac{1}{2}(1+I)n_b ,\;\; n_p = \frac{1}{2}(1-I)n_b 
\end{split}
\end{equation}
with the following definition of the isospin asymmetry factor $F_N(I)$:
\begin{align}\label{asymm-fact}
    F_N(I) = \frac{1}{2}[(1+I)^N + (1-I)^N]
\end{align}

Similarly, using Eq.~\ref{eq_H} for $\mathcal{H}_0$, $\mathcal{H}_3$, and $\mathcal{H}_{\text{eff}}$, we obtain 
\begin{align}\label{H0-1}
    \mathcal{H}_0 = \frac{1}{8}t_0n_b^2[2(x_0+2)-(2x_0+1)F_2(I)]
\end{align}

\begin{align}\label{H3-1}
    \mathcal{H}_3 = \frac{1}{48} t_3 n_b^{\iota+2} [2(x_3+2) - (2x_3+1)F_2(I)]
\end{align}

\begin{align}\label{Heff-1}
    \mathcal{H}_{\text{eff}} = \frac{3}{40}\left( \frac{3\pi^2}{2} \right)^{\frac{2}{3}}n_b^{\frac{8}{3}} 
    \Big\{ [t_1(x_1+2)+t_2(x_2+2)]F_{5/3}(I) \notag \\
    + \frac{1}{2}[t_2(2x_2+1)-t_1(2x_1+1)]F_{8/3}(I) \Big\}
\end{align}

Combining the above expressions, the total energy density of the system can be expressed as 
\begin{equation}\label{energy-density}
\begin{split}
    \epsilon_b =& \; \mathcal{K} + \mathcal{H}_0 + \mathcal{H}_3 + \mathcal{H}_{\text{eff}} \\ 
    =& \frac{3}{5}\frac{\hbar^2}{2m}\left( \frac{3\pi^2}{2} \right)^{\frac{2}{3}}n_b^{\frac{5}{3}}F_{5/3}(I) \\
    +& \frac{1}{8}t_0n_b^2[2(x_0+2)-(2x_0+1)F_2(I)] \\ 
    +& \frac{1}{48} t_3 n_b^{\iota+2} [2(x_3+2) - (2x_3+1)F_2(I)] \\ 
    +& \frac{3}{40}\left( \frac{3\pi^2}{2} \right)^{\frac{2}{3}}n_b^{\frac{8}{3}} \Big\{ [t_1(x_1+2)+t_2(x_2+2)]F_{5/3}(I) \\
    +& \frac{1}{2}[t_2(2x_2+1)-t_1(2x_1+1)]F_{8/3}(I) \Big\}
\end{split}
\end{equation}
and the energy per nucleon is given by
\begin{equation}\label{E/A}
\begin{split}
    \frac{E}{A}(n_b, I) =& \frac{\epsilon_b}{n_b} \\
    =& \frac{3}{5}\frac{\hbar^2}{2m}\left( \frac{3\pi^2}{2} \right)^{\frac{2}{3}}n_b^{\frac{2}{3}}F_{5/3}(I) \\
    +& \frac{1}{8}t_0n_b[2(x_0+2)-(2x_0+1)F_2(I)] \\
    +& \frac{1}{48} t_3 n_b^{\iota+1} [2(x_3+2) - (2x_3+1)F_2(I)] \\
    +& \frac{3}{40}\left( \frac{3\pi^2}{2} \right)^{\frac{2}{3}}n_b^{\frac{5}{3}}
    \Big\{ [t_1(x_1+2)+t_2(x_2+2)]F_{5/3}(I) \\
    +& \frac{1}{2}[t_2(2x_2+1)-t_1(2x_1+1)]F_{8/3}(I) \Big\}
\end{split}
\end{equation}

\begin{figure}[h!]
    \centering
    \includegraphics[width=0.45\textwidth]{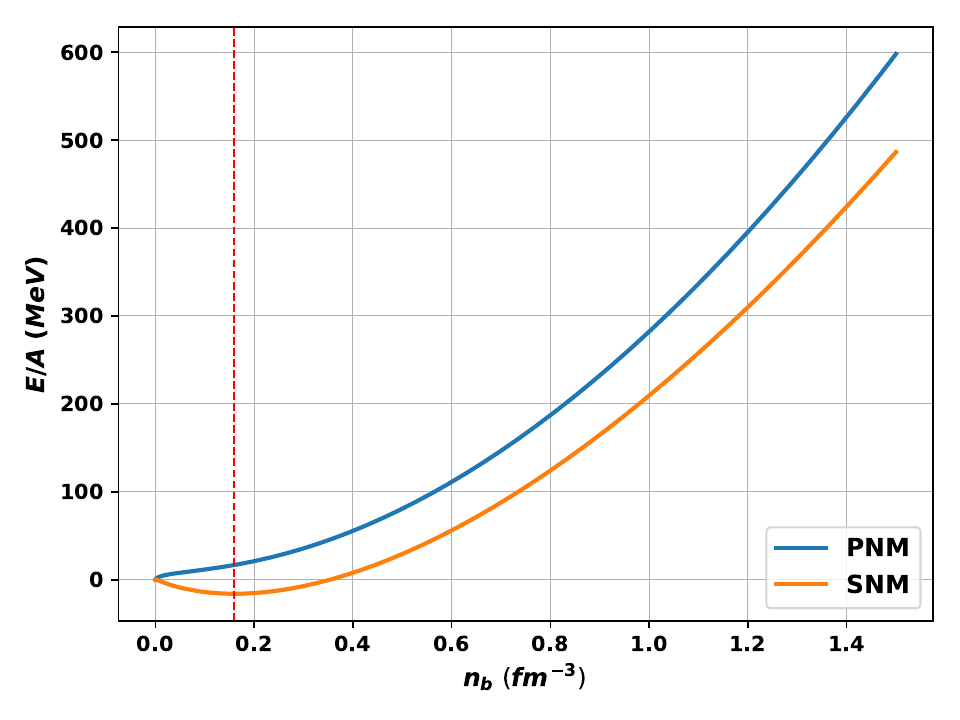}
    \caption{Energy per nucleon as a function of the baryon number density ($n_{b}$) for the Skyrme model corresponding to the SLy4 EOS, shown for symmetric nuclear matter (orange) and pure neutron matter (blue). The vertical red dashed line marks the nuclear saturation density ($n_{b0} \approx 0.1595 \text{ fm}^{-3}$).}
\label{fig:sly4_EbyA}
\end{figure}

Fig.~\ref{fig:sly4_EbyA} shows the energy per nucleon ($\frac{E}{A}$) as a function of the baryon number density ($n_b$) determined from Eq.~\ref{E/A}. We plot both the isospin-symmetric ($I=0$) nuclear matter (henceforth, SNM) and the maximally isospin-asymmetric case ($I=1$) representing pure neutron matter (henceforth, PNM).

\subsection{Symmetry energy at different orders}\label{sym_en_diff_order_}

\begin{figure}[htbp]
    \centering
    \includegraphics[width=0.5\textwidth]{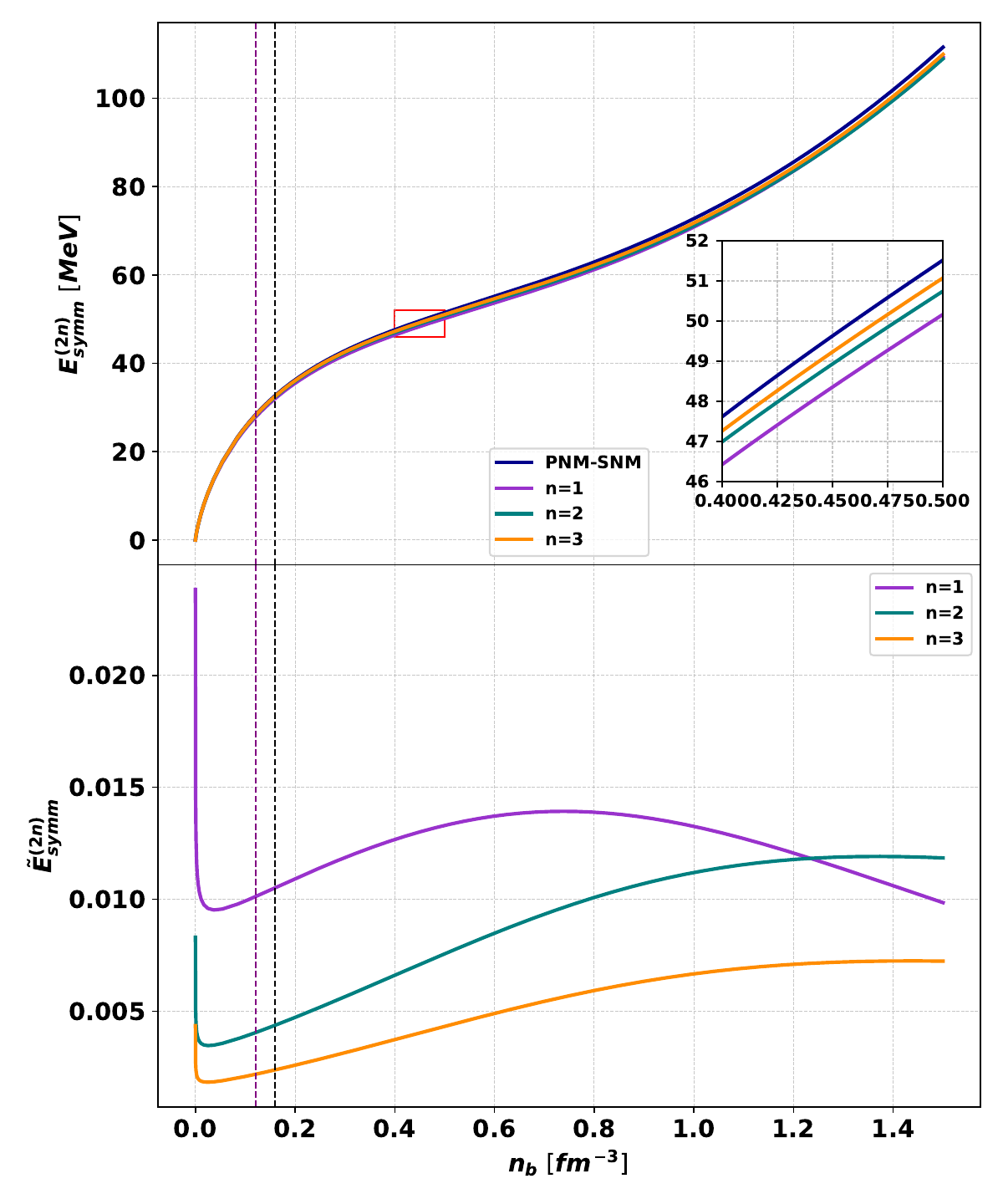}
    \caption{{\it Top panel:} Symmetry energy ($E_{\text{sym}}^{(2n)}$) of homogeneous nuclear matter over a wide range of baryon number density $n_{b}$ shown for different correction orders $n = 1, 2, 3$, and compared with the exact value $E_{\text{sym}}^{(\text{exact})} \equiv E_{\text{PNM}} - E_{\text{SNM}}$ for the SLy4 parametrization. Detail regarding Eq.~\ref{asymm-energy_i} is provided in Sec.~\ref{sym_en_diff_order}. {\it Bottom panel:} Fractional differences between the symmetry energy at various correction orders $E_{\text{sym}}^{(2n)}$ and its exact value $E_{\text{sym}}^{(\text{exact})}$. The red and olive colored vertical dashed lines mark saturation ($n_{b0} \approx 0.1595 \text{ fm}^{-3}$) and muon onset ($n_b = 0.121004~\text{fm}^{-3}$) density, respectively.}
\label{fig:sym_en_diff_order_sly4}
\end{figure}

The Taylor series expansion of $\frac{E}{A}(n_b, I)$ about $I = 0$ for powers of the isospin asymmetry parameter $I$ at a fixed $n_b$ can be expressed as 
\begin{equation}
\begin{aligned}\label{E-by-A_taylor_skyrme}
    \frac{E}{A}(n_b, I) = \frac{E}{A}(n_b, I=0) + a_s^{(2)}(n_b)I^2 + a_s^{(4)}(n_b)I^4 + \\
    a_s^{(6)}(n_b)I^6 + ... + a_s^{(2n)}(n_b)I^{2n} ... 
\end{aligned}
\end{equation}
with 
\begin{equation}\label{asymm-O(1)}
\begin{split}
    a_s^{(2)}(n_b) =& \frac{1}{2}\frac{\partial^2 [\frac{E}{A}(n_b)]}{\partial I^2} \bigg|_{I=0} \\
    =& \frac{1}{3}\frac{\hbar^2}{2m}\left( \frac{3\pi^2}{2} \right)^{\frac{2}{3}} 
    n_b^{\frac{2}{3}} - \frac{1}{8}t_0(2x_0+1)n_b \\
    -& \frac{1}{24}\left( \frac{3\pi^2}{2} \right)^{\frac{2}{3}} (3t_1x_1 - t_2(4+5x_2))n_b^{\frac{5}{3}} \\
    -& \frac{1}{48}t_3(2x_3+1)n_b^{\iota+1} \\
\end{split}
\end{equation}

\begin{equation}\label{asymm-O(2)}
\begin{aligned}
    a_s^{(4)}(n_b) =& \frac{1}{4!}\frac{\partial^4 [\frac{E}{A}(n_b)]}{\partial I^4} \bigg|_{I=0} \\
    =& \frac{1}{81}\frac{\hbar^2}{2m}\left( \frac{3\pi^2}{2} \right)^{\frac{2}{3}} n_b^{\frac{2}{3}} \\
    +& \frac{1}{648}\left( \frac{3\pi^2}{2} \right)^{\frac{2}{3}}(3t1(1+x_1)+t_2(1-x_2))n_b^{\frac{5}{3}}
\end{aligned}
\end{equation}

\begin{equation}\label{asymm-O(3)}
\begin{aligned}
    a_s^{(6)}(n_b) =& \frac{1}{6!}\frac{\partial^6 [\frac{E}{A}(n_b)]}{\partial I^6} \bigg|_{I=0} \\
    =& \frac{7}{2187}\frac{\hbar^2}{2m}\left( \frac{3\pi^2}{2} \right)^{\frac{2}{3}} n_b^{\frac{2}{3}} \\
    +& \frac{7}{87480}\left( \frac{3\pi^2}{2} \right)^{\frac{2}{3}}(3t1(4+3x_1)+t_2(8+x_2))n_b^{\frac{5}{3}}
\end{aligned}
\end{equation}

\begin{figure}[htbp]
    \centering
    \includegraphics[width=0.5\textwidth]{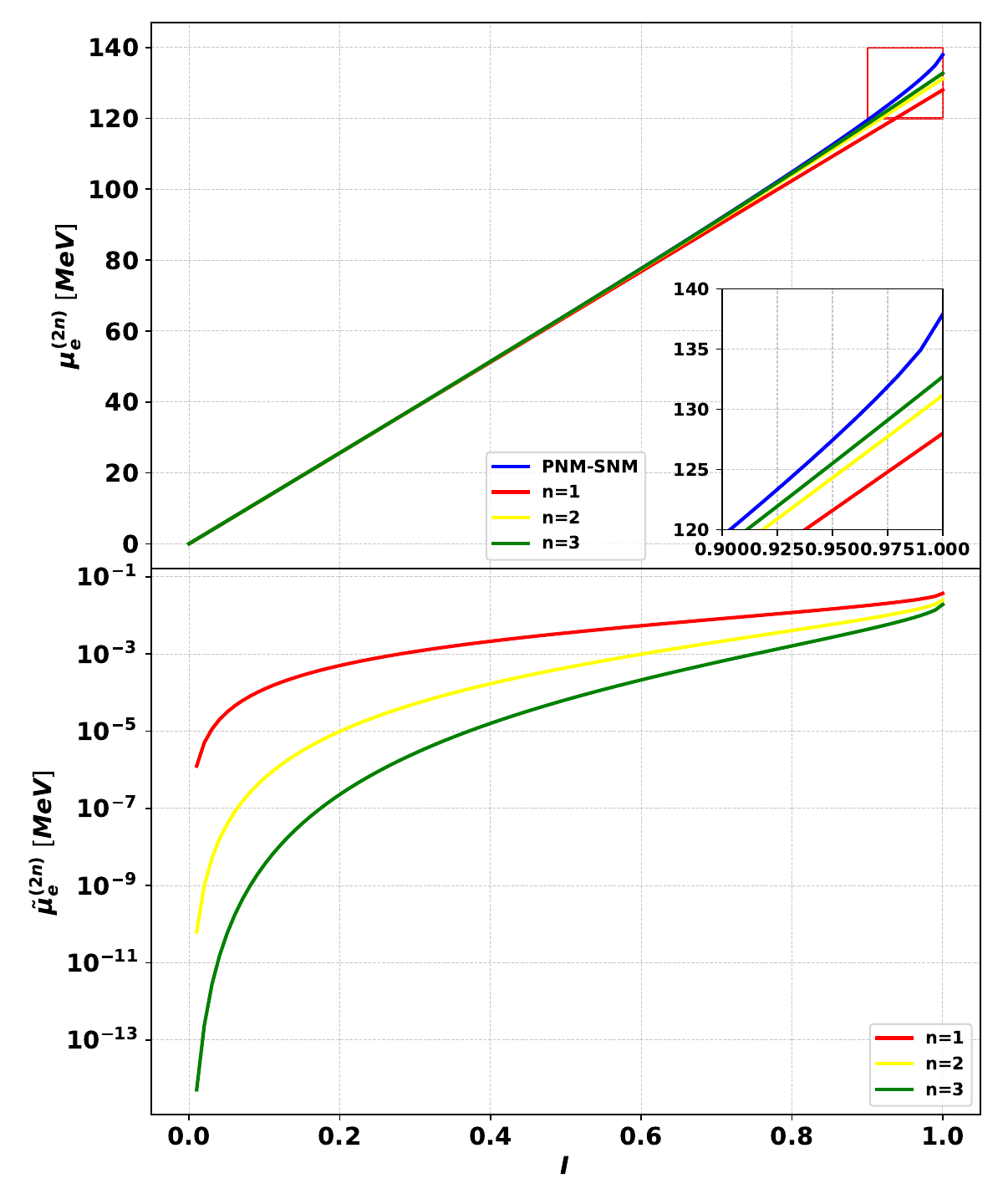}
    \caption{The difference between neutron and proton chemical potentials ($\Delta\mu_{np} \equiv \mu_n - \mu_p$) for various values of the isospin asymmetry parameter at the nuclear saturation density ($n_{b0}$) using SLy4 interaction parameters corresponding to different correction orders $\mathcal{O}(I^{2n})$, see Eq.~\ref{chemical-potential-order-corr}, ~\ref{chemical-potential-exact}.} 
    \label{fig:sly4_mu_diff}
\end{figure}

The nuclear symmetry energy parameter ($E_{\text{sym}}$), or \emph{symmetry energy} for short, measures the excess energy in pure neutron matter (PNM) relative to isospin-symmetric nuclear matter (SNM). This corresponds to the difference in ground-state energy values between PNM and SNM at a fixed baryon density $n_b$. Formally, the symmetry energy ($E_{\text{sym}}$) of nuclear matter is defined in literature using two distinct approaches: (i) as the difference between the average single-particle energy of PNM and SNM at a fixed baryon number density ($n_{b}$), namely,
\begin{equation}\label{sym_energy_exact-order} 
    E_{\text{sym}}(n_b) = \frac{E}{A}(n_b, I=1) - \frac{E}{A}(n_b, I=0) 
\end{equation}
(see~\citet{diff_sym_en_LattimerGroup_prc2024}); or (ii) expansionally via the curvature term
\begin{equation}\label{sym_energy_quadratic-order} 
     E_{\text{sym}}(n_b) = \frac{1}{2}\left.\frac{\partial^2 \left[\frac{E}{A}(n_b)\right]}{\partial I^2}\right|_{I=0}
\end{equation}
(see~\citet{Chabanat_etal_1997NuPhA, diff_sym_en_LattimerGroup_prc2024}). Henceforth, we refer to the first definition as the exact calculation of symmetry energy, denoted by $E_{\text{sym}}^{(\text{exact})}$, and the second definition as the $n=1$ order symmetry energy expansion, denoted more generally by $E_{\text{sym}}^{(2n)}$ as elaborated below.

According to definition (i), the symmetry energy at different baryon densities accurate up to $\mathcal{O}(I^{2n})$ can be expressed as:
\begin{equation}\label{asymm-energy_i}
\begin{split}
    E_{\text{sym}}^{(2n)}(n_b) &= \frac{E}{A}(n_b, I=1) - \frac{E}{A}(n_b, I=0) \\
    &= a_s^{(2)}(n_b) + a_s^{(4)}(n_b) + a_s^{(6)}(n_b) + \dots + a_s^{(2n)}(n_b)
\end{split}
\end{equation}
Therefore, in the limiting case where $n \to \infty$, we recover the exact symmetry energy:
\begin{equation}
    E_{\text{sym}}^{(\text{exact})} = \lim_{n \to \infty} E_{\text{sym}}^{(2n)}
\end{equation}

Conversely, according to definition (ii), the symmetry energy is truncated as:
\begin{equation}
    E_{\text{sym}}^{(2)}(n_b) = \left.\frac{1}{2}\frac{\partial^2 \left[\frac{E}{A}(n_b)\right]}{\partial I^2}\right|_{I=0} = a_s^{(2)}(n_b)
\end{equation}
which represents only the leading-order term in the $I^{2n}$ series expansion (see Eq.~\ref{asymm-energy_i}).

In Fig.~\ref{fig:sym_en_diff_order_sly4}, we illustrate the variation of symmetry energy with baryon number density across several correction orders of $\mathcal{O}(I^{2n})$ derived from Eq.~\ref{E-by-A_taylor_skyrme}.

\section{Equation of state for cold $\beta$-equilibrated nucleonic matter}\label{beta_quilb_matter}

We consider our system describing neutron star matter to consist of neutrons ($n$), protons ($p$), electrons ($e$), and muons ($\mu$) only, denoted in short as $npe\mu$ matter. The total energy density $\epsilon$, which includes the rest-mass energy of the constituent particles, is composed of a nucleonic (i.e., baryonic) contribution $\epsilon_b(n_n,n_p)$ and a leptonic contribution $\epsilon_l$ due to electrons and muons. It is given by
\begin{equation}
\begin{aligned}\label{total_en_den}
    \epsilon(n_n,n_p,n_e,n_{\mu}) =& [n_n m_n c^2 + n_p m_p c^2 + \varepsilon_b(n_n, n_p)] \\
    +& \epsilon_e(n_e) + \epsilon_{\mu}(n_{\mu}) \\
    =& \epsilon_b(n_n, n_p) + \epsilon_e(n_e) + \epsilon_{\mu}(n_{\mu})
\end{aligned}
\end{equation}

where $\epsilon_e(n_e)$ and $\epsilon_{\mu}(n_{\mu})$ are the relativistic total energy densities for electrons and muons, respectively, and $\epsilon_b(n_n, n_p)$ explicitly incorporates the interaction energy density contribution from the baryonic sector\footnote{We use $\rho$ to denote the mass density and $\epsilon$ the energy density. In the present analysis, both quantities are reported in units of $\rm MeV$ $\rm fm^{-3}$, and are therefore used interchangeably in some figures and discussions where only the density scale is relevant.}

The charge neutrality condition of the $npe\mu$ matter requires
\begin{align}\label{eqm1_charge_neutrality}
    n_p = n_e + n_{\mu}
\end{align}
Furthermore, the equilibrium condition for $npe\mu$ matter with respect to weak interactions imposes constraints involving the chemical potentials of the different particle species: 
\begin{align}\label{eqm1_chemical_pot}
    \mu_n = \mu_p + \mu_e
\end{align}
\begin{align}\label{eqm2}
    \mu_{\mu} = \mu_e
\end{align}
where the chemical potential $\mu_j$ of a particle species $j$ is defined as 
\begin{align}\label{chemical-potential-0}
    \mu_j = \left(\frac{\partial \epsilon}{\partial n_j}\right)_{n_{i \neq j}}  
\end{align}
These relations combine to yield
\begin{align}\label{chemical-potential-1}
    \mu_{\mu} = \mu_e = \mu_n - \mu_p
\end{align}

Now, the chemical potential of the neutron ($\mu_n$) can be expressed using the chain rule as
\begin{equation}\label{chem_pot_neutron_exact} 
\begin{split}
    \mu_n =& \bigg(\frac{\partial \epsilon}{\partial n_n}\bigg)_{n_p, n_e, n_\mu} = \bigg(\frac{\partial \epsilon_b}{\partial n_n}\bigg)_{n_p} = \bigg(\frac{\partial (n_b \frac{E}{A})}{\partial n_n}\bigg)_{n_p} \\
    =& n_b \bigg(\frac{\partial (\frac{E}{A})}{\partial n_n}\bigg)_{n_p} + \bigg(\frac{\partial n_b}{\partial n_n}\bigg)_{n_p}\frac{E}{A}(n_b, I) \\
    =& n_b \bigg(\frac{\partial (\frac{E}{A})}{\partial I}\bigg)_{n_b} \bigg(\frac{\partial I}{\partial n_n}\bigg)_{n_p} + n_b \bigg(\frac{\partial (\frac{E}{A})}{\partial n_b}\bigg)_I \bigg(\frac{\partial n_b}{\partial n_n}\bigg)_{n_p} \\
    +& \frac{E}{A}(n_b, I) \\
    =& n_b \bigg(\frac{\partial (\frac{E}{A})}{\partial I}\bigg)_{n_b} \bigg(\frac{\partial I}{\partial n_n}\bigg)_{n_p} + n_b \bigg(\frac{\partial (\frac{E}{A})}{\partial n_b}\bigg)_I + \frac{E}{A}(n_b, I) 
\end{split}
\end{equation}
and similarly, the chemical potential of the proton ($\mu_p$) expands to
\begin{equation}
\begin{aligned}\label{chem_pot_proton_exact}  
    \mu_p =& \bigg(\frac{\partial \epsilon}{\partial n_p}\bigg)_{n_n, n_e, n_\mu} = \bigg(\frac{\partial \epsilon_b}{\partial n_p}\bigg)_{n_n} = \bigg(\frac{\partial (n_b \frac{E}{A})}{\partial n_p}\bigg)_{n_n} \\
    =& n_b \bigg(\frac{\partial (\frac{E}{A})}{\partial n_p}\bigg)_{n_n} + \bigg(\frac{\partial n_b}{\partial n_p}\bigg)_{n_n}\frac{E}{A}(n_b, I) \\
    =& n_b \bigg(\frac{\partial (\frac{E}{A})}{\partial I}\bigg)_{n_b} \bigg(\frac{\partial I}{\partial n_p}\bigg)_{n_n} + n_b \bigg(\frac{\partial (\frac{E}{A})}{\partial n_b}\bigg)_I \bigg(\frac{\partial n_b}{\partial n_p}\bigg)_{n_n} \\
    +& \frac{E}{A}(n_b, I) \\
    =& n_b \bigg(\frac{\partial (\frac{E}{A})}{\partial I}\bigg)_{n_b} \bigg(\frac{\partial I}{\partial n_p}\bigg)_{n_n} + n_b \bigg(\frac{\partial (\frac{E}{A})}{\partial n_b}\bigg)_I + \frac{E}{A}(n_b, I) \\
\end{aligned}
\end{equation}

\begin{figure}[htbp]
    \centering
    \includegraphics[width=0.5\textwidth]{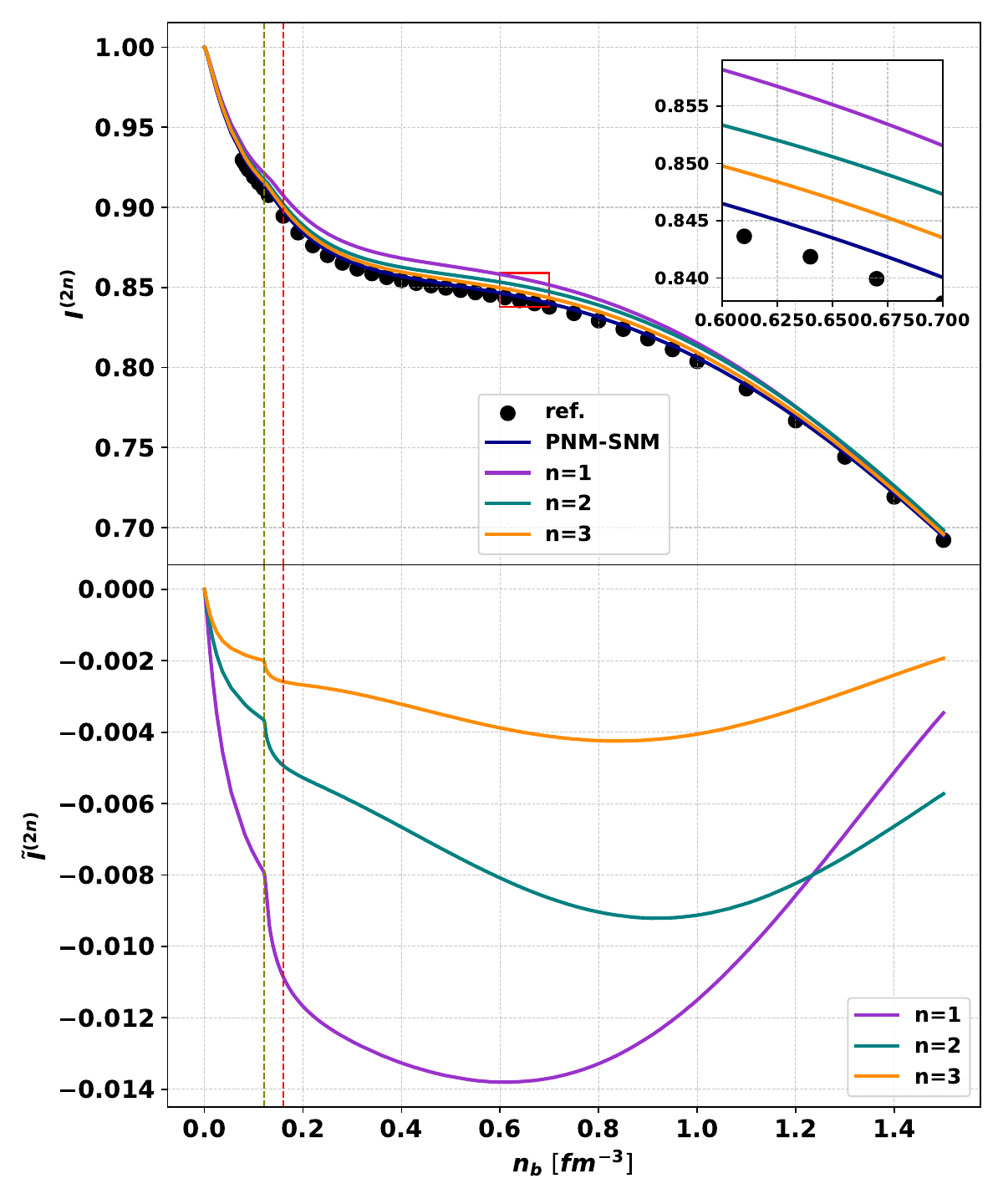}
    \caption{The isospin asymmetry parameter $I$ (defined in Eq.~\ref{isospin-n-p}) at $\beta$-equilibrium shown over the relevant range of baryon density $n_{b} \in [10^{-6}, 1.5] \text{ fm}^{-3}$, comparing various orders of corrections to the symmetry energy $E_{\text{sym}}^{(2n)}$ against the exact evaluation for the SLy4 EOS. The black dots denote the previous results reported in~\citet{DouchinHaensel_AA2001}. The red- and olive-colored vertical dashed lines mark saturation ($n_{b0} \approx 0.1595 \text{ fm}^{-3}$) and muon onset ($n_b = 0.121004~\text{fm}^{-3}$) density, respectively.}
    \label{fig:sly4_asymm}
\end{figure}

\subsection{Symmetry energy at different orders of beta-equilibrated matter}\label{sym_en_diff_order}

As mentioned in Sec.~\ref{sym_en_diff_order_}, following convention (i), the symmetry energy can be defined as the difference between the energy per particle of PNM and SNM. A Taylor series expansion of $\frac{E}{A}(n_b, I)$ in powers of the isospin asymmetry parameter ($I$) about $I = 0$ (i.e., symmetric nuclear matter where $n_n = n_p$) yields
\begin{equation}
\begin{aligned}\label{E/A-taylor}
    \frac{E}{A}(n_b, I) =& \frac{E}{A}(n_b, I=0) + a_s^{(2)}(n_b) I^2 + a_s^{(4)}(n_b) I^4 \\
    +& a_s^{(6)}(n_b) I^6 + ... + a_s^{(2n)}(n_b)I^{2n} + ... 
\end{aligned}
\end{equation}
Thus, we obtain the derivative
\begin{equation}
\begin{aligned}\label{chemical-potential-n-3}
    \bigg(\frac{\partial [\frac{E}{A}(n_b, I)]}{\partial I}\bigg)_{n_b} =& 2a_s^{(2)}(n_b)I + 4a_s^{(4)}(n_b)I^3 + 6a_s^{(6)}(n_b)I^5 +\\ 
    & ... + 2na_s^{(2n)}(n_b)I^{2n-1} + ... 
\end{aligned}
\end{equation}
along with the following partial density gradients:
\begin{equation}
\begin{aligned}\label{chemical-potential-n-4}
    \bigg( \frac{\partial I}{\partial n_n} \bigg)_{n_p} =& \bigg( \frac{\partial \big[\frac{(n_n - n_p)}{(n_n + n_p)}\big]}{\partial n_n} \bigg)_{n_p} = \frac{(n_n + n_p)-(n_n -n_p)}{(n_n+n_p)^2} \\
    =& \frac{2 n_p}{n_b^2} \\
    \bigg( \frac{\partial I}{\partial n_p} \bigg)_{n_n} =& \bigg( \frac{\partial \big[\frac{(n_n - n_p)}{(n_n + n_p)}\big]}{\partial n_p} \bigg)_{n_n} = \frac{-(n_n + n_p)-(n_n - n_p)}{(n_n+n_p)^2} \\
    =& -\frac{2 n_n}{n_b^2}
\end{aligned}
\end{equation}

The difference between the neutron and proton chemical potentials, $\Delta\mu_{np} \equiv \mu_n - \mu_p$, can be evaluated using Eq.~\ref{chem_pot_neutron_exact} and Eq.~\ref{chem_pot_proton_exact} as
\begin{equation}\label{chemical-potential-n-final} 
\begin{split}
    & (\mu_n - \mu_p) \\
    & = n_b \bigg(\frac{\partial (\frac{E}{A})}{\partial I}\bigg)_{n_b} \bigg[ \bigg(\frac{\partial I}{\partial n_n}\bigg)_{n_p} - \bigg(\frac{\partial I}{\partial n_p}\bigg)_{n_n} \bigg] \\
    & = n_b \bigg[ 2a_s^{(2)}(n_b)I + 4a_s^{(4)}(n_b)I^3 + 6a_s^{(6)}(n_b)I^5 +\\ 
    & ... + 2na_s^{(2n)}(n_b)I^{2n-1} + ... \bigg] \bigg[\frac{2 (n_p + n_n)}{n_b^2} \bigg] \\
    & = 2 \bigg[ 2a_s^{(2)}(n_b)I + 4a_s^{(4)}(n_b)I^3 + 6a_s^{(6)}(n_b)I^5 +\\ 
    & ... + 2na_s^{(2n)}(n_b)I^{2n-1} + ... \bigg] \\
    & = 4a_s^{(2)}(n_b)I + 8a_s^{(4)}(n_b)I^3 + 12a_s^{(6)}(n_b)I^5 +\\ 
    & ... + 4na_s^{(2n)}(n_b)I^{2n-1} + ... \\
\end{split}
\end{equation}

Thus, the difference between the nucleon chemical potentials can be written in a compact form up to the $\mathcal{O}(I^{2n})$ correction order as 
\begin{equation}\label{chemical-potential-order-corr}
    \Delta\mu_{np}^{(2n)} = (\mu_n - \mu_p)^{(2n)} = \sum_{k = 1}^{n} 4k a_s^{(2k)}(n_b)I^{2k-1}
\end{equation}

Note that the nuclear symmetry energy is defined as the energy difference between pure neutron matter and symmetric nuclear matter, $E_{\text{sym}}(n_b) = \frac{E}{A}(n_b, I=1) - \frac{E}{A}(n_b, I=0)$, as per our definition-(i) (see Sec.~\ref{sym_en_diff_order}), which matches the exact evaluation of $(\mu_n - \mu_p)$ at full asymmetry. Moreover, the exact expression for $(\mu_n - \mu_p)^{(\text{exact})}$ can be derived directly from Eq.~\ref{E/A} and Eq.~\ref{asymm-fact} without any expansion approximations:
\begin{equation}\label{chemical-potential-exact}
\begin{split}
    & (\mu_n - \mu_p)^{(exact)} \\
    =& n_b \bigg(\frac{\partial (\frac{E}{A})}{\partial I}\bigg)_{n_b} \bigg[ \bigg(\frac{\partial I}{\partial n_n}\bigg)_{n_p} - \bigg(\frac{\partial I}{\partial n_p}\bigg)_{n_n} \bigg] \\
    =& 2 \bigg(\frac{\partial (\frac{E}{A})}{\partial I}\bigg)_{n_b} \\
    =& \frac{3}{5}\frac{\hbar^2}{2m}\left( \frac{3\pi^2}{2} \right)^{\frac{2}{3}}n_b^{\frac{2}{3}} G_{5/3}(I) \\
    -& \frac{1}{8}t_0n_b(2x_0+1) G_2(I) - \frac{1}{48} t_3 n_b^{\iota+1}(2x_3+1) G_2(I) \\ 
    +& \frac{3}{40}\left( \frac{3\pi^2}{2} \right)^{\frac{2}{3}}n_b^{\frac{5}{3}} 
    \{ [t_1(x_1+2)+t_2(x_2+2)] G_{5/3}(I) \\
    +& \frac{1}{2}[t_2(2x_2+1)-t_1(2x_1+1)] G_{8/3}(I) \}
\end{split}
\end{equation}
where
\begin{equation}\label{H(m,I)}
\begin{split}
    G_N(I) &= 2 \left(\frac{\partial [F_N(I)]}{\partial I}\right)_{n_b} \\
    &= N\left[(1+I)^{N-1} - (1-I)^{N-1}\right] \\
    &= 2N F_{N-1}(I)
\end{split}
\end{equation}

For a fixed baryon number density $n_b$, the isospin asymmetry parameter $I$ is obtained numerically by solving the coupled conditions for charge neutrality and $\beta$-equilibrium of $npe\mu$ matter. The solution is determined self-consistently using an iterative root-finding procedure applied to Eq.~\ref{chemical-potential-order-corr} (or Eq.~\ref{chemical-potential-exact} for the exact treatment), together with the charge neutrality Eq.~\ref{eqm1_charge_neutrality} and the leptonic equilibrium relations Eq.~\ref{chemical-potential-1}. Details of the numerical convergence tests are presented in Appendix~\ref{app_solver}.

\subsection{Constructing leptonic sector for the beta-equilibrated matter}\label{leptonic_sector_beta-equilb} 

Baryon number conservation, $n_b = n_n + n_p$, and charge neutrality, $n_p = n_e + n_{\mu}$, are maintained throughout the interior of neutron stars. Subject to these constraints, the isospin asymmetry parameter $I$ at a fixed baryon number density $n_b$ can be self-consistently determined for stellar matter under $\beta$-equilibrium (see Eq.~\ref{chemical-potential-1}). The baryonic pressure of this matter is then given by
\begin{align}\label{baryonic_pressure}
    P_b(I, n_b) = n_b^2 \frac{\partial \left(\frac{E}{A}\right)}{\partial n_b}
\end{align}

For the leptonic sector, the pressure contribution from a single lepton species $l$ is expressed as
\begin{align}\label{leptonic_pressure}
    P_l = n_l^2 \frac{\partial \left(\frac{\epsilon_l}{n_l}\right)}{\partial n_l} = n_l \frac{\partial \epsilon_l}{\partial n_l} - \epsilon_l = n_l \mu_l - \epsilon_l
\end{align}
where the total energy density of a relativistic fermion species is defined by the standard integration:
\begin{align}\label{relativistic_fermion}
    \epsilon = g \int_0^{k_f} \sqrt{\hbar^2 k^2 + m^2} \frac{d^3 k}{(2\pi)^3}
\end{align}
For a spin degeneracy of $g=2$ and introducing the dimensionless parameter $x = \frac{\hbar k_f}{m}$, this expression simplifies to
\begin{align}\label{relativistic_fermion-1}
    \epsilon = \frac{m^4 c^5}{8\pi^2 \hbar^3}\left[ x\sqrt{1+x^2}(2x^2 + 1) - \ln\left(x + \sqrt{1+x^2}\right) \right]
\end{align}

\begin{figure*}[htbp]
    \raggedright (a)\hspace*{\columnwidth}(b)\\[-0.5cm]
    \includegraphics[clip,width=\columnwidth]{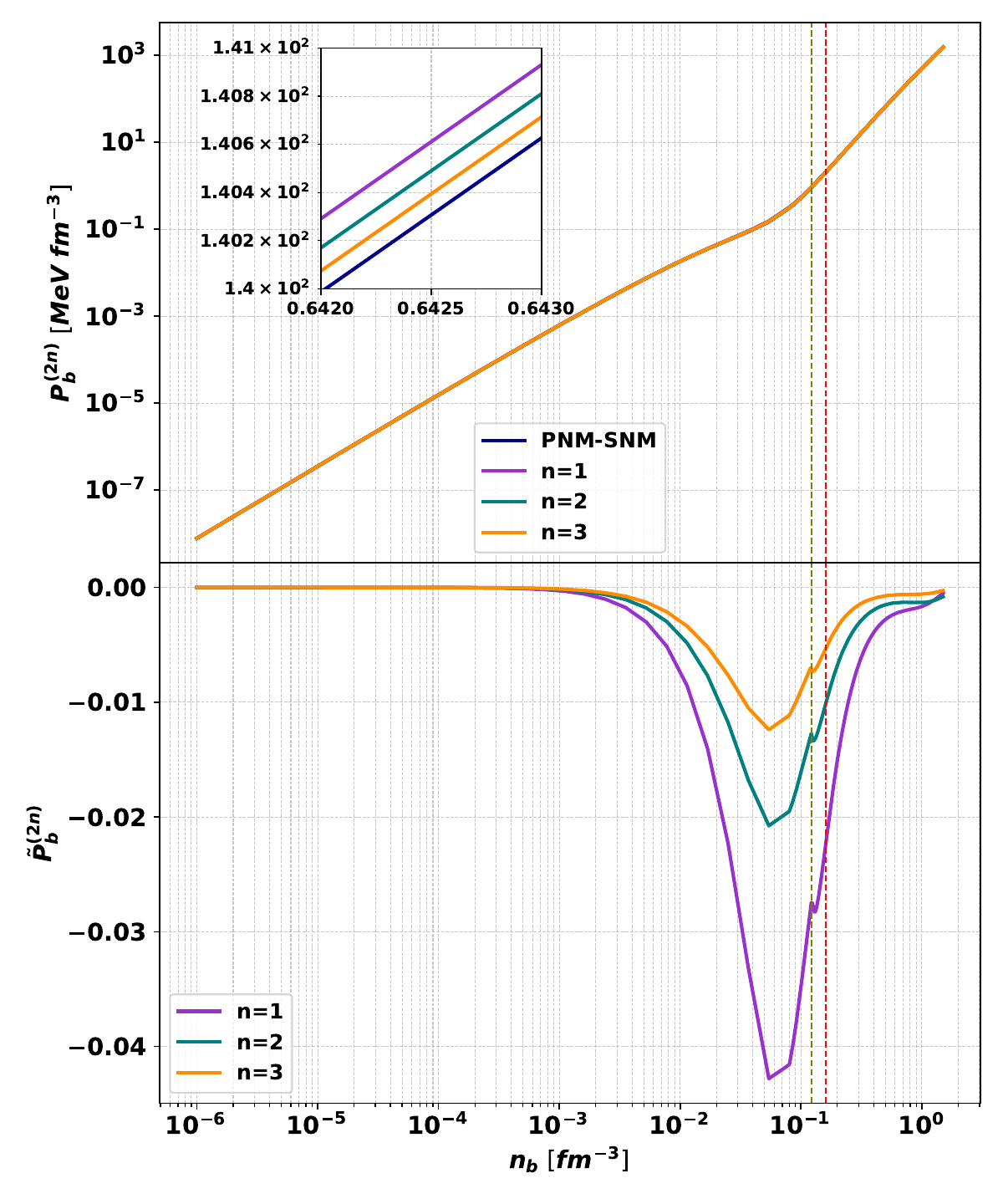}
    \includegraphics[clip,width=\columnwidth]{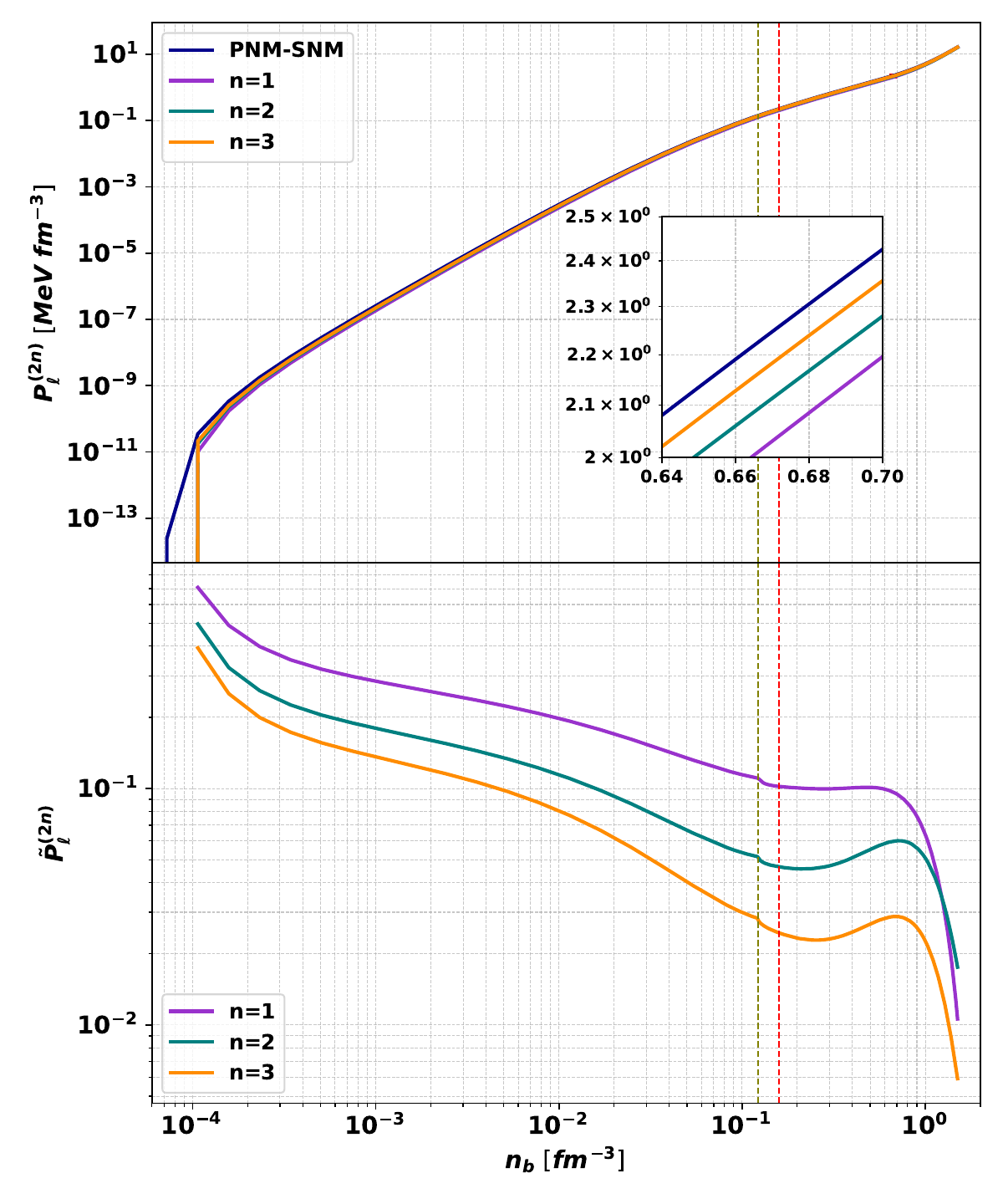} \\
    \caption{Variations in the baryonic pressure $P_{b}^{(2n)}$ (in \textit{top-left panel}) and the leptonic pressure $P_{l}^{(2n)}$ (in \textit{top-right panels}) of neutron star matter at $\beta$-equilibrium, evaluated using different correction orders $\mathcal{O}(I^{2n})$ of the symmetry energy $E_{\text{sym}}^{(2n)}$ for the SLy4 Skyrme parametrization, are presented. The corresponding lower panels show the fractional differences $\tilde{P}_{b}^{(2n)}$ and $\tilde{P}_{l}^{(2n)}$ plotted over a wide range of baryon number densities ($n_{b}$) relevant to neutron star interiors. The dashed vertical line in red color denotes the saturation density, and the dashed vertical line in olive color denotes the onset density for muon production.}
\label{fig:sly4_qulb_pres}
\end{figure*}

By imposing the conditions for $\beta$-equilibrium, charge neutrality, and baryon number conservation simultaneously, the total pressure of the neutron star matter at a given baryon number density $n_b$ becomes
\begin{align}\label{beta_qulb_pressure}
    P(n_b) = P_b(I, n_b) + P_{e}(n_{e}) + P_{\mu}(n_{\mu})
\end{align}

Equations~\ref{total_en_den} and \ref{beta_qulb_pressure} constitute the $\beta$-equilibrated equation of state (EOS) that describes cold neutron star matter in realistic astrophysical scenarios.

\subsection{Fractional deviation in different physical quantities}\label{frac_dev_phys_qunt} 

Here, we establish a few useful notations. Throughout this work, we compute various physical quantities (denoted generically as $\theta$) corresponding to different orders of isospin $\mathcal{O}(I^{2n})$ corrections in the symmetry energy $E_{\text{sym}}^{(2n)}$, labeled as $\theta^{(2n)}$. We thoroughly investigate both the qualitative and quantitative deviations in $\theta^{(2n)}$ arising from higher-order $\mathcal{O}(I^{2n})$ corrections (for $n = 1, 2, 3$), along with their \emph{corresponding exact values} $\theta^{(\text{exact})}$ derived using definition-(i) of the exact symmetry energy $E_{\text{sym}}^{(\text{exact})}$ (see Eq.~\ref{sym_energy_exact-order}). To evaluate these deviations, we define a quantitative fractional difference measure $\tilde{\theta}^{(2n)}$ as
\begin{equation}\label{frac_diff_theta-tilde}
    \tilde{\theta}^{(2n)} = \frac{\left[\theta^{(\text{exact})} - \theta^{(2n)}\right]}{\theta^{(\text{exact})}}
\end{equation}

This quantity $\tilde{\theta}^{(2n)}$ measures the fractional deviation of an approximated physical property relative to its exact counterpart $\theta^{(\text{exact})}$. For instance, following this notation, we illustrate the baryon density dependence of the symmetry energy deviations $\tilde{E}_{\text{sym}}^{(2n)}$ in Fig.~\ref{fig:sym_en_diff_order_sly4} for the case of the Skyrme interaction parametrization corresponding to the SLy4 EOS.

\section{Characteristics of higher order isospin corrections in the matter composition}\label{sec:higher_order_isospin}

\subsection{Higher order isospin corrections: the SLy4 case}\label{subsec:higher_order_isospin_sl4}

We have investigated the effect of higher order corrections of nuclear symmetry energy $E_{sym}^{(2n)}$ on various compositional and physical quantities in details. Here we summarize our key findings for the case of Skyrme interaction model parameters corresponding to SLy4 equation of state (EOS). 

\begin{table}[H]
\centering
\resizebox{\columnwidth}{!}{%
\begin{tabular}{l|ccccccccc}
\toprule
 EOS        & $t_0$ & $t_1$ & $t_2$ & $t_3$ & $x_0$ & $x_1$ & $x_2$ & $x_3$ & $\tau$ \\
\midrule
SLy4        & $-2488.91$ & $486.82$ & $-546.39$ & $13777.0$ & $0.834$ & $-0.3438$ & $-1.0$ & $1.354$ & $1/6$ \\
\bottomrule
\end{tabular}%
}
\caption{The parameter values of the Skyrme interaction model, as described in Section~\ref{sec:skyrme_model}, corresponding to the SLy4 EOS is explicitly provided in this table.}
\label{tab:skyrme_param_sly4}
\end{table}

\begin{figure*}[htbp]
    \raggedright (a)\hspace*{\columnwidth}(b)\\[-0.5cm]
    \includegraphics[clip,width=\columnwidth]{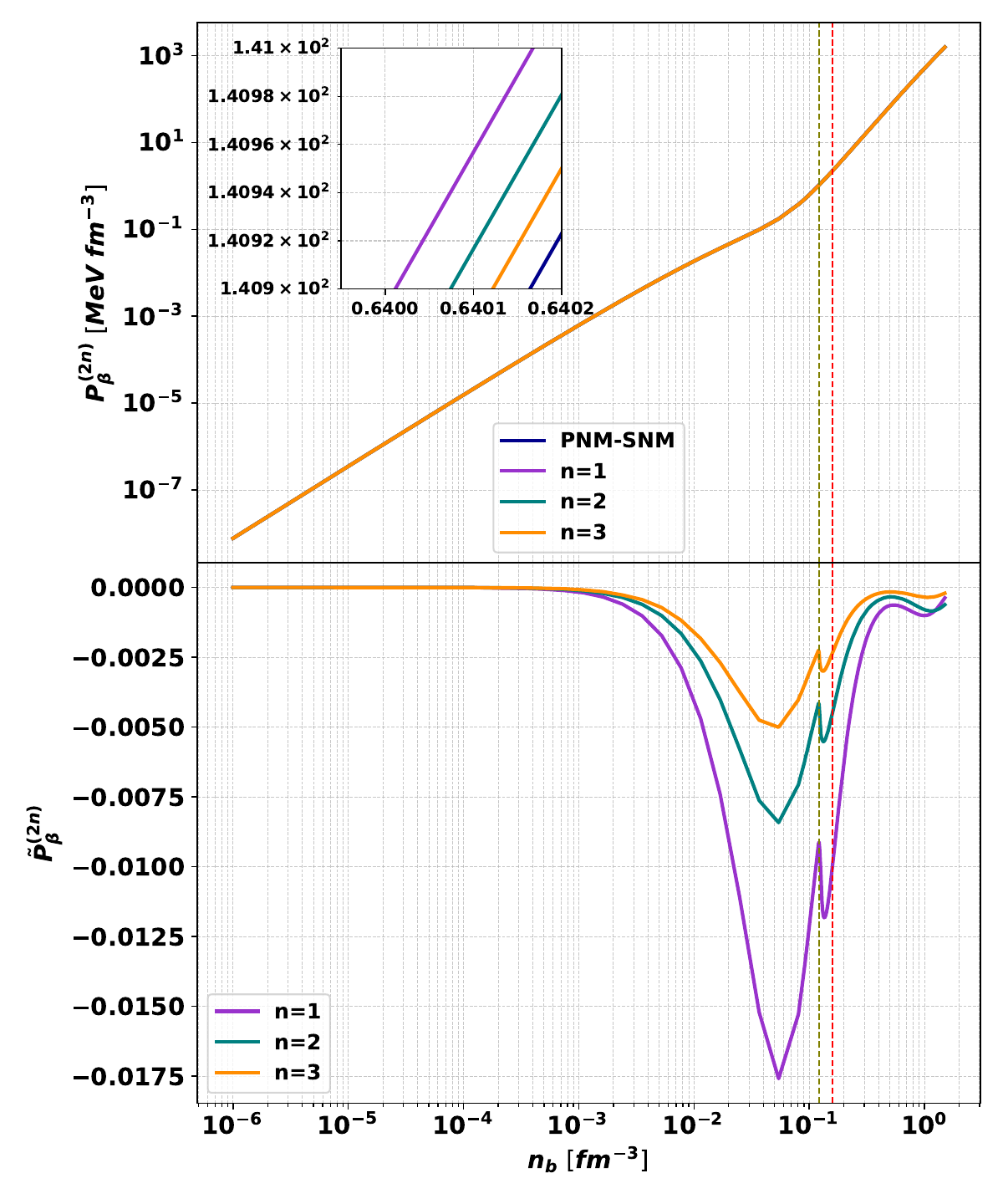} 
    \includegraphics[clip,width=\columnwidth]{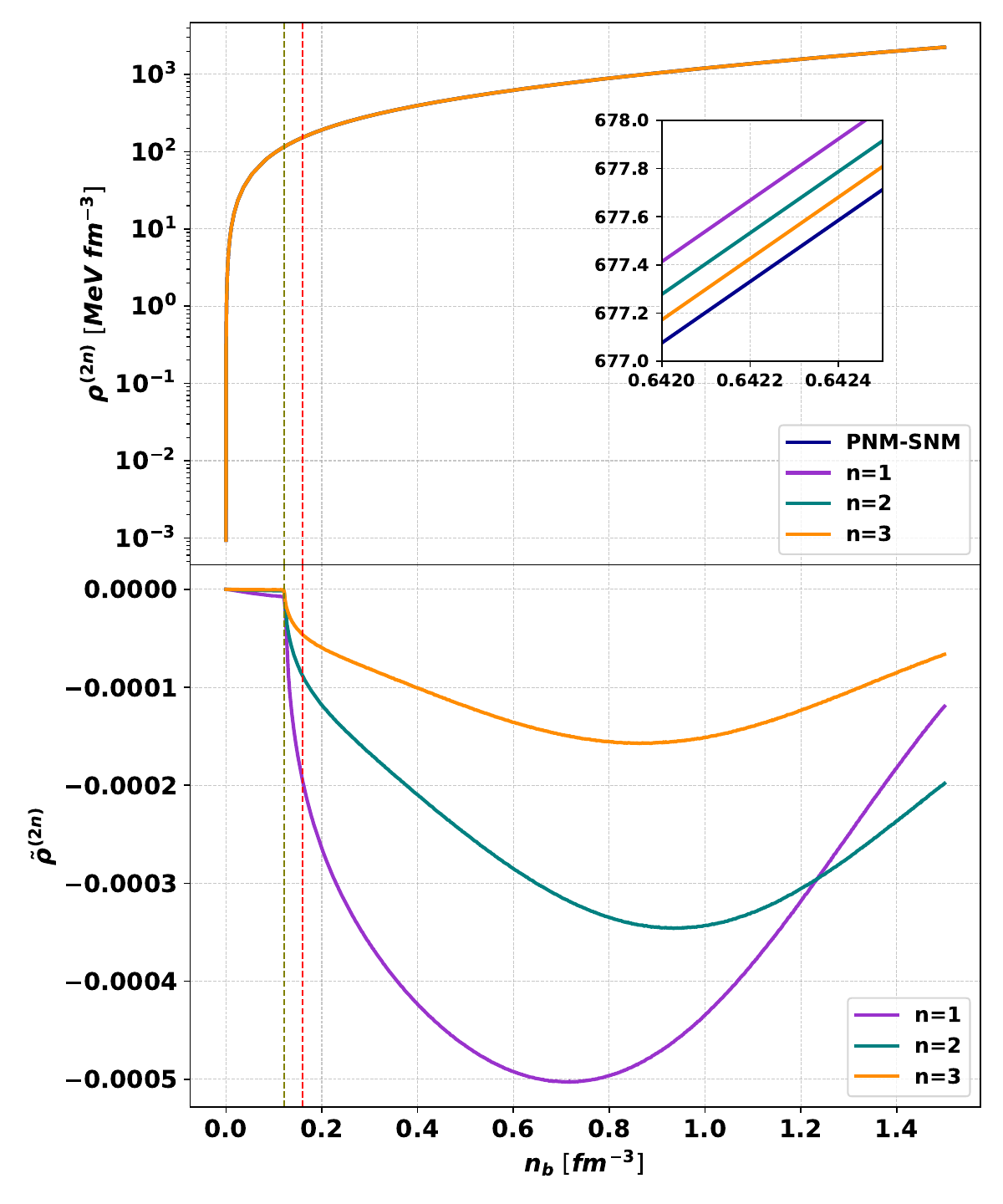} \\ 
    \caption{Total pressure $P_{\beta}^{(2n)}$ (in \textit{top-left panel}) and total energy density $\rho^{(2n)}$ (in \textit{top-right panel}) are presented over the same number density ($n_{b}$) range for $\beta$-equilibriated neutron star matter for different correction orders $\mathcal{O}(I^{2n})$ of  SLy4 Skyrme Parameters. The corresponding lower panels show the fractional differences $\tilde{P}_{\beta}^{(2n)}$ and $\tilde{\rho}^{(2n)}$ over the same $n_{b}$ range. 
    }\label{fig:sly4_p_beta}
\end{figure*}

\begin{figure}[!h]
    \centering
    \includegraphics[width=\columnwidth]{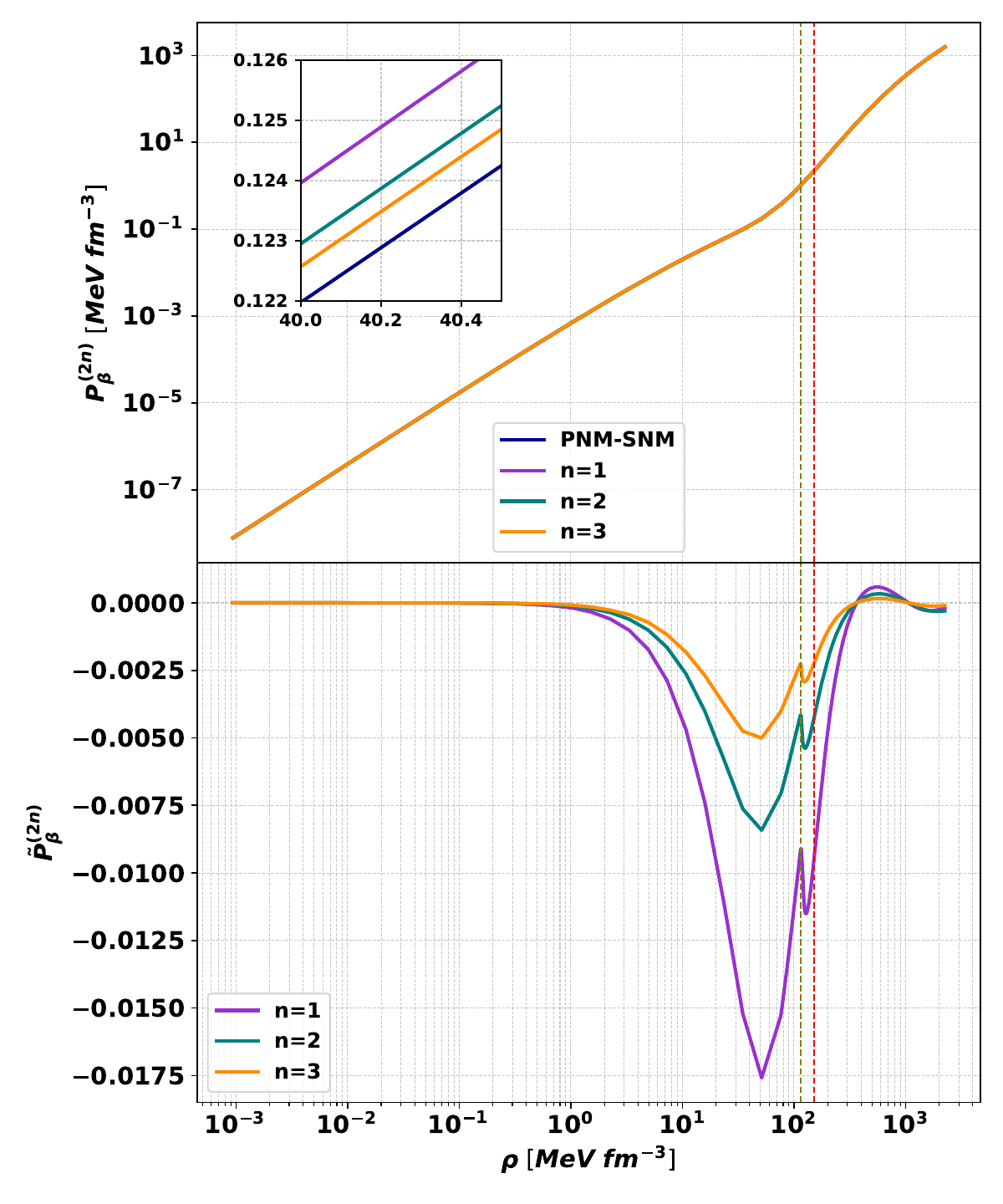}
    \caption{This figure illustrates the equation of state (EOS) obtained for neutron star matter in $\beta$-equilibrium corresponding to SLy4 Skyrme parameters at various correction orders $\mathcal{O}(I^{2n})$ for expansions of symmetry energy ($E_{\text{sym}})$: $exact$, $n=1$, $n=2$, and $n=3$. The bottom-panel presents corresponding fractional differences for various orders from the exact estimation.}
\label{fig:sly4_betaeqm_eos_wref}
\end{figure}

\begin{figure}[!h]
    \centering
    \includegraphics[width=0.5\textwidth]{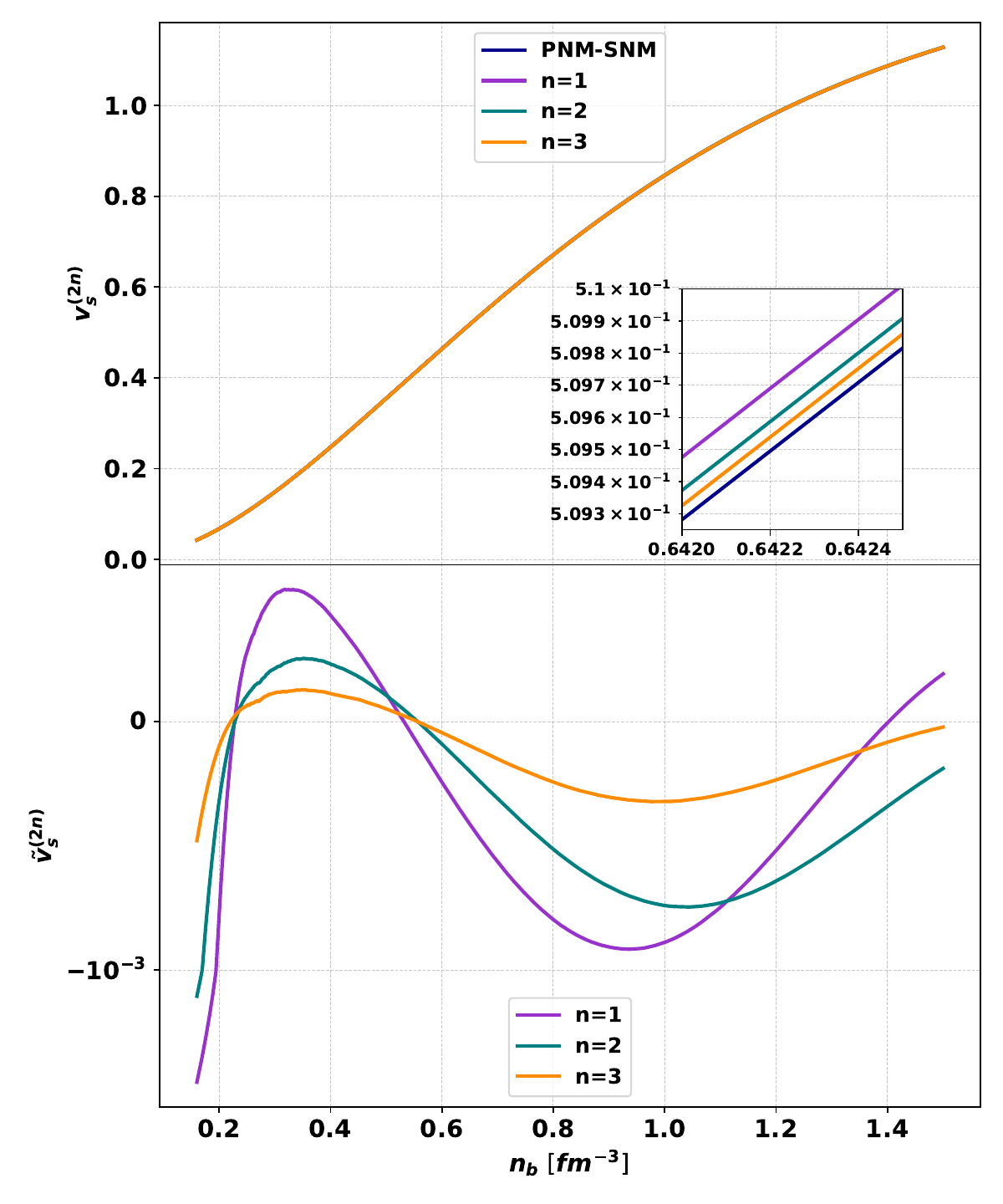}
    \caption{ Velocity of sound variation with baryon number density ($n_b$) at various corrections for SLy4 Skyrme Parameters.}
\label{fig:sly4_vs}
\end{figure}

Firstly, we have estimated the change in symmetry energy $E_{sym}^{(2n)}$ at different order of corrections in isospin asymmetry parameter $\mathcal{O} (I^{(2n)})$ from the exact value $E_{sym}^{(exact)}$. The effect of higher-order isospin corrections to the symmetry energy for the SLy4 Skyrme EOS over the baryon density range $n_b \leq 1.5~{\rm fm}^{-3}$, relevant for neutron star matter from sub-saturation to the central densities of the massive stars, shows a systematic convergence toward the exact result with increasing order in the isospin expansion for various physical quantities as demonstrated in Fig.~\ref{fig:sym_en_diff_order_sly4} -- \ref{fig:sly4_vs}. While the symmetry energy itself remains approximated reasonably well at lower values of isospin asymmetry $I$, the deviations in the neutron--proton chemical potential difference become significant in highly neutron-rich matter, reaching $\gtrsim 7\%$ near $I\sim1$ for the quadratic approximation (see Fig.~\ref{fig:sly4_mu_diff}). 

As illustrated in Fig.~\ref{fig:sly4_qulb_pres} for $\beta$-equilibrated matter, the baryonic pressure receives relatively modest corrections, with the largest deviation occurring near sub-saturation density, where $\tilde{P}_{b}^{(2)} \gtrsim 4\%$, and decreasing rapidly at higher densities. In contrast, leptonic quantities remain substantially more sensitive to higher-order corrections. The leptonic pressure exhibits deviations exceeding $70\%$ at very low densities and remains at the $\sim10\%$ level around and above saturation density for the quadratic approximation (see Fig.~\ref{fig:sly4_qulb_pres}, right-panel). Similar trends are observed for the electron and muon number densities, with the largest deviations appearing in the muonic sector due to its stronger sensitivity to the chemical equilibrium condition (check ~\ref{app_aadn_plots}). 

Despite these sizable corrections in individual leptonic observables, the total $\beta$-equilibrium EOS remains comparatively stable. The fractional corrections to the total pressure remain below $\sim1\%$ at saturation density and decrease further at higher densities (Fig.~\ref{fig:sly4_p_beta}), while the total energy density differs from the exact result only at the level of $\sim10^{-4}$. Likewise, the sound speed receives only very small corrections, indicating that the causal structure and bulk stiffness of the EOS remain essentially unaffected by higher-order isospin terms. The resulting $\beta$-equilibrium EOS (Fig.~\ref{fig:sly4_betaeqm_eos_wref}) and sound-speed (Fig.~\ref{fig:sly4_vs}) profiles therefore demonstrate that, although higher-order isospin corrections can strongly modify composition-sensitive quantities, their impact on the global macroscopic EOS relevant for neutron-star structure is comparatively small for the SLy4 parametrization.

\subsection{Higher order isospin corrections: population of Skyrme models}\label{population_study}

\begin{figure*}[htbp]
    \raggedright (a)\hspace*{\columnwidth}(b)\\[-0.5cm]
    \includegraphics[clip,width=\textwidth]{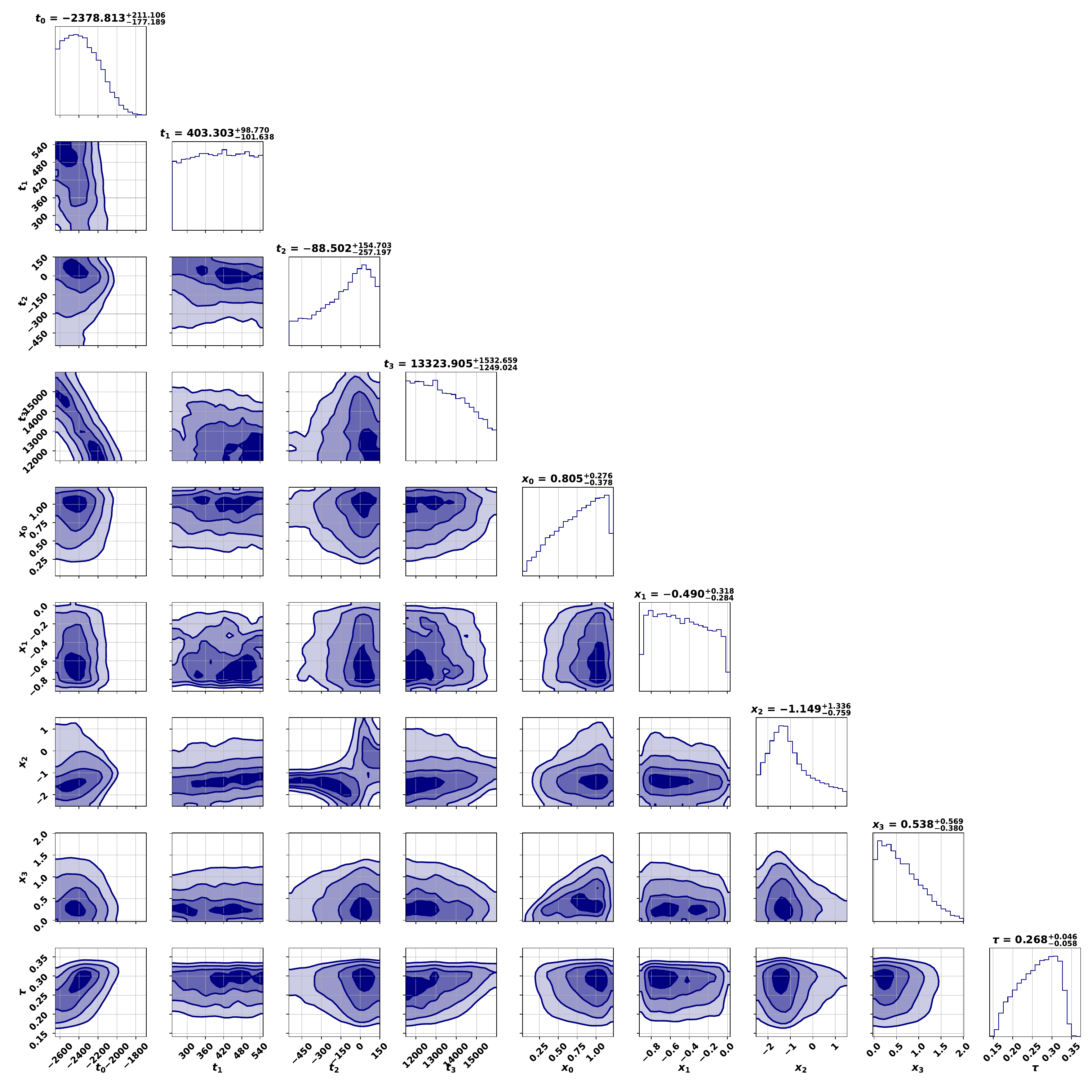} \\ 
    \caption{The population of Skyrme interaction model has been studied over the entire parameters space $\vec{\theta}$, see subsection~\ref{population_study} for details. The ranges corresponding to each of the 9 parameters are quoted in Table~\ref{tab:skyrme_param_range}. 
    }\label{fig:pop_skyrme_physical_dist}
\end{figure*}

\begin{table}[H]
\centering
\resizebox{\columnwidth}{!}{%
\begin{tabular}{l|ccccccccc}
\toprule
 Parameters & $t_0$ & $t_1$ & $t_2$ & $t_3$ & $x_0$ & $x_1$ & $x_2$ & $x_3$ & $\tau$ \\
\midrule
Lower Limit & $-2650$ & $250$ & $-550$ & $11500$ & $0.05$ & $-0.9$ & $-2.5$ & $0.0$ & $1/6$ \\
Upper Limit & $-1250$ & $550$ & $150$ & $16000$ & $1.2$ & $0.0$ & $1.5$ & $2.0$ & $1/3$ \\
\bottomrule
\end{tabular}%
}
\caption{Upper and lower limit of the Skyrme parameters used to sample and constrain to get a distribution of physical EoS. }
\label{tab:skyrme_param_range}
\end{table}

SLy4 is one of the well known EOSs within the Skyrme family, that is widely used by the nuclear physics and astrophysics community. While it is a representative EOS that satisfy all the well established constraints from nuclear physics experiments and astronomical observations, it is not a unique choice within the Skyrme family. Thus, we have thoroughly studied the population of Skyrme interaction model~\cite{Chabanat_etal_1998NuPhA, DouchinHaensel_AA2001} covering a wide parameter space~\cite{diff_sym_en_LattimerGroup_prc2024} as presented in Table~\ref{tab:skyrme_param_range}. Each of the samples are drawn from this multi-dimensional Skyrme parameter space $\vec{\theta}$ to represent the baryonic sector governed by the Skyrme model as delineated in Section~\ref{sec:skyrme_model} (also see Eq.~\ref{E/A-taylor}). Each of them was then subjected to satisfy the $\beta$-equilibrium and charge neutrality conditions between the leptonic and baryonic sector, as outlined in subsection~\ref{leptonic_sector_beta-equilb}. This results in a cold EOS that determines the stellar structure of a neutron star. Each of these cold EOSs are used to construct stable stellar structures of a neutron star by solving Tolman–Oppenheimer–Volkoff (TOV) equations corresponding to a unique sample in the Skyrme parameter space $\vec{\theta}$. The samples are drawn uniformly in $\vec{\theta}$ within the corresponding ranges in Table~\ref{tab:skyrme_param_range} for studying various properties of the neutron stars in this population of Skyrme EOS models. 

We consider an EOS, generated from each sample, to be a valid one provided it satisfies a few necessary conditions: 
\begin{itemize}
    \item the corresponding Skyrme interaction model meet the constraint on nuclear saturation density $n_{b_0} \in [0.14, 0.17]$ fm$^{-3}$, from several experiments on finite nuclei~\cite{Horowitz_etal_PREX_PRC2020, Drischler_etal_PRL2019, Chen_etal_PRC2014} 

    \item the stellar structure constructed from the EOS corresponding to the underlying Skyrme interaction model should support stable neutron stars having gravitational mass $\geq 2M_{\odot}$~\cite{Demorest_etal_nat2010, Antoniodis_sci2013, Romani_etal_apj2022}. These two conditions put a constraints on the multi-dimensional $\vec{\theta}$ parameter space. 

    \item the EOS for $\beta$-equilibrated matter corresponding to each EOS follow (i) the monotonicity condition for the density ($\rho$) Vs pressure ($P$) relationship, i.e., $\frac{d\rho}{dP} \geq 0 \; \; \forall n_{b}$, to ensure thermodynamic stability; and (ii) the causality condition $\frac{d\rho}{dP} \leq 1 \; \forall n_{b}$, throughout the interior of a neutron star having mass below the corresponding TOV-limit 

    \item the exact symmetry energy, arising from correction of isospin asymmetry parameter ($I$), is positive semi-definite, i.e., $E_{sym}^{(exact)} \geq 0 \; \forall n_{b}$, throughout the neutron star within the TOV-limit 
\end{itemize}

Consideration of these conditions has lead to the 9-dimensional distribution of Skyrme model parameters as depicted in Figure~\ref{fig:pop_skyrme_physical_dist}. This corner-plot presents all the 36 combinations of pair-wise correlations of the Skyrme parameter, as well as 9 individual marginalized 1-D distributions. Any randomly drawn sample from this 9-D joint distribution will meet all the aforementioned physical conditions necessary for a valid EOS. We have performed detailed investigations over a population of randomly drawn large number ($\sim 85,000$) of samples covering the entire Skyrme parameter space $\vec{\theta}$, corresponding to their internal composition and physical properties relevant for the astrophysical neutron stars. We present the statistical characteristics of these population studies and summarize the key outcomes below. 

\begin{figure}[!h]
    \centering
    \includegraphics[width=\columnwidth]{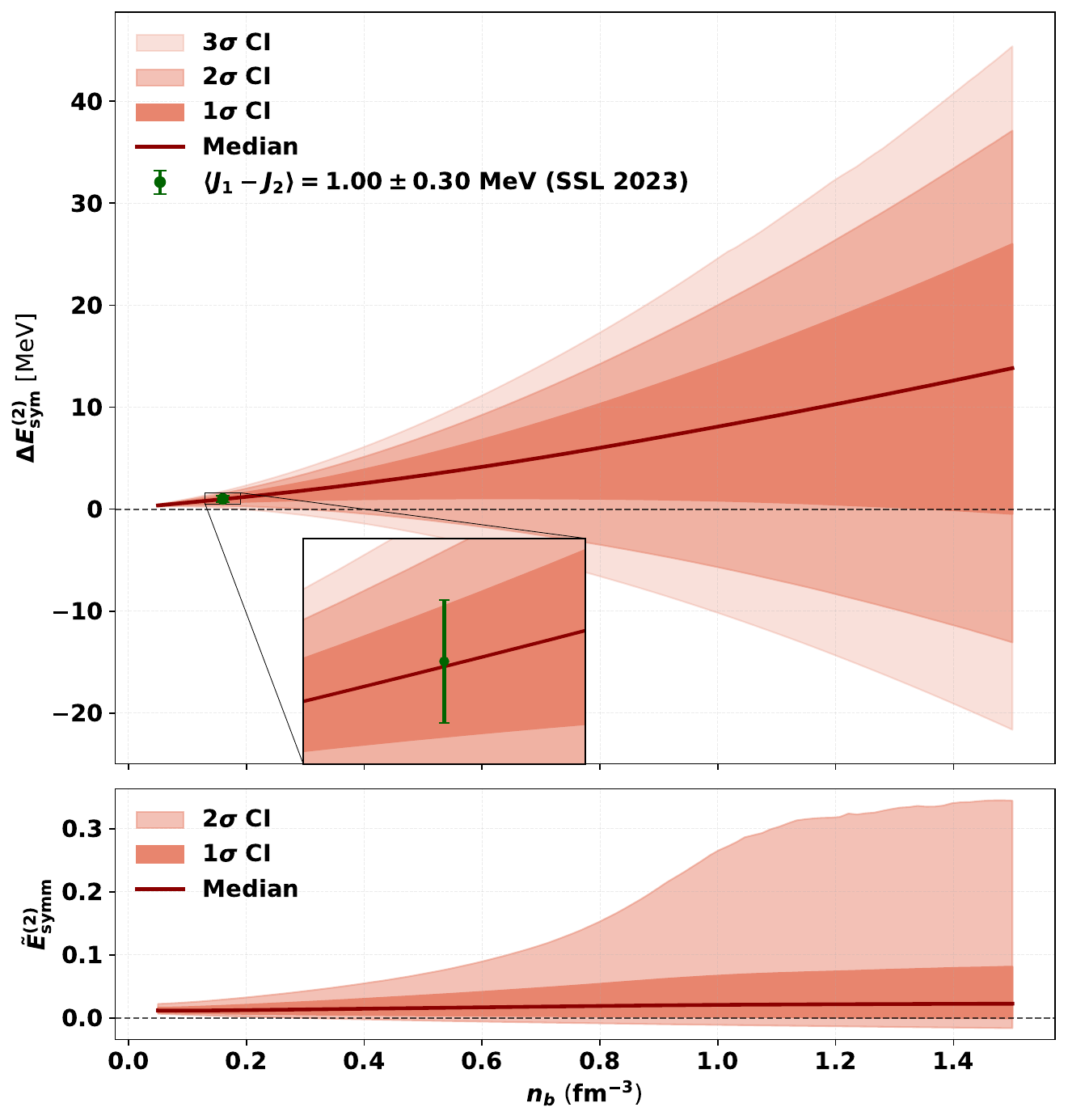}
    \caption{\textit{Top panel} shows $1\sigma,~2\sigma,~3\sigma$ intervals for difference in symmetry energy shown for Skyrme parameters in Fig. \ref{fig:pop_skyrme_physical_dist}. \textit{Bottom panel} shows the corresponding fractional differences for the same. For comparison, we also mark the result reported by~\citet{diff_sym_en_LattimerGroup_prc2024}. 
    }
    \label{fig:diff_symmE}
\end{figure}

\begin{figure}[htbp]
    \includegraphics[clip,width=\columnwidth]{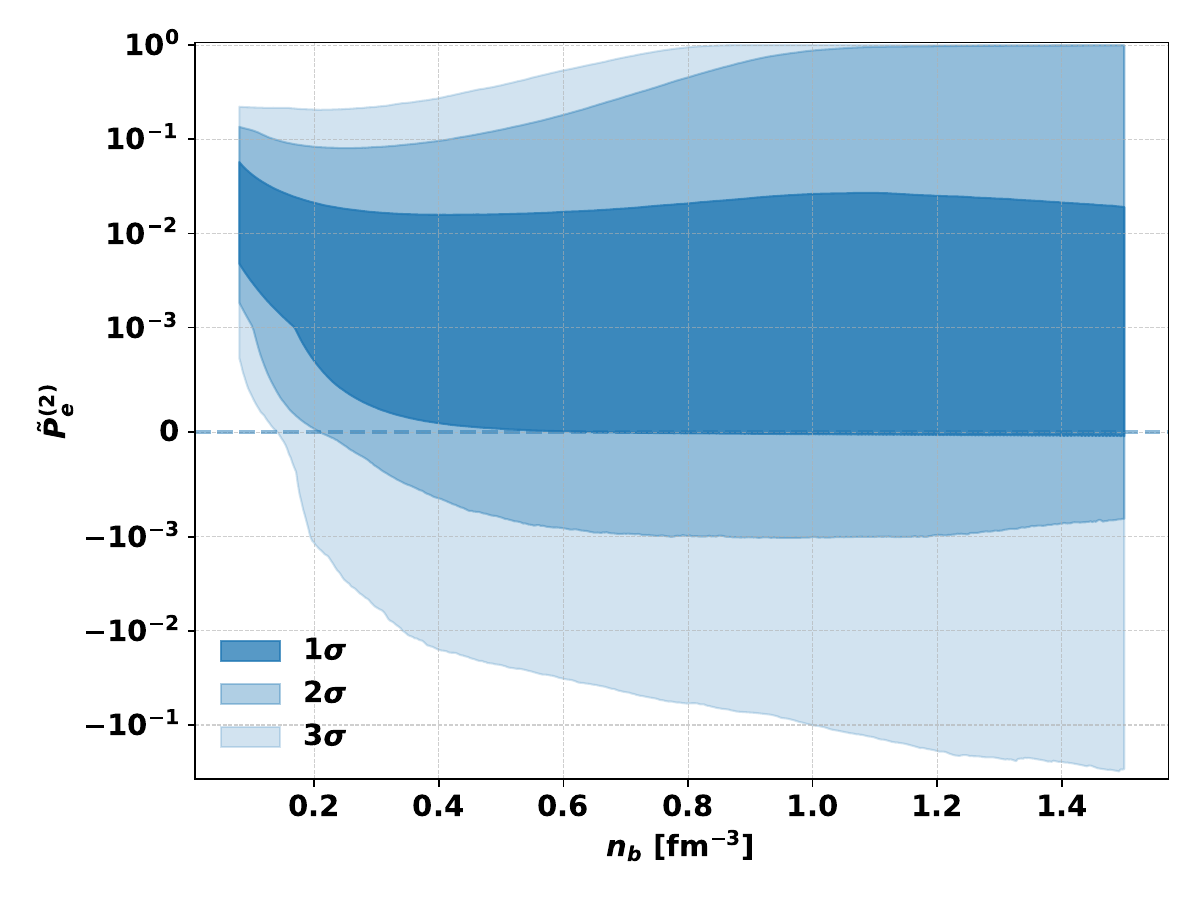} \\
    \includegraphics[clip,width=\columnwidth]{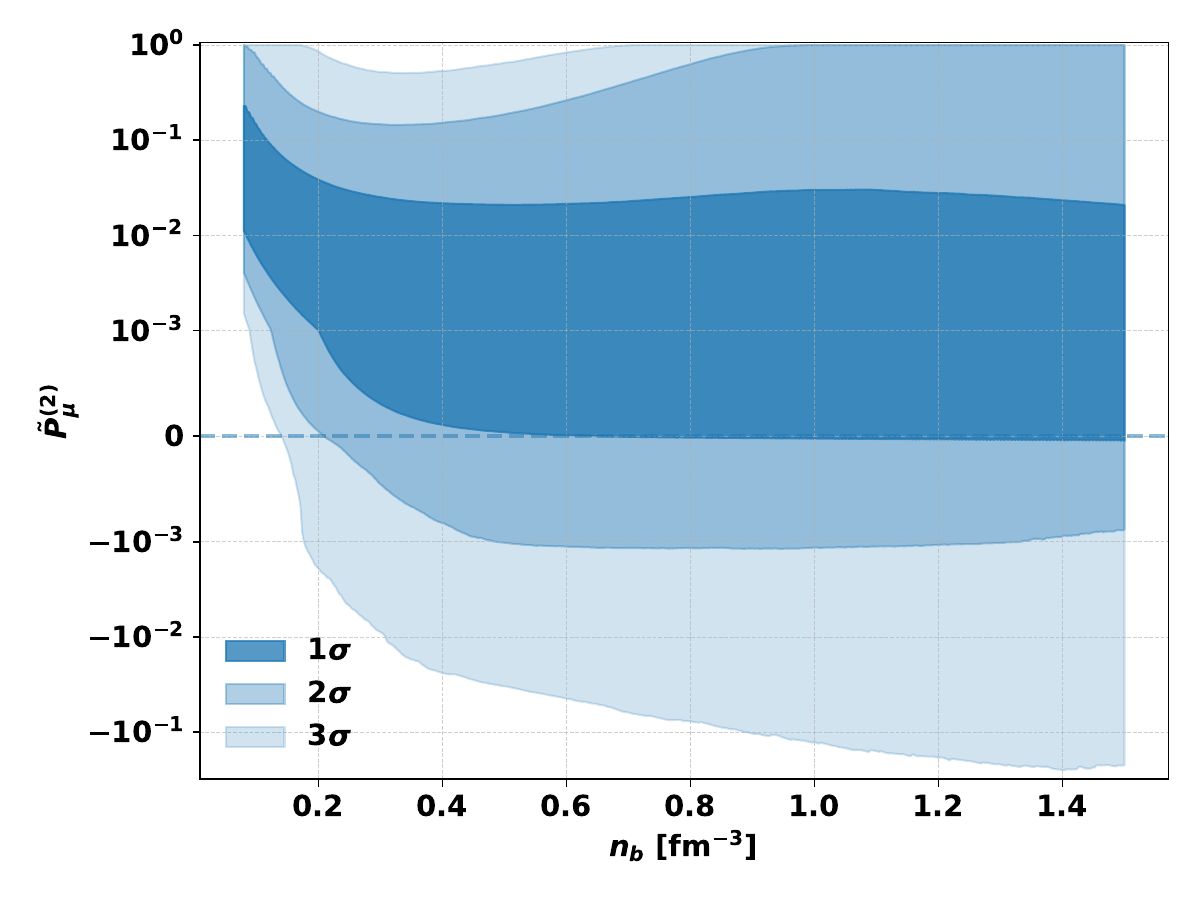} \\
    \includegraphics[clip,width=\columnwidth]{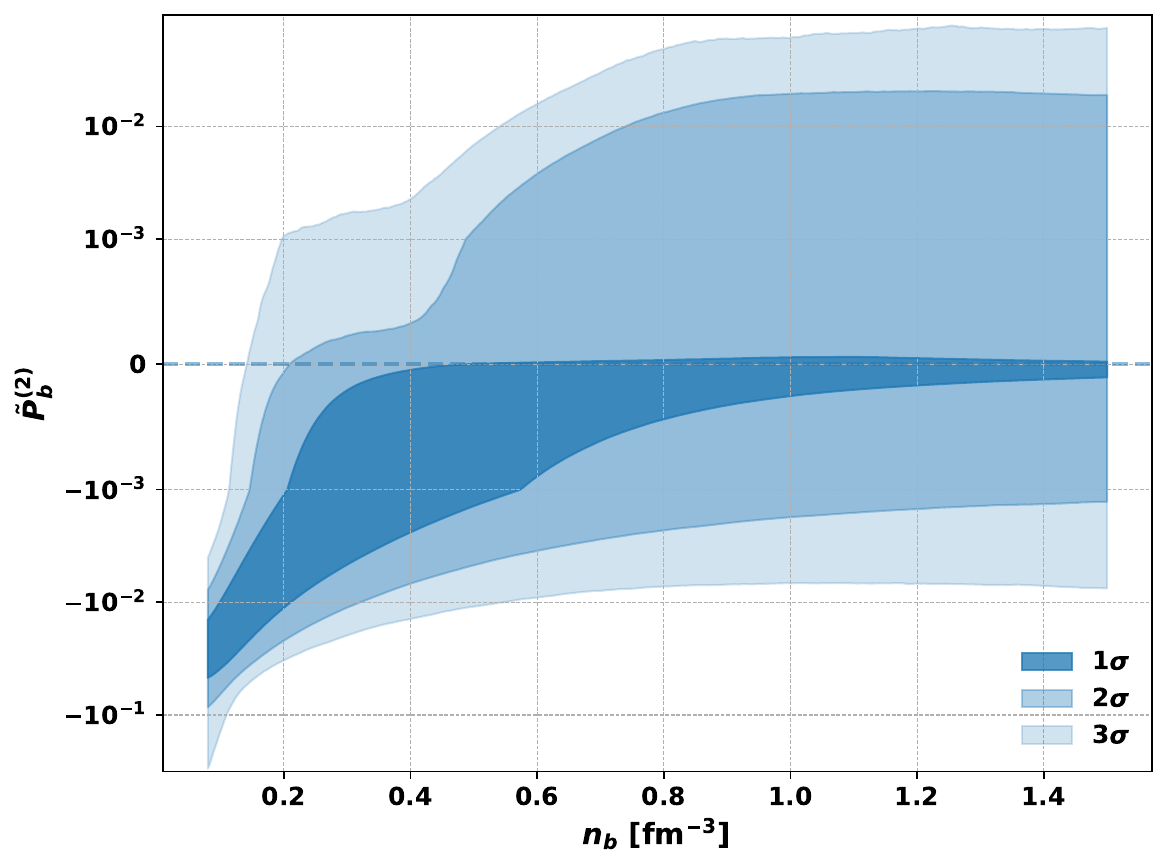} \\
    \caption{This figure demonstrates the overall statistical behaviour of fractional difference in pressure between exact and quadratic order corrections in symmetry energy $E_{sym}$ (see Eq.~\ref{frac_diff_theta-tilde}) for the population of Skyrme EOSs over the entire parameter space $\vec{\theta}$. The fractional deviations in pressure $\tilde{P}_{e}^{(2)}$ for electrons (top-panel), $\tilde{P}_{\mu}^{(2)}$ for muons (middle-panel) and $\tilde{P}_{b}^{(2)}$ for baryons (bottom-panel) are shown for the model populations over the density range $n_{b} \in [0.08, 1.5] \, \rm{fm}^{-3}$ at $1\sigma$ (dark-shading), $2\sigma$ (medium-shading) and $3\sigma$ (light-shading) levels.
    }\label{fig:popn_pres_e_mu_b}
\end{figure}

\begin{figure}[htbp]
    \includegraphics[clip,width=\columnwidth]{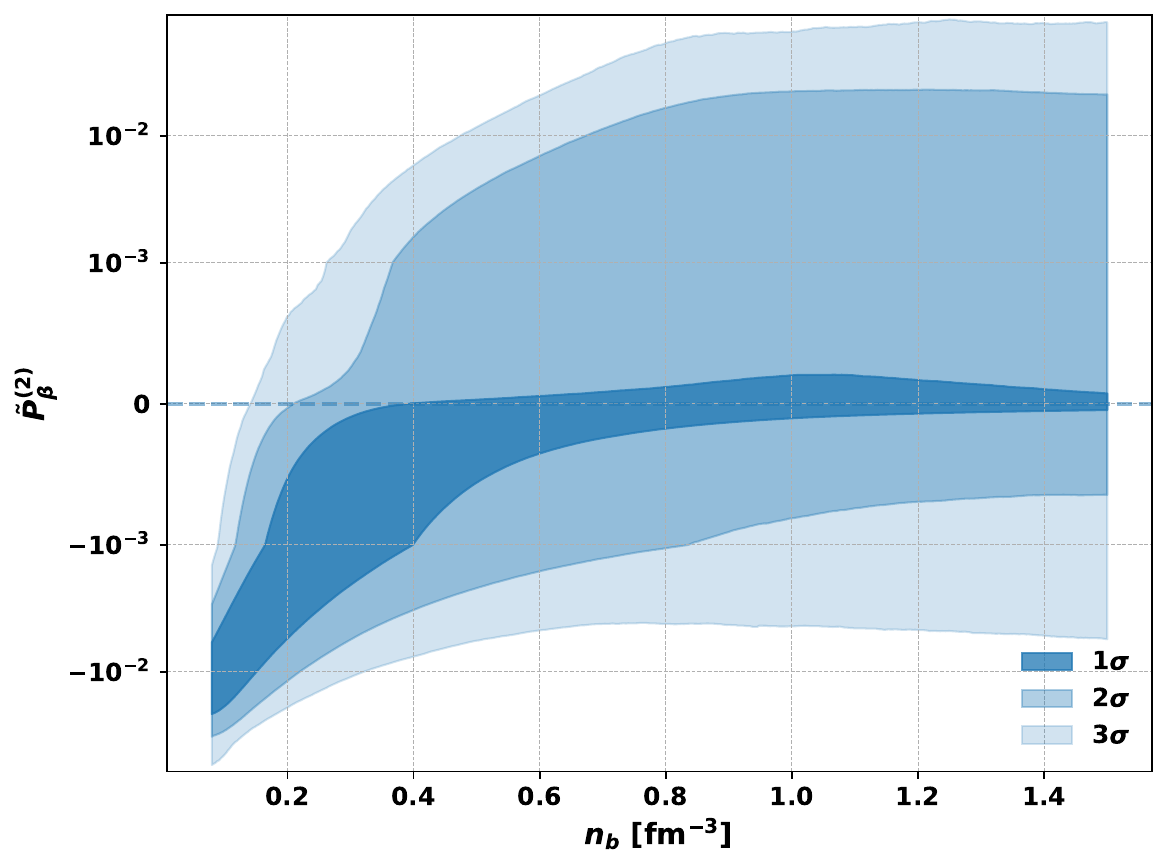} \\
    \includegraphics[clip,width=\columnwidth]{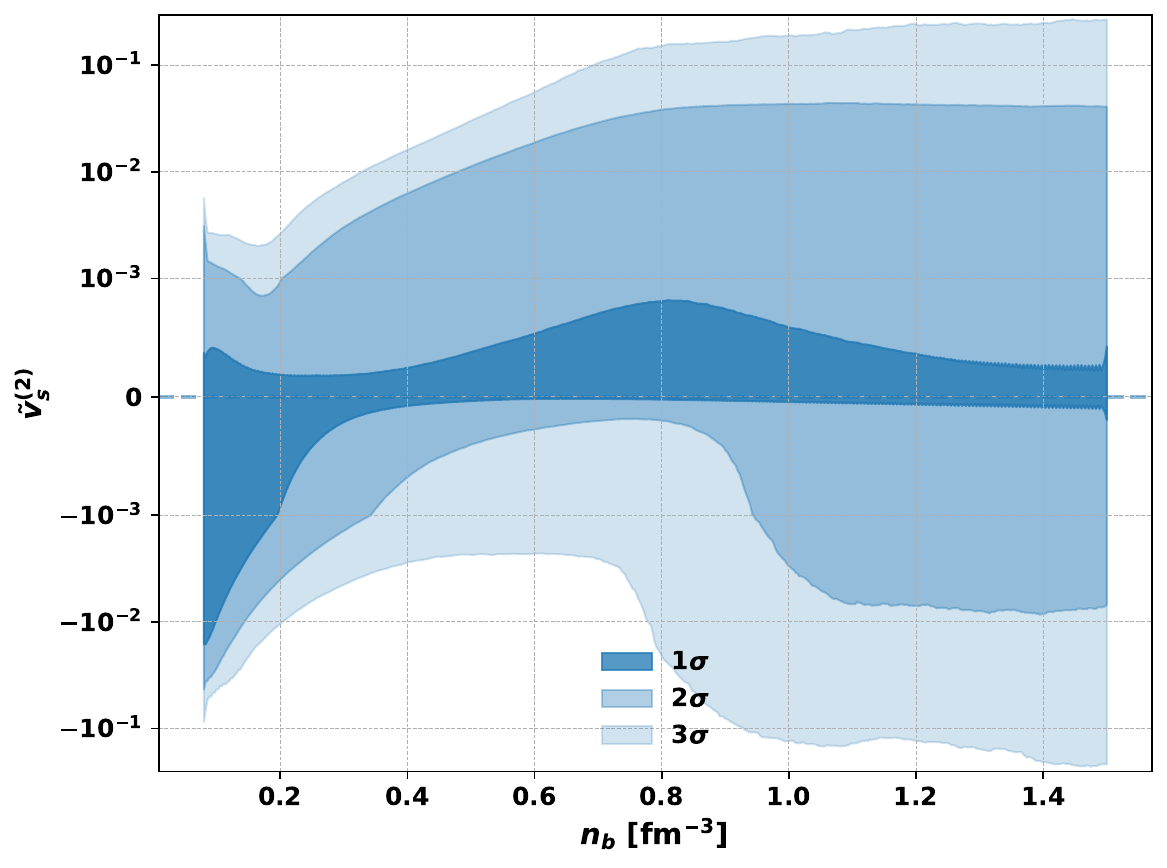} \\
    \caption{This figure demonstrates the overall statistical behaviour of fractional difference in $\beta$-equilibrium matter pressure and sound speed between exact and quadratic order corrections in symmetry energy $E_{sym}$ (see Eq.~\ref{frac_diff_theta-tilde}) for the population of Skyrme EOSs over the entire parameter space $\vec{\theta}$. The fractional deviations in $\beta$-equilibrium matter pressure $\tilde{P}_{\beta}^{(2)}$ (top-panel) and, sound speed $\tilde{v}_{s}^{(2)}$ (bottom-panel) are shown for the model populations over the density range $n_{b} \in [0.08, 1.5] \, \rm{fm}^{-3}$ at $1\sigma$ (dark-shading), $2\sigma$ (medium-shading) and $3\sigma$ (light-shading) levels. 
    }\label{fig:popn_pres_beta_eqlb}
\end{figure}

\begin{figure}[htbp]
    \includegraphics[clip,width=\columnwidth]{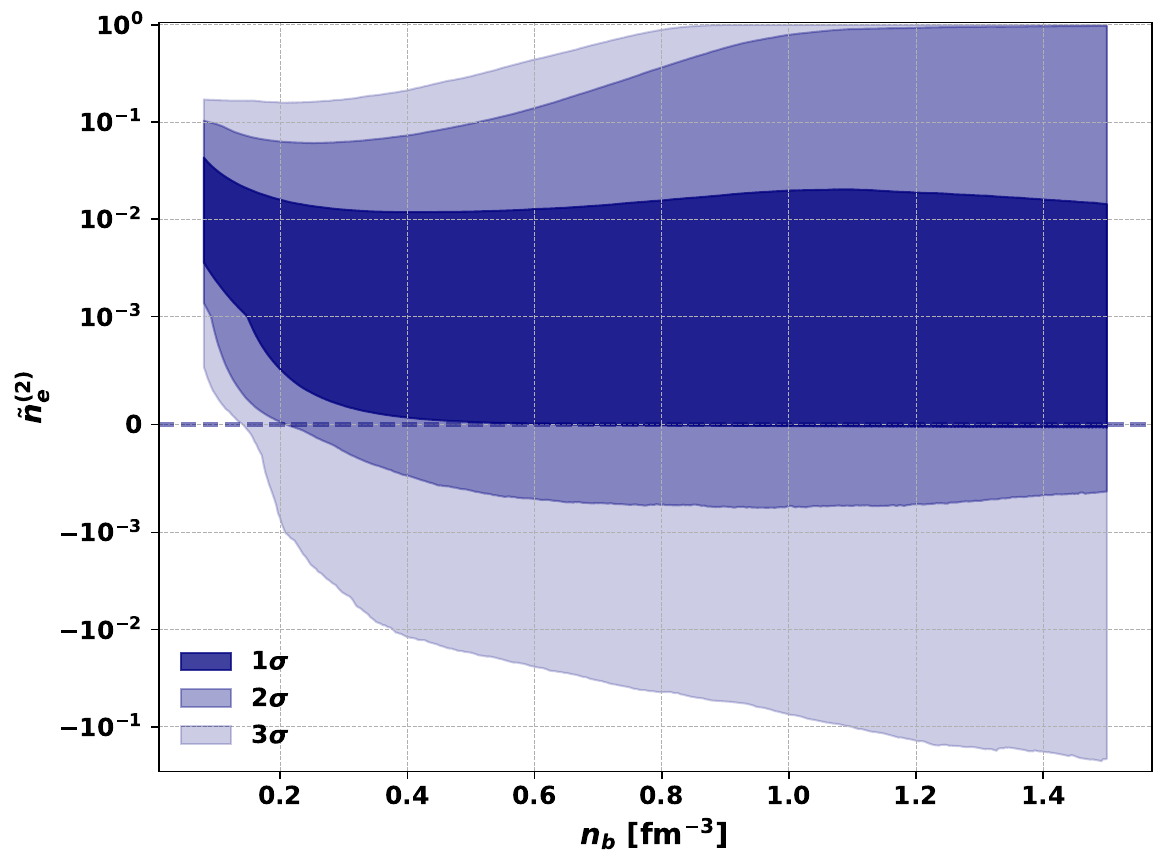} \\
    \includegraphics[clip,width=\columnwidth]{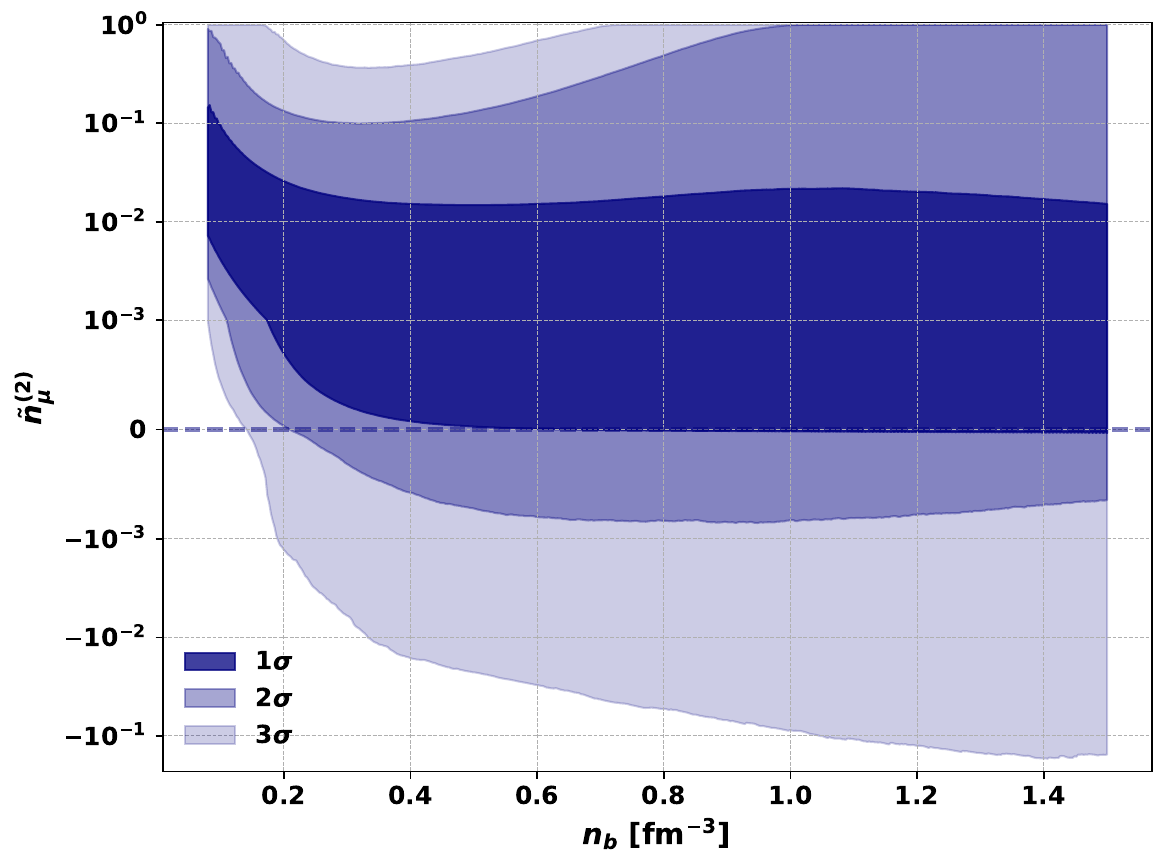} \\
    \caption{This figure, similar to Figure~\ref{fig:popn_pres_e_mu_b}, demonstrates the overall statistical behaviour of the fractional differences of number density between exact and quadratic order correction in $E_{sym}$ for the population of the Skyrme models over $\vec{\theta}$. The fractional deviations in number density $\tilde{n}_{e}^{(2)}$ for electrons (top-panel), $\tilde{n}_{\mu}^{(2)}$ for muons (bottom-panel) are shown for the model populations over the same density range $n_{b} \in [0.08, 1.5] \, \rm{fm}^{-3}$ at $1\sigma$, $2\sigma$ and $3\sigma$ levels. 
    }\label{fig:popn_numden_e_mu}
\end{figure}

The statistical behaviour of the most relevant thermodynamic quantities for higher-order isospin corrections over the full population of physically valid Skyrme EOSs is summarized in Figs.~\ref{fig:diff_symmE}--\ref{fig:popn_numden_e_mu} and Tables~\ref{tab:delta_Esym}--\ref{tab:tilde_non_pressure_quantities}. Across the allowed Skyrme parameter space $\vec{\theta}$, we find that the deviation between the exact and quadratic-order treatment of the symmetry energy increases systematically with density. While the corrections remain modest around saturation density, they become increasingly important in the supra-nuclear density regime relevant for the inner core of massive neutron stars.

Firstly, the difference between the exact and quadratic-order symmetry energy, $\Delta E_{\rm sym}$, grows monotonically with baryon density. Around half-saturation density ($n_b \approx 0.08~{\rm fm^{-3}}$), the median deviation is only $\sim 0.5$ MeV, with a relatively narrow $1\sigma$ interval. Near saturation density ($n_b \approx 0.16~{\rm fm^{-3}}$), the deviation remains below $\sim 1$ MeV for most EOSs (see Table~\ref{tab:delta_Esym}). However, beyond $2n_0$ ($n_b=0.32~{\rm fm^{-3}}$), the spread broadens considerably as depicted in Fig.~\ref{fig:diff_symmE}, reaching median values of several MeV and extending to $\gtrsim 10$ MeV in the high-density regime. At the highest density $n_b=1.5~{\rm fm^{-3}}$ explored here, the median difference reaches $\sim 14$ MeV, while the $90\%$ confidence interval spans a much wider range, indicating substantial EOS-to-EOS variation at extreme densities. This behaviour demonstrates that higher-order isospin contributions become increasingly relevant in the dense inner core region of neutron stars, particularly the massive ones.

\begin{table*}[]
\centering
\renewcommand{\arraystretch}{2.35}
\resizebox{\textwidth}{!}{%
\begin{tabular}{|c|c|c|c|c|c|c|c|c|c|}
\hline
 $\mathbf{n_b} [\mathbf{fm^{-3}}]$ 
 & $0.08$ 
 & $0.16$ 
 & $0.32$ 
 & $0.48$ 
 & $0.64$ 
 & $0.80$ 
 & $0.96$ 
 & $1.28$ 
 & $1.50$ \\
\hline

$\Delta E_{\rm sym}$ 
& $0.53_{0.42(0.35)}^{0.62(0.68)}$
& $0.98_{0.63(0.41)}^{1.27(1.46)}$
& $1.98_{0.88(0.19)}^{2.92(3.52)}$
& $3.15_{1.00(-0.36)}^{5.01(6.19)}$
& $4.50_{1.03(-1.18)}^{7.49(9.40)}$
& $6.02_{1.00(-2.23)}^{10.35(13.12)}$
& $7.67_{0.84(-3.50)}^{13.53(17.29)}$
& $11.20_{0.20(-6.75)}^{20.65(26.62)}$
& $13.82_{-0.56(-9.46)}^{26.08(33.84)}$ \\

\hline
\end{tabular}%
}
\caption{
Difference in symmetry energy between exact and quadratic order expansion. Median values with $1\sigma$ intervals (subscript/superscript) and $90\%$ confidence intervals (in parentheses) are shown for the difference in symmetry energy for the respective case for the Skyrme samples shown in Fig. ~\ref{fig:pop_skyrme_physical_dist}.} 
\label{tab:delta_Esym}
\end{table*}

A similar trend is observed for the symmetry-energy expansion parameters listed in Table~\ref{tab:delta_symm_params}. Here, $J$, $L$, $K_{\rm sym}$, and $Q_{\rm sym}$ denote, respectively, the symmetry energy at saturation density and its slope, curvature, and skewness parameters with respect to density~\cite{BALDO_ppnp2016_203, Lattimer_particles2023, Raithel_ApJ2019}. The corrections to $J$ and $L$ remain relatively small, typically at the level of $\mathcal{O}(1)$ MeV, whereas the higher-order coefficients $K_{\rm sym}$ and $Q_{\rm sym}$ exhibit much broader distributions. This indicates that the quadratic approximation reproduces the symmetry energy and its low-order density derivatives near saturation accurately, but is progressively less faithful for the higher-order density derivatives that control the supra-nuclear extrapolation. Overall, consistency is observed with the broadening of deviations in $\Delta E_{sym}^{(2)}$ as demonstrated in Fig.~\ref{fig:diff_symmE}. For completeness, an alternative characterization of the beyond-quadratic contribution using the $\eta$ parametrization of Ref.~\cite{Steiner_prc2006} is presented in Appendix~\ref{app_eta_beyond_quadratic}. 

\begin{table}[]
\centering
\renewcommand{\arraystretch}{2.35}
\resizebox{\columnwidth}{!}{%
\begin{tabular}{|c|c|c|c|}
\hline
 $\Delta J$ [MeV] 
 & $\Delta L$ [MeV] 
 & $\Delta K_{\rm sym}$ [MeV] 
 & $\Delta Q_{\rm sym}$ [MeV] \\
\hline

$0.94_{0.61(0.41)}^{1.22(1.41)}$
& $2.64_{1.02(0.00)}^{4.05(4.96)}$
& $1.16_{-2.05(-4.10)}^{3.98(5.76)}$
& $2.95_{0.15(-1.63)}^{6.15(8.21)}$

\\
\hline
\end{tabular}%
}
\caption{
Difference in symmetry energy parameters between exact and quadratic order of the symmetry energy. Median values with $1\sigma$ intervals (subscript/superscript) and $90\%$ confidence intervals (in parentheses) are shown for the difference in symmetry energy parameters for the respective case for the Skyrme samples shown in Fig. ~\ref{fig:pop_skyrme_physical_dist}. 
}
\label{tab:delta_symm_params}
\end{table}

\begin{table*}[]
\centering
\scriptsize
\setlength{\tabcolsep}{6pt}
\renewcommand{\arraystretch}{2.35}

\resizebox{\textwidth}{!}{%
\begin{tabular}{|c|c|c|c|c|c|}
\hline

$\mathbf{n_b} \; [\mathrm{fm^{-3}}]$ & $\tilde{P}_e^{(2)} \; [\times 10^{-2}]$ & $\tilde{P}_\mu^{(2)} \; [\times 10^{-2}]$ & $\tilde{P}_{\ell}^{(2)} \; [\times 10^{-2}]$ & $\tilde{P}_b^{(2)} \; [\times 10^{-3}]$ & $\tilde{P}_\beta^{(2)} \; [\times 10^{-3}]$ \\
\hline

0.08 & $1.597_{0.477(0.253)}^{5.709(10.546)}$ & $4.386_{1.115(0.566)}^{22.177(65.127)}$ & $1.798_{0.555(0.297)}^{5.802(10.546)}$ & $-26.900_{-46.797(-66.836)}^{-14.324(-9.673)}$ & $-12.685_{-21.422(-27.984)}^{-5.874(-3.746)}$ \\ \hline

0.16 & $0.551_{0.113(0.044)}^{2.562(5.908)}$ & $0.964_{0.181(0.070)}^{5.443(15.297)}$ & $0.643_{0.131(0.051)}^{2.999(6.801)}$ & $-7.403_{-17.884(-26.807)}^{-2.526(-1.109)}$ & $-3.416_{-8.869(-14.125)}^{-1.090(-0.478)}$ \\ \hline

0.32 & $0.198_{0.019(-0.007)}^{1.635(4.976)}$ & $0.268_{0.025(-0.010)}^{2.406(8.019)}$ & $0.222_{0.021(-0.008)}^{1.866(5.752)}$ & $-1.091_{-4.011(-7.581)}^{-0.154(0.056)}$ & $-0.483_{-1.761(-3.614)}^{-0.060(0.110)}$ \\ \hline

0.48 & $0.111_{0.004(-0.021)}^{1.586(6.695)}$ & $0.138_{0.005(-0.026)}^{2.097(9.533)}$ & $0.122_{0.004(-0.023)}^{1.767(7.577)}$ & $-0.303_{-1.547(-3.482)}^{-0.002(0.230)}$ & $-0.121_{-0.620(-1.541)}^{0.024(1.194)}$ \\ \hline

0.64 & $0.074_{0.000(-0.024)}^{1.738(10.984)}$ & $0.088_{0.000(-0.028)}^{2.175(14.883)}$ & $0.079_{0.000(-0.025)}^{1.905(12.281)}$ & $-0.106_{-0.764(-2.006)}^{0.019(1.382)}$ & $-0.043_{-0.301(-0.883)}^{0.066(3.530)}$ \\ \hline

0.80 & $0.054_{-0.001(-0.023)}^{2.078(20.639)}$ & $0.062_{-0.001(-0.026)}^{2.524(27.540)}$ & $0.057_{-0.001(-0.024)}^{2.253(22.912)}$ & $-0.045_{-0.439(-1.337)}^{0.034(4.578)}$ & $-0.018_{-0.176(-0.590)}^{0.119(7.130)}$ \\ \hline

0.96 & $0.041_{-0.002(-0.021)}^{2.535(39.150)}$ & $0.046_{-0.002(-0.023)}^{2.946(50.037)}$ & $0.043_{-0.002(-0.022)}^{2.732(43.050)}$ & $-0.022_{-0.280(-0.996)}^{0.053(8.165)}$ & $-0.009_{-0.113(-0.435)}^{0.188(10.321)}$ \\ \hline

1.28 & $0.023_{-0.003(-0.017)}^{2.370(60.328)}$ & $0.026_{-0.003(-0.018)}^{2.635(71.444)}$ & $0.024_{-0.003(-0.018)}^{2.517(65.167)}$ & $-0.007_{-0.146(-0.651)}^{0.040(8.341)}$ & $-0.004_{-0.061(-0.291)}^{0.143(10.210)}$ \\ \hline

1.50 & $0.017_{-0.004(-0.017)}^{1.903(59.455)}$ & $0.019_{-0.004(-0.018)}^{2.078(68.413)}$ & $0.018_{-0.004(-0.018)}^{2.008(63.646)}$ & $-0.004_{-0.104(-0.533)}^{0.021(7.465)}$ & $-0.003_{-0.044(-0.246)}^{0.073(9.029)}$ \\ \hline

\end{tabular}
}
\caption{
Fractional differences in various thermodynamic quantities (pressures) at quadratic and exact order of symmetry energy. Median values with $1\sigma$ intervals (subscript/superscript) and $90\%$ confidence intervals (in parentheses) are shown for the fractional difference in various thermodynamic quantities (pressures) for the respective case for the Skyrme samples shown in Fig. ~\ref{fig:pop_skyrme_physical_dist}. 
}
\label{tab:tilde_all_pressure_quantities}
\end{table*}

\begin{table*}[]
\centering
\scriptsize
\setlength{\tabcolsep}{6pt}
\renewcommand{\arraystretch}{2.35}

\resizebox{\textwidth}{!}{%
\begin{tabular}{|c|c|c|c|c|c|}
\hline
$\mathbf{n_b} \; [\mathrm{fm^{-3}}]$ & $\tilde{n}_e^{(2)} \; [\times 10^{-2}]$ & $\tilde{n}_\mu^{(2)} \; [\times 10^{-2}]$ & $\tilde{n}_{\ell}^{(2)} \; [\times 10^{-2}]$ & $\tilde{\epsilon}^{(2)} \; [\times 10^{-3}]$ & $\tilde{v}_s^{(2)} \; [\times 10^{-3}]$ \\
\hline

0.08 & $1.200_{0.358(0.190)}^{4.313(8.019)}$ & $2.808_{0.722(0.369)}^{14.376(47.925)}$ & $1.586_{0.460(0.241)}^{5.069(8.019)}$ & $-0.083_{-0.126(-0.147)}^{-0.028(-0.006)}$ & $-4.301_{-16.204(-28.351)}^{0.373(2.180)}$ \\ \hline

0.16 & $0.413_{0.085(0.033)}^{1.928(4.465)}$ & $0.648_{0.123(0.047)}^{3.606(10.133)}$ & $0.493_{0.099(0.038)}^{2.404(5.737)}$ & $-0.087_{-0.226(-0.328)}^{-0.025(-0.011)}$ & $-0.223_{-1.961(-5.006)}^{0.224(0.570)}$ \\ \hline

0.32 & $0.149_{0.014(-0.005)}^{1.229(3.755)}$ & $0.187_{0.017(-0.007)}^{1.662(5.507)}$ & $0.165_{0.015(-0.006)}^{1.396(4.380)}$ & $-0.054_{-0.274(-0.556)}^{-0.007(0.003)}$ & $-0.000_{-0.169(-0.654)}^{0.189(1.737)}$ \\ \hline

0.48 & $0.083_{0.003(-0.016)}^{1.192(5.065)}$ & $0.098_{0.003(-0.019)}^{1.475(6.685)}$ & $0.090_{0.003(-0.017)}^{1.311(5.702)}$ & $-0.038_{-0.325(-0.847)}^{-0.002(0.008)}$ & $0.006_{-0.042(-0.192)}^{0.345(4.928)}$ \\ \hline

0.64 & $0.055_{0.000(-0.018)}^{1.306(8.357)}$ & $0.063_{0.000(-0.020)}^{1.546(10.591)}$ & $0.059_{0.000(-0.019)}^{1.411(9.278)}$ & $-0.028_{-0.387(-1.308)}^{-0.000(0.009)}$ & $0.004_{-0.016(-0.085)}^{0.603(10.550)}$ \\ \hline

0.80 & $0.040_{-0.001(-0.017)}^{1.562(15.918)}$ & $0.045_{-0.001(-0.019)}^{1.805(20.023)}$ & $0.042_{-0.001(-0.018)}^{1.669(17.587)}$ & $-0.021_{-0.465(-1.910)}^{0.001(0.009)}$ & $0.008_{-0.024(-0.073)}^{0.808(16.749)}$ \\ \hline

0.96 & $0.031_{-0.001(-0.015)}^{1.908(31.105)}$ & $0.033_{-0.001(-0.017)}^{2.121(38.123)}$ & $0.032_{-0.001(-0.016)}^{2.026(34.293)}$ & $-0.016_{-0.516(-2.350)}^{0.001(0.008)}$ & $0.012_{-0.039(-0.120)}^{0.654(17.538)}$ \\ \hline

1.28 & $0.017_{-0.002(-0.013)}^{1.783(50.015)}$ & $0.019_{-0.002(-0.013)}^{1.911(57.869)}$ & $0.018_{-0.002(-0.013)}^{1.869(54.624)}$ & $-0.009_{-0.473(-2.914)}^{0.001(0.007)}$ & $0.014_{-0.066(-0.233)}^{0.293(15.070)}$ \\ \hline

1.50 & $0.013_{-0.003(-0.013)}^{1.430(49.190)}$ & $0.014_{-0.003(-0.013)}^{1.513(55.369)}$ & $0.013_{-0.003(-0.013)}^{1.490(52.840)}$ & $-0.007_{-0.396(-3.030)}^{0.002(0.007)}$ & $0.030_{-0.197(-0.581)}^{0.426(12.649)}$ \\ \hline

\end{tabular}
}
\caption{
Fractional differences in various thermodynamic quantities (lepton number densities, total energy density, sound speed) at quadratic and exact order of symmetry energy. Median values with $1\sigma$ intervals (subscript/superscript) and $90\%$ confidence intervals (in parentheses) are shown for the fractional difference in various thermodynamic quantities (densities and sound speed) for the respective case for the Skyrme samples shown in Fig. ~\ref{fig:pop_skyrme_physical_dist}. 
}
\label{tab:tilde_non_pressure_quantities}
\end{table*}

The fractional deviations in the leptonic sector are presented in Figs.~\ref{fig:popn_pres_e_mu_b} and~\ref{fig:popn_numden_e_mu}. For electrons, the quadratic approximation generally reproduces both the pressure and number density at the percent level accuracy around saturation density. The median fractional deviations decrease from a few percent at sub-saturation density to below $\sim 1\%$ around $2n_0$, before developing increasingly broad tails at higher density. In particular, although the median corrections remain small even at large density, the $90\%$ confidence intervals widen substantially (see top-panel of Fig.~\ref{fig:popn_pres_e_mu_b}), reaching several tens of percent for a non-negligible subset of EOSs above $\sim 5n_0$. 

Muon quantities exhibit significantly larger deviations than electrons throughout the density range $n_{b} \in [0.08, 1.5] \, \rm{fm}^{-3}$. Around $0.08~{\rm fm^{-3}}$, the $1\sigma$ deviations in the muon pressure and number density already reach $\sim 15$--$25\%$, with even larger excursions at the $90\%$ level. Although the median corrections decrease at intermediate densities, the statistical spread increases strongly again in the high-density regime (see middle-panel of Fig.~\ref{fig:popn_pres_e_mu_b}). This enhanced sensitivity originates from the strong dependence of the muon population on the isospin asymmetry parameter $I$ and therefore on the detailed density-dependence of the symmetry energy. 

In contrast, the baryonic and total $\beta$-equilibrated matter pressures are considerably less sensitive to higher-order isospin corrections. The fractional difference in baryonic pressure, $\tilde{P}_b^{(2)}$, remains negative and typically below the percent level over most of the density range. Near saturation density, the median deviation is only $\approx -0.4\%$, and even at supra-nuclear densities the corrections remain comparatively small. The total pressure of $\beta$-equilibrated matter, $\tilde{P}_\beta^{(2)}$, follows a similar trend, indicating that the dominant bulk thermodynamic properties of neutron-star matter are relatively robust against higher-order corrections in the symmetry energy expansion.

The macroscopic thermodynamic quantities exhibit similarly small corrections. The fractional deviation in the total energy density remains at the level of $\lesssim 10^{-3}$--$10^{-2}$ over the full density range, while the sound speed shows deviations that are typically below one percent even at high density. This demonstrates that although higher-order isospin corrections can substantially modify the detailed leptonic composition of dense matter, their impact on the global stiffness and causal structure of the EOS remains comparatively modest for the population of physically viable Skyrme models considered here.

Overall, these population studies show that the quadratic approximation to the nuclear symmetry energy $E^{(2)}_{sym}$ remains reasonably accurate near and moderately above saturation density for bulk neutron-star observables. However, the approximation becomes increasingly inadequate for describing the detailed composition and isospin-sensitive properties of dense matter in the inner core region, where higher-order corrections can lead to substantial EOS-dependent deviations.

\subsection{Effect of higher order corrections for muon appearance \& d-URCA process}\label{muon_appearance_den} 
\subsubsection{Muon appearance}\label{muon_appear}

Matter in the interior of a neutron star is governed by the conditions of charge neutrality and $\beta$-equilibrium. As the density increases, the electron chemical potential rises, and once it reaches the muon rest mass, it becomes energetically favorable to populate a muon Fermi sea. In equilibrium the muons satisfy $\mu_\mu = \mu_e$ (Eq.~\ref{eqm2}). The muon onset is thus set by the condition
\begin{equation}\label{eqm1}
    \mu_{e} \geq m_{\mu} c^{2} ,
\end{equation}
where $m_{\mu}$ is the muon rest mass, and is therefore controlled by the electron chemical potential.

Consequently, the onset of muons within the stellar core depends directly on the electron number density $n_{e}$. This leptonic profile is inherently coupled to the isospin asymmetry parameter $I$ at a given baryon density $n_{b}$, determined by the specific vector of Skyrme interaction parameters $\vec{\theta}$. The critical baryon number density at which muons first appear, the muon onset threshold, $n^{(\mu)}_{b}$ is therefore highly sensitive to higher-order corrections in the isospin asymmetry expansion, $I^{2n}$.

\begin{figure}[!h]
    \centering
    \includegraphics[width=\columnwidth]{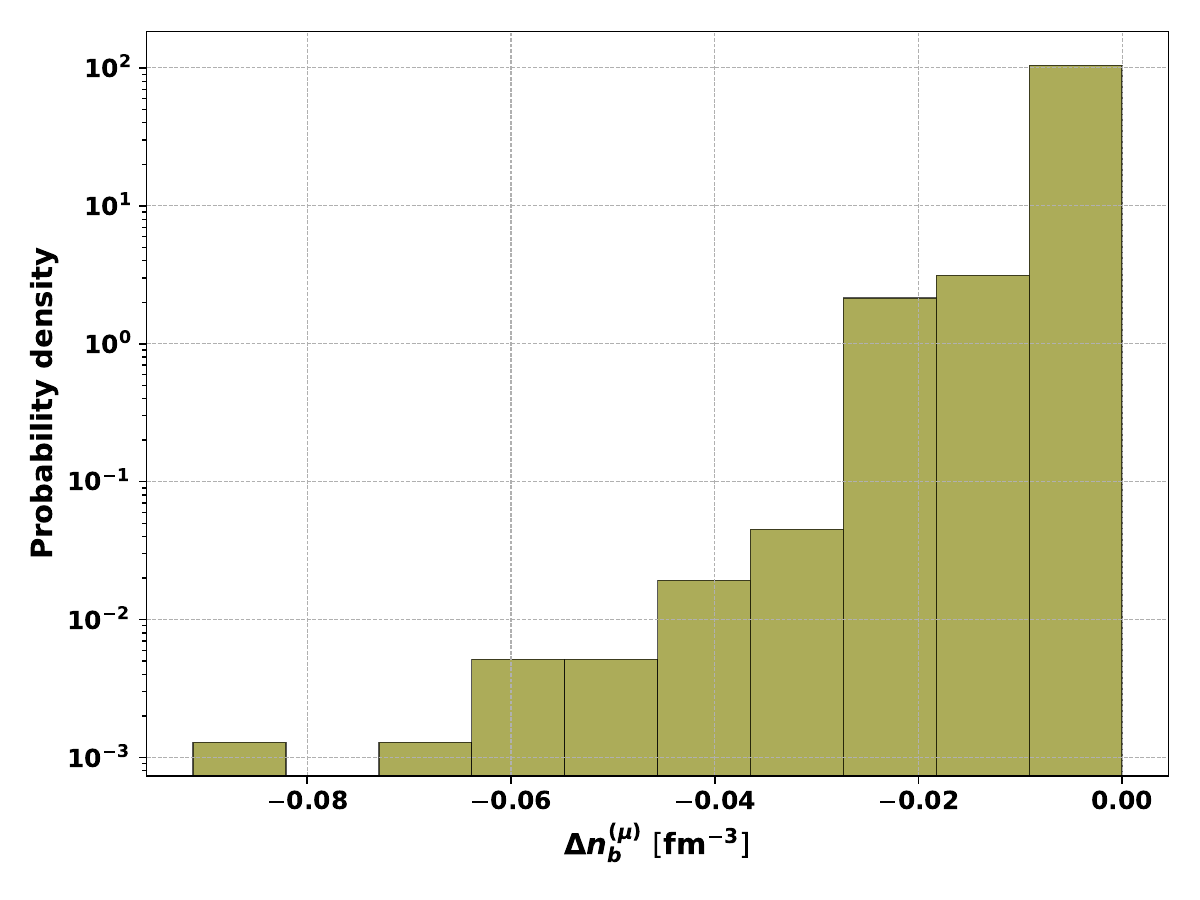}
    \caption{Difference in the muon onset baryon density, $n_{b}^{(\mu)}$, calculated using the exact versus the quadratically truncated symmetry energy, shown across the sampled Skyrme population from Fig.~\ref{fig:pop_skyrme_physical_dist}.}
    \label{fig:diff_muon_onset_threshold_baryon_ndensity}
\end{figure}

We investigate the impact of successive higher-order terms by evaluating the quadratic ($n=1$), quartic ($n=2$), hexic ($n=3$), and exact forms of the nuclear symmetry energy, $E_{\text{sym}}$, for $\beta$-equilibrated SLy4 matter. For the benchmark SLy4 EOS, the numerical convergence across these truncation orders is detailed in Appendix~\ref{appendix} (see Sec.~\ref{app_solver_tol}). Furthermore, this numerical convergence has been rigorously verified for self-consistent, $\beta$-equilibrated solutions containing neutrons, protons, electrons, and muons across the entire sampled parameter space $\vec{\theta}$ of the Skyrme model population under both quadratic and exact treatments of $E_{\text{sym}}$. 

For the population of the Skyrme models over $\vec{\theta}$, the difference in muon onset is shown in Fig.~\ref{fig:diff_muon_onset_threshold_baryon_ndensity}. The shift $\Delta n_b^{(\mu)} = n_b^{(\mu,exact)} - n_b^{(\mu,2)}$ is non-positive across the entire population, indicating that the exact treatment never delays muon appearance relative to the quadratic approximation and, for a subset of EOSs, brings muons in at a lower baryon density. The effect is negligible for the overwhelming majority of models: $95.12\%$ lie within $|\Delta n_b^{(\mu)}| \leq 0.01$~fm$^{-3}$ and $99.98\%$ within $0.04$~fm$^{-3}$, with no EOS exceeding a maximum excursion of $0.091$~fm$^{-3}$. 

\begin{figure}[!h]
    \centering
    \hspace{0.25cm}
    \includegraphics[width=0.93\columnwidth]{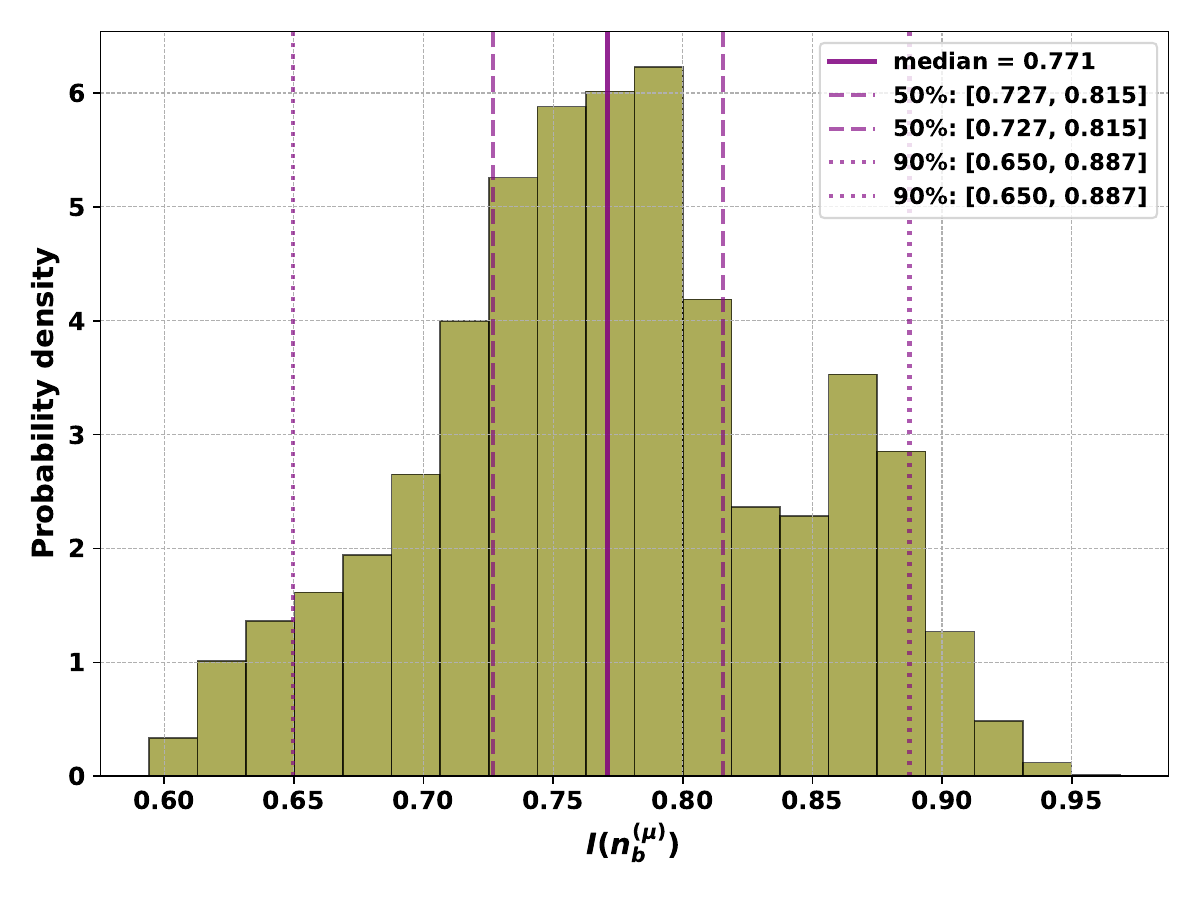}
    \includegraphics[width=\columnwidth]{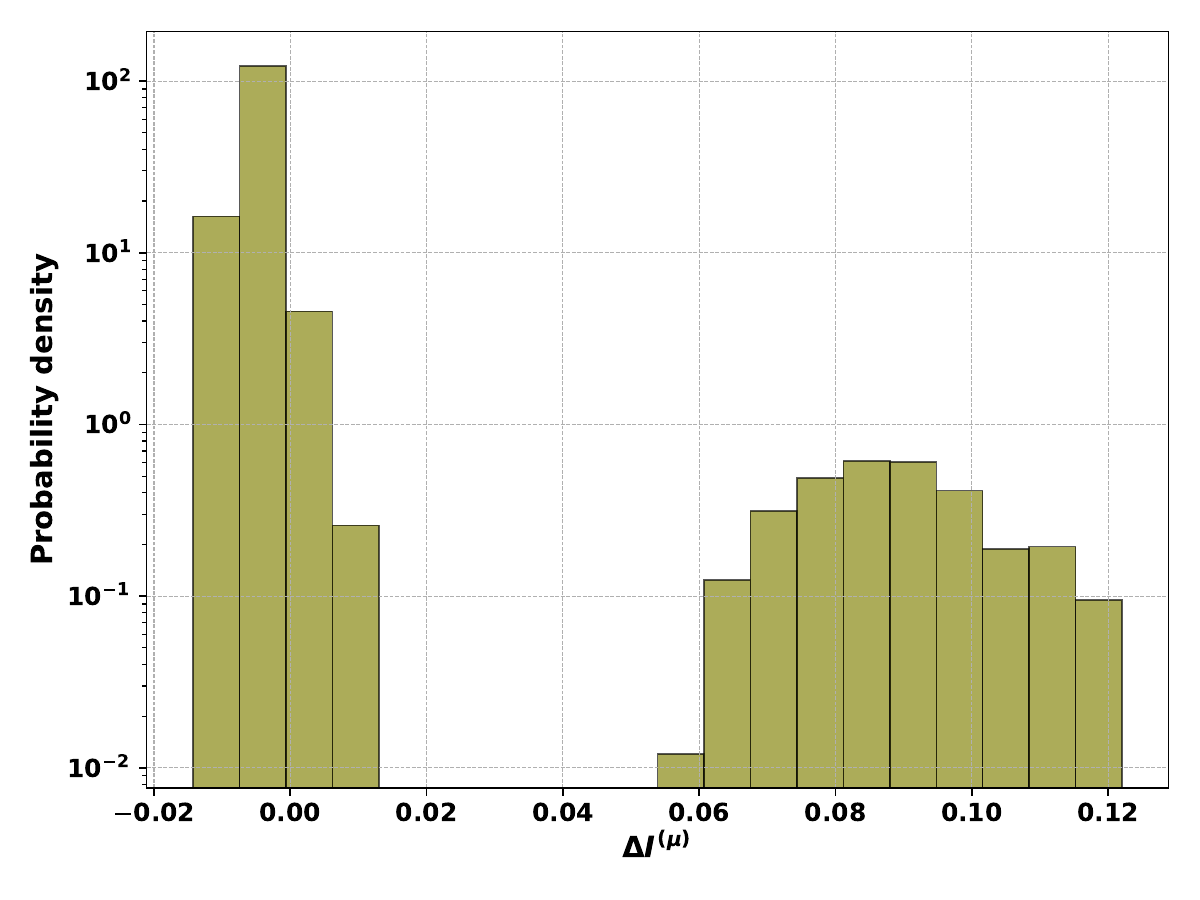}
    \caption{The nuclear asymmetry parameter $I = \frac{n_n-n_p}{n_n+n_p}$ at muon onset for exact symmetry energy (top panel) and the corresponding difference in the nuclear asymmetry parameter at muon onset at exact and quadratic truncated symmetry energy (bottom panel) for the Skyrme samples shown in Fig. ~\ref{fig:pop_skyrme_physical_dist}.}
    \label{fig:muon_onset_asymmetry_histograms}
\end{figure}

\subsubsection{Direct URCA process}\label{subsection:dUrca}

For a cold neutron star relevant for most of the astronomically old populations, Fermi momenta ($k_F$) of different constituent particle species of the neutron star matter determines the onset of direct-Urca process~\cite{Lattimer_prl1991, Steiner_prc2006, LattimerLim_apj2013}. This involves weak-interactions of either electron or muon species in one of the following channels 

\begin{align}\label{eqm1b}
    n \rightarrow p + e + \bar{\nu}_{e} \\
    n \rightarrow p + \mu + \bar{\nu}_{\mu} \\
    p + e \rightarrow n + \nu_{e} \\
    p + \mu \rightarrow n + \nu_{\mu}
\end{align}

\begin{figure}[!h]
    \centering
    \includegraphics[width=\columnwidth]{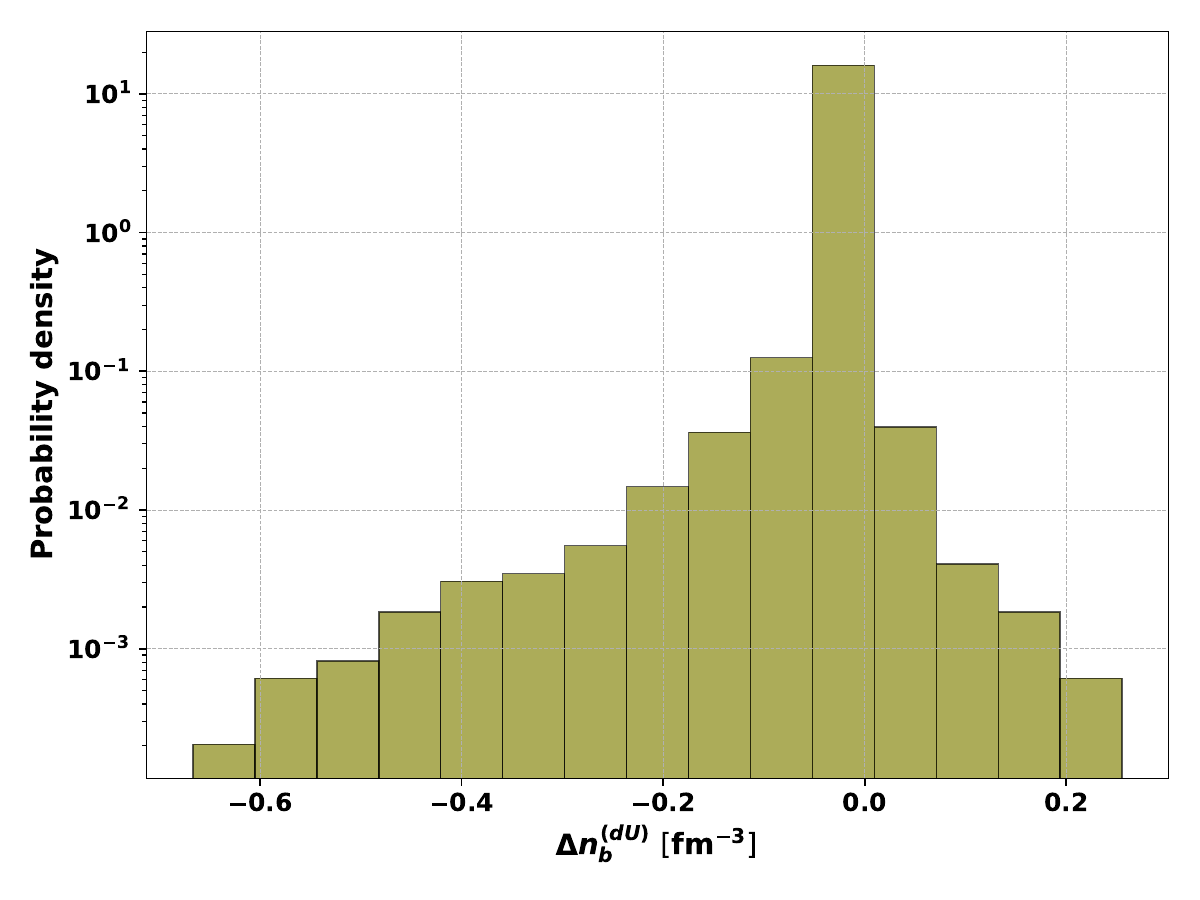}
    \caption{Difference in critical density $n_b^{(dU)}$ for onset of direct Urca processes at exact and quadratic truncated symmetry energy for the Skyrme samples shown in Fig. ~\ref{fig:pop_skyrme_physical_dist}. The onset is evaluated from the Fermi-momentum triangle condition described in the subsection~\ref{subsection:dUrca}.}
    \label{fig:diff_urca_threshold_baryon_ndensity}
\end{figure}

The direct-Urca process can proceed only if energy and momentum are simultaneously conserved among the respective participants of the strongly degenerate matter. At zero temperature, momentum conservation requires that the neutron, proton, and lepton Fermi momenta satisfy the triangle inequality $k_{Fn} \leq k_{Fp} + k_{Fl}$, i.e.\ that the three Fermi momenta be able to form a closed triangle (for the electron channel $l \equiv e$ and for the muon channel $l \equiv \mu$). Equivalently, using Heron's formula, the corresponding triangle area is real and positive, $s(s-k_{Fn})(s-k_{Fp})(s-k_{Fl}) > 0$ with $s = (k_{Fn}+k_{Fp}+k_{Fl})/2$. The lowest baryon density at which this condition is first met defines the direct-Urca onset $n_b^{(dU)}$ in the respective channel. 

The lowest threshold value of baryon number density corresponding to the onset of direct-Urca process, denoted here by $n_{b}^{(dU)}$, in either of these four channels has been computed for each of the valid $\beta$-equilibrated NS EOSs corresponding to the Skyrme model parameters in $\{\theta\}$. However, this critical value $n_{b}^{(dU)}$ for the onset of direct-Urca process depends on the specific order of isospin corrections to the density dependent symmetry energy $E_{sym}(n_{b})$ for a fixed set of values of the Skyrme interaction parameters $\theta_{i}$. The difference in baryon number density for the onset of direct-Urca between the exact order and quadratic order, denoted here as $\Delta n_{b}^{(dU)} = n_{b}^{(dU, exact)} - n_{b}^{(dU, 2)}$, has been computed for the entire population of all the $\sim 85,000$ Skyrme-based EOSs in $\{\theta\}$. 

The distribution of the critical density shift for $d$-Urca onset, $\Delta n_b^{(dU)}$, is inherently asymmetric. While ranging from $-0.67~\mathrm{fm}^{-3}$ to $0.26~\mathrm{fm}^{-3}$, the distribution is sharply peaked near a median of $-0.0009~\mathrm{fm}^{-3}$ (Figure~\ref{fig:diff_urca_threshold_baryon_ndensity}). The corresponding $1\sigma$, $2\sigma$, and $3\sigma$ credible intervals are evaluated as $[-0.0041, -0.0003]$, $[-0.0301, -0.0001]$, and $[-0.1982, 0.0259]~\mathrm{fm}^{-3}$, which translate to $[-0.0255, -0.0017]$, $[-0.1879, -0.0003]$, and $[-1.2385, 0.1620]\,n_{\mathrm{sat}}$ (where $n_{\mathrm{sat}} = 0.16~\mathrm{fm}^{-3}$). Only a minute fraction of the valid Skyrme EOSs exhibit substantial deviations: approximately $0.52\%$, $0.14\%$, $0.062\%$, $0.028\%$, $0.005\%$, and $0.003\%$ yield $|\Delta n_b^{(dU)}| > 0.1, 0.2, 0.3, 0.4, 0.5,$ and $0.6~\mathrm{fm}^{-3}$, respectively, with no EOS exceeding $0.8~\mathrm{fm}^{-3}$. This behavior indicates that the exact treatment favors a lower baryon density threshold for $d$-Urca onset on average than the quadratic-order correction. This shift ultimately implies enhanced thermal cooling rates for astrophysical neutron stars relative to prior estimates. 

This downward shift in the $d$-Urca density threshold carries important structural and thermal implications. Physically, a lower critical density allows the $d$-Urca process to operate across a wider radial profile within the stellar interior, encompassing a larger core volume. Because the efficiency of global neutrino cooling is tied to the total mass of the active $d$-Urca core, this spatial extension drives a more rapid thermal decline. While a precise quantification of the cooling rates depends heavily on the explicit mapping of local thresholds to non-equilibrium transport coefficients, the geometric expansion of the $d$-Urca active zone consistently supports an accelerated cooling scenario for the exact symmetry energy framework. 

\begin{figure}[!h]
    \centering
    \includegraphics[width=\columnwidth]{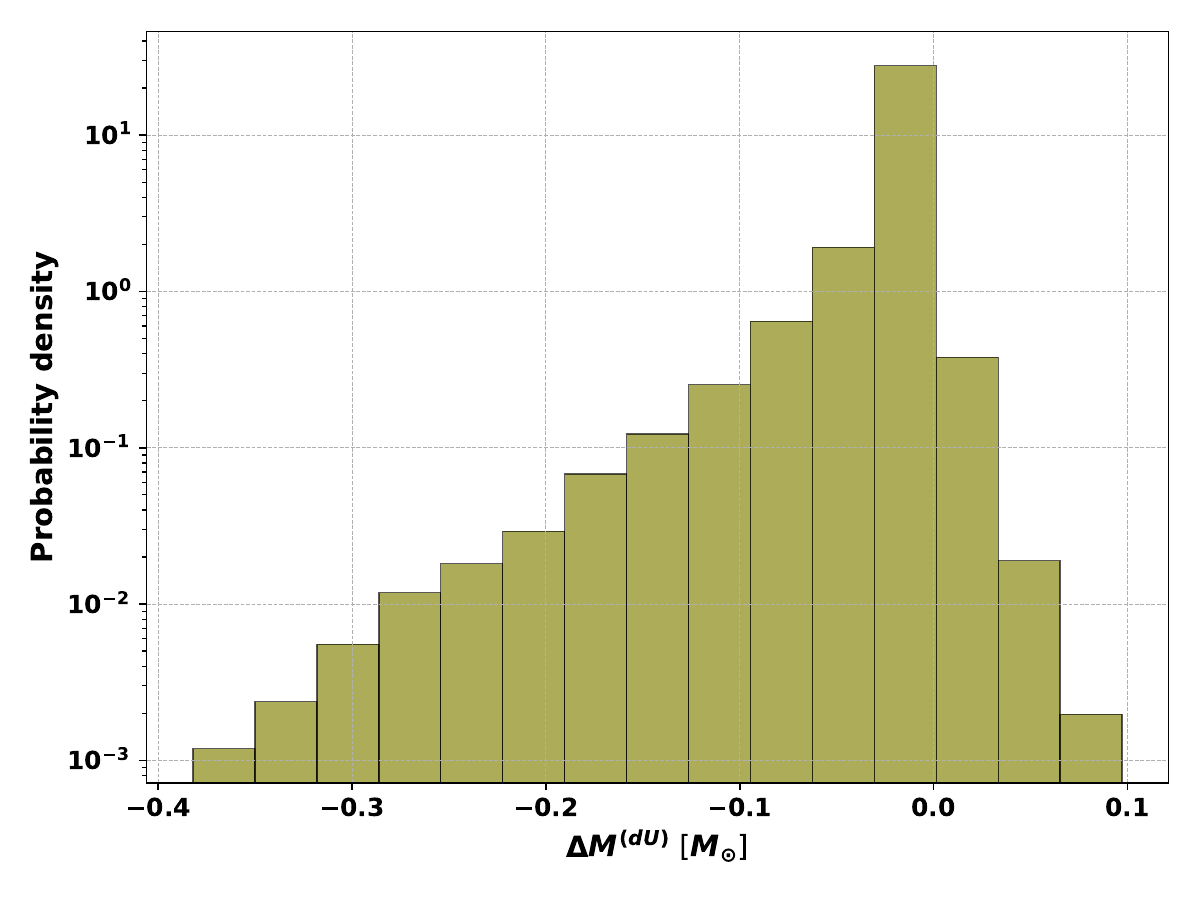}
    \caption{ Difference in total mass of a critical neutron star $M^{(dU)}$ where the onset of direct Urca processes are there at the central densities of the star at exact and quadratic truncated symmetry energy for the Skyrme samples shown in Fig. ~\ref{fig:pop_skyrme_physical_dist}.}
    \label{fig:diff_urca_threshold_mass}
\end{figure}

\begin{figure}[htbp]
    \includegraphics[clip,width=\columnwidth]{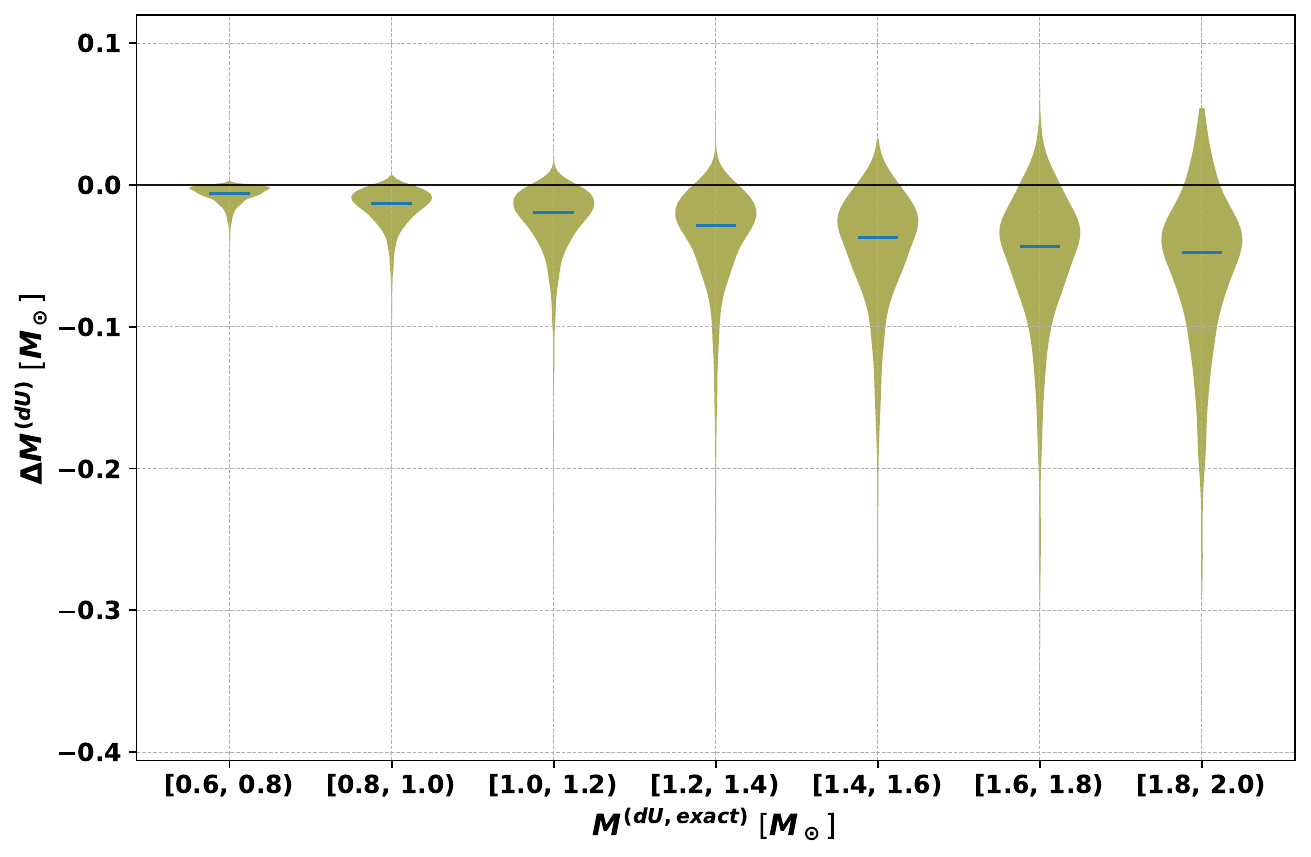}
    \caption{Distribution of the difference in gravitational mass with direct-Urca onset at the center of the star $\Delta M^{(dU)} = M^{(dU,exact)} - M^{(dU,2)}$ within bins of the exact onset mass $M^{(dU,exact)}$. A complementary two-dimensional view of the same distribution, using logarithmic hexagonal binning, is shown in Fig.~\ref{fig:diff_urca_threshold_mass_hexbin} of Appendix~\ref{app_dUrca_onset_deviation}.} 
    \label{fig:delta_M_histograms_by_M}
\end{figure}

The shift in the direct-Urca onset density alters the lowest neutron star mass capable of sustaining rapid interior cooling. We examine this behavior by calculating the minimum mass threshold under the quadratic and exact frameworks ($M^{(dU,2)}$ and $M^{(dU,\mathrm{exact})}$). Figure~\ref{fig:diff_urca_threshold_mass} displays the distribution of the differences, $\Delta M^{(dU)} = M^{(dU,\mathrm{exact})} - M^{(dU,2)}$, across the full $\sim\!85{,}000$ Skyrme-based EOS population in $\{\theta\}$. While the distribution is strongly localized around a median of $-0.003~M_\odot$, the $1\sigma$, $2\sigma$, and $3\sigma$ credible intervals span $[-0.019, 0.000]$, $[-0.080, 0.000]$, and $[-0.218, 0.025]~M_\odot$, respectively. Extreme variations are statistically rare; approximately $26.5\%$, $15.5\%$, $5.3\%$, $1.5\%$, $0.18\%$, and $0.019\%$ of valid models yield $|\Delta M^{(dU)}| > 0.01, 0.02, 0.05, 0.10, 0.20,$ and $0.30\,M_\odot$, respectively, with no model exceeding $0.40\,M_\odot$ and an absolute upper bound of $0.38\,M_\odot$. These findings highlight that accounting for exact isospin corrections to the symmetry energy $E_{\mathrm{sym}}$ is critical for reliably modeling direct-Urca cooling behaviors in observed neutron stars. 

Figure~\ref{fig:delta_M_histograms_by_M} presents the distribution of the direct-Urca threshold mass difference, $\Delta M^{(dU)} = M^{(dU, exact)} - M^{(dU, 2)}$, conditioned on the exact threshold mass, $M^{(dU, exact)}$. To construct this conditional representation, the ensemble of valid Skyrme equations of state (EOSs) is partitioned into successive intervals of width $\Delta M = 0.2\,M_\odot$ over the range $M^{(dU, exact)} \in [1.0, 2.2)\,M_\odot$, with the normalized distribution of $\Delta M^{(dU)}$ evaluated independently within each bin. This approach effectively isolates the dependence of the quadratic-order truncation error on the stellar mass at which direct-Urca onset occurs, quantifying the expected modeling variance across the EOS population for specific mass regimes. Across all mass intervals, the distributions remain heavily concentrated near $\Delta M^{(dU)} \simeq 0$. However, they exhibit a systematic negative skewness, demonstrating that the quadratic-order approximation consistently overestimates the threshold mass compared to the exact treatment.

\section{Summary and Discussion}

In this work, we have systematically investigated the role of higher-order isospin corrections to the nuclear symmetry energy within the framework of Skyrme-like effective nuclear interactions. While the conventional quadratic, or parabolic, approximation to the symmetry energy is widely employed in studies of neutron-rich matter and neutron-star equations of state (EOSs), the extreme isospin asymmetry realized in neutron star interiors motivates a careful examination of the validity of this approximation and the importance of higher-order contributions.

We first analyzed the behaviour of higher-order corrections for the SLy4 Skyrme parametrization and subsequently extended the study to a broad population of physically viable Skyrme EOSs sampled over the multidimensional parameter space of the interaction. The sampled EOSs were constrained to satisfy basic nuclear-matter requirements, thermodynamic stability, causality, positive-definite symmetry energy, and the observational requirement of supporting neutron stars with masses above $2\,M_\odot$.

Our analysis demonstrates that higher-order isospin corrections become increasingly important with increasing baryon density and isospin asymmetry. In particular, deviations between the exact symmetry energy and the conventional quadratic approximation systematically grow in the supra-nuclear density regime relevant to the inner cores of massive neutron stars. While the symmetry energy itself remains reasonably approximated near saturation density, the neutron-proton chemical potential difference exhibits appreciable deviations in highly neutron-rich matter. Consequently, several composition-sensitive quantities, including the proton fraction, electron and muon number densities, and leptonic pressures, receive substantial corrections under $\beta$-equilibrium conditions.

Despite these sizeable modifications in the microscopic composition of dense matter, the global thermodynamic properties of the EOS remain comparatively stable. We find that the total pressure, energy density, and sound speed receive only relatively small corrections from higher-order isospin terms for most physically viable EOSs in the Skyrme population considered here. This indicates that the conventional quadratic approximation captures the bulk stiffness and macroscopic structure of neutron-star matter reasonably well, even though it may not accurately describe the detailed composition and chemical equilibrium conditions at high density.

The population study further reveals significant EOS-to-EOS variation in the magnitude of higher-order corrections, particularly at supra-nuclear densities. The spread in the higher-order symmetry-energy coefficients, such as $K_{\rm sym}$ and $Q_{\rm sym}$, suggests that uncertainties associated with the poorly constrained isovector sector of the nuclear interaction become increasingly amplified in dense matter. These effects may have important implications for neutron-star cooling, neutrino emissivity, transport properties, threshold conditions for rapid cooling processes, and other composition-sensitive observables.

The present work is limited to Skyrme-like effective interactions and cold degenerate matter composed of nucleons and leptons under $\beta$-equilibrium. Extensions to finite-temperature matter, hyperonic or deconfined degrees of freedom, and relativistic mean-field or microscopic many-body approaches would provide important complementary perspectives on the role of higher-order isospin effects in dense matter, and are left for future considerations. Future multimessenger observations of neutron stars, combined with improved experimental and theoretical constraints on the nuclear symmetry energy, may further help constrain the higher-order isovector structure of the nuclear EOS.

Overall, our results demonstrate that higher-order isospin corrections, although subdominant for the bulk neutron-star EOS, can play a quantitatively important role in determining the microscopic composition and thermodynamic equilibrium of dense neutron-rich matter in neutron-star interiors.

Beyond the effects of higher-order symmetry energy corrections on the bulk and compositional properties of cold, dense, $\beta$-equilibrated nuclear matter, the canonical description of an old neutron star these corrections may also prominently impact out-of-equilibrium transport properties. While large compositional discrepancies observed across a marginal subset of the Skyrme EOS population qualitatively hint at these transport effects, a rigorous quantitative description of these quantities remains beyond the scope of this work.

Because higher-order isospin corrections significantly modify the neutron-proton chemical potential difference, they are expected to influence $\beta$-equilibration rates~\cite{Alford_etal_univ2021} and the associated bulk viscosity of dense matter~\cite{Sawyer_prd1989, Haensel_etal_AA2000, Jones_prd2001}, both of which are fundamentally governed by departures from chemical equilibrium. Furthermore, the explicit dependence of thermal conductivity~\cite{Baiko_aa2001, ShterninYakovlev_prd2007} and shear viscosity~\cite{ShterninYakovlev_prd2008} on electron and muon fractions suggests that these transport coefficients may also be modulated by higher-order isospin contributions, tracking the deviations observed in the leptonic number densities.

The sensitivity of the equilibrium stellar composition to higher-order isospin corrections may carry critical implications for binary neutron-star mergers as well. Notably, the electron fraction of the dynamic and wind ejecta serves as a key governing parameter for $r$-process nucleosynthesis and the resulting kilonova light curves~\cite{Siegel_NatRP2022, Metzger_lrr2020}. 

\acknowledgements 

A.M. acknowledges support from the DST-SERB Start-up Research Grant No. SRG/2020/001290. A.M. thanks the Institute for Nuclear Theory at the University of Washington for its kind hospitality and useful discussion. 

\bibliography{references}

@ARTICLE{LattimerPrakash_2004Sci,
       author = {{Lattimer}, J.~M. and {Prakash}, M.},
        title = "{The Physics of Neutron Stars}",
      journal = {Science},
         year = 2004,
        month = apr,
       volume = {304},
       number = {5670},
        pages = {536-542},
          doi = {10.1126/science.1090720},
archivePrefix = {arXiv},
       eprint = {astro-ph/0405262},
 primaryClass = {astro-ph},
       adsurl = {https://ui.adsabs.harvard.edu/abs/2004Sci...304..536L}
}

@ARTICLE{LattimerPrakash_2007PR,
       author = {{Lattimer}, James M. and {Prakash}, Madappa},
        title = "{Neutron star observations: Prognosis for equation of state constraints}",
      journal = {Physics Reports},
         year = 2007,
        month = apr,
       volume = {442},
       number = {1-6},
        pages = {109-165},
          doi = {10.1016/j.physrep.2007.02.003},
archivePrefix = {arXiv},
       eprint = {astro-ph/0612440},
 primaryClass = {astro-ph},
       adsurl = {https://ui.adsabs.harvard.edu/abs/2007PhR...442..109L}
}

@ARTICLE{Busza_etal_HIC_2018ARNPS,
       author = {{Busza}, Wit and {Rajagopal}, Krishna and {van der Schee}, Wilke},
        title = "{Heavy Ion Collisions: The Big Picture and the Big Questions}",
      journal = {Annual Review of Nuclear and Particle Science},
         year = 2018,
        month = oct,
       volume = {68},
       number = {1},
        pages = {339-376},
          doi = {10.1146/annurev-nucl-101917-020852},
archivePrefix = {arXiv},
       eprint = {1802.04801},
 primaryClass = {hep-ph},
       adsurl = {https://ui.adsabs.harvard.edu/abs/2018ARNPS..68..339B}
}

@ARTICLE{Walecka_1975PLB,
       author = {{Walecka}, J.~D.},
        title = "{Equation of state for neutron matter at finite T in a relativistic mean-field theory}",
      journal = {Physics Letters B},
         year = 1975,
        month = oct,
       volume = {59},
       number = {2},
        pages = {109-112},
          doi = {10.1016/0370-2693(75)90678-4},
       adsurl = {https://ui.adsabs.harvard.edu/abs/1975PhLB...59..109W}
}

@ARTICLE{WaleckaSerot_1979PLB,
       author = {{Serot}, Brian D. and {Walecka}, J.~D.},
        title = "{Properties of finite nuclei in a relativistic quantum field theory}",
      journal = {Physics Letters B},
         year = 1979,
        month = nov,
       volume = {87},
       number = {3},
        pages = {172-176},
          doi = {10.1016/0370-2693(79)90957-2},
       adsurl = {https://ui.adsabs.harvard.edu/abs/1979PhLB...87..172S}
}

@article{Bethe_1971AnnuRev,
   author = "Bethe, H A",
   title = "Theory of Nuclear Matter", 
   journal= "Annual Review of Nuclear and Particle Science",
   year = "1971",
   volume = "21",
   number = "Volume 21, ",
   pages = "93-244",
   doi = "https://doi.org/10.1146/annurev.ns.21.120171.000521",
   url = "https://www.annualreviews.org/content/journals/10.1146/annurev.ns.21.120171.000521",
   publisher = "Annual Reviews",
   issn = "1545-4134",
   type = "Journal Article",
  }

@article{Lee_etal_1998PRC,
  title = {Nuclear symmetry energy},
  author = {Lee, C.-H. and Kuo, T. T. S. and Li, G. Q. and Brown, G. E.},
  journal = {Phys. Rev. C},
  volume = {57},
  issue = {6},
  pages = {3488--3491},
  numpages = {0},
  year = {1998},
  month = {Jun},
  publisher = {American Physical Society},
  doi = {10.1103/PhysRevC.57.3488},
  url = {https://link.aps.org/doi/10.1103/PhysRevC.57.3488}
}

@article{GonzalezBoquera_etal_2017PRC,
  title = {Higher-order symmetry energy and neutron star core-crust transition with Gogny forces},
  author = {Gonzalez-Boquera, C. and Centelles, M. and Vi\~nas, X. and Rios, A.},
  journal = {Phys. Rev. C},
  volume = {96},
  issue = {6},
  pages = {065806},
  numpages = {22},
  year = {2017},
  month = {Dec},
  publisher = {American Physical Society},
  doi = {10.1103/PhysRevC.96.065806},
  url = {https://link.aps.org/doi/10.1103/PhysRevC.96.065806}
}

@article{KatayamaSaito_2013PRC,
  title = {Properties of dense, asymmetric nuclear matter in Dirac-Brueckner-Hartree-Fock approach},
  author = {Katayama, Tetsuya and Saito, Koichi},
  journal = {Phys. Rev. C},
  volume = {88},
  issue = {3},
  pages = {035805},
  numpages = {11},
  year = {2013},
  month = {Sep},
  publisher = {American Physical Society},
  doi = {10.1103/PhysRevC.88.035805},
  url = {https://link.aps.org/doi/10.1103/PhysRevC.88.035805}
}

@article{Sammarruca_2014article,
author = {Sammarruca, Francesca},
year = {2014},
month = {10},
pages = {851-879},
title = {Recent Advances in Microscopic Approaches to Nuclear Matter and Symmetry Energy},
volume = {6},
journal = {Symmetry},
doi = {10.3390/sym6040851}
}

@ARTICLE{APR_1998PRC,
       author = {{Akmal}, A. and {Pandharipande}, V.~R. and {Ravenhall}, D.~G.},
        title = "{Equation of state of nucleon matter and neutron star structure}",
      journal = {\prc},
         year = 1998,
        month = sep,
       volume = {58},
       number = {3},
        pages = {1804-1828},
          doi = {10.1103/PhysRevC.58.1804},
archivePrefix = {arXiv},
       eprint = {nucl-th/9804027},
 primaryClass = {nucl-th},
       adsurl = {https://ui.adsabs.harvard.edu/abs/1998PhRvC..58.1804A}
}

@article{PandharipandeWiringa_RMP1979,
  title = {Variations on a theme of nuclear matter},
  author = {Pandharipande, V. R. and Wiringa, R. B.},
  journal = {Rev. Mod. Phys.},
  volume = {51},
  issue = {4},
  pages = {821--861},
  numpages = {0},
  year = {1979},
  month = {Oct},
  publisher = {American Physical Society},
  doi = {10.1103/RevModPhys.51.821},
  url = {https://link.aps.org/doi/10.1103/RevModPhys.51.821}
}

@article{Skyrme_1958,
title = {The effective nuclear potential},
journal = {Nuclear Physics},
volume = {9},
number = {4},
pages = {615-634},
year = {1958},
issn = {0029-5582},
doi = {https://doi.org/10.1016/0029-5582(58)90345-6},
url = {https://www.sciencedirect.com/science/article/pii/0029558258903456},
author = {T.H.R. Skyrme}
}

@article{VautherinBrink_1972PRC,
  title = {Hartree-Fock Calculations with Skyrme's Interaction. I. Spherical Nuclei},
  author = {Vautherin, D. and Brink, D. M.},
  journal = {Phys. Rev. C},
  volume = {5},
  issue = {3},
  pages = {626--647},
  numpages = {0},
  year = {1972},
  month = {3},
  publisher = {American Physical Society},
  doi = {10.1103/PhysRevC.5.626},
  url = {https://link.aps.org/doi/10.1103/PhysRevC.5.626}
}

@ARTICLE{Chabanat_etal_1997NuPhA,
       author = {{Chabanat}, E. and {Bonche}, P. and {Haensel}, P. and {Meyer}, J. and {Schaeffer}, R.},
        title = "{A Skyrme parametrization from subnuclear to neutron star densities}",
      journal = {Nuclear Physics A},
         year = 1997,
        month = feb,
       volume = {627},
        pages = {710-746},
          doi = {10.1016/S0375-9474(97)00596-4},
       adsurl = {https://ui.adsabs.harvard.edu/abs/1997NuPhA.627..710C}
}

@article{Chabanat_etal_1998NuPhA,
title = {A Skyrme parametrization from subnuclear to neutron star densities Part II. Nuclei far from stabilities},
journal = {Nuclear Physics A},
volume = {635},
number = {1},
pages = {231-256},
year = {1998},
issn = {0375-9474},
doi = {https://doi.org/10.1016/S0375-9474(98)00180-8},
url = {https://www.sciencedirect.com/science/article/pii/S0375947498001808},
author = {E. Chabanat and P. Bonche and P. Haensel and J. Meyer and R. Schaeffer}
}

@ARTICLE{DouchinHaensel_AA2001,
       author = {{Douchin}, F. and {Haensel}, P.},
        title = "{A unified equation of state of dense matter and neutron star structure}",
      journal = {Astrophysics and Astronomy},
         year = 2001,
        month = dec,
       volume = {380},
        pages = {151-167},
          doi = {10.1051/0004-6361:20011402},
archivePrefix = {arXiv},
       eprint = {astro-ph/0111092},
 primaryClass = {astro-ph},
       adsurl = {https://ui.adsabs.harvard.edu/abs/2001A&A...380..151D}
}

@article{Horowitz_etal_PREX_PRC2020,
  title = {Insights into nuclear saturation density from parity-violating electron scattering},
  author = {Horowitz, C. J. and Piekarewicz, J. and Reed, Brendan},
  journal = {Phys. Rev. C},
  volume = {102},
  issue = {4},
  pages = {044321},
  numpages = {6},
  year = {2020},
  month = {Oct},
  publisher = {American Physical Society},
  doi = {10.1103/PhysRevC.102.044321},
  url = {https://link.aps.org/doi/10.1103/PhysRevC.102.044321}
}

@article{Drischler_etal_PRL2019,
  title = {Chiral Interactions up to Next-to-Next-to-Next-to-Leading Order and Nuclear Saturation},
  author = {Drischler, C. and Hebeler, K. and Schwenk, A.},
  journal = {Phys. Rev. Lett.},
  volume = {122},
  issue = {4},
  pages = {042501},
  numpages = {6},
  year = {2019},
  month = {Jan},
  publisher = {American Physical Society},
  doi = {10.1103/PhysRevLett.122.042501},
  url = {https://link.aps.org/doi/10.1103/PhysRevLett.122.042501}
}

@article{Chen_etal_PRC2014,
  title = {Building relativistic mean field models for finite nuclei and neutron stars},
  author = {Chen, Wei-Chia and Piekarewicz, J.},
  journal = {Phys. Rev. C},
  volume = {90},
  issue = {4},
  pages = {044305},
  numpages = {17},
  year = {2014},
  month = {Oct},
  publisher = {American Physical Society},
  doi = {10.1103/PhysRevC.90.044305},
  url = {https://link.aps.org/doi/10.1103/PhysRevC.90.044305}
}

@ARTICLE{diff_sym_en_LattimerGroup_prc2024,
       author = {{Sun}, Boyang and {Bhattiprolu}, Saketh and {Lattimer}, James M.},
        title = "{Compiled properties of nucleonic matter and nuclear and neutron star models from nonrelativistic and relativistic interactions}",
      journal = {\prc},
         year = 2024,
        month = may,
       volume = {109},
       number = {5},
          eid = {055801},
        pages = {055801},
          doi = {10.1103/PhysRevC.109.055801},
archivePrefix = {arXiv},
       eprint = {2311.00843},
 primaryClass = {nucl-th},
       adsurl = {https://ui.adsabs.harvard.edu/abs/2024PhRvC.109e5801S}
}

@ARTICLE{Forbes_etal_prd2019,
       author = {{Forbes}, Michael McNeil and {Bose}, Sukanta and {Reddy}, Sanjay and {Zhou}, Dake and {Mukherjee}, Arunava and {De}, Soumi},
        title = "{Constraining the neutron-matter equation of state with gravitational waves}",
      journal = {\prd},
         year = 2019,
        month = oct,
       volume = {100},
       number = {8},
          eid = {083010},
        pages = {083010},
          doi = {10.1103/PhysRevD.100.083010},
       adsurl = {https://ui.adsabs.harvard.edu/abs/2019PhRvD.100h3010F}
}

@ARTICLE{Demorest_etal_nat2010,
       author = {{Demorest}, P.~B. and {Pennucci}, T. and {Ransom}, S.~M. and {Roberts}, M.~S.~E. and {Hessels}, J.~W.~T.},
        title = "{A two-solar-mass neutron star measured using Shapiro delay}",
      journal = {\nat},
         year = 2010,
        month = oct,
       volume = {467},
       number = {7319},
        pages = {1081-1083},
          doi = {10.1038/nature09466},
archivePrefix = {arXiv},
       eprint = {1010.5788},
 primaryClass = {astro-ph.HE},
       adsurl = {https://ui.adsabs.harvard.edu/abs/2010Natur.467.1081D}
}

@ARTICLE{Antoniodis_sci2013,
       author = {{Antoniadis}, John and {Freire}, Paulo C.~C. and {Wex}, Norbert and {Tauris}, Thomas M. and {Lynch}, Ryan S. and {van Kerkwijk}, Marten H. and {Kramer}, Michael and {Bassa}, Cees and {Dhillon}, Vik S. and {Driebe}, Thomas and {Hessels}, Jason W.~T. and {Kaspi}, Victoria M. and {Kondratiev}, Vladislav I. and {Langer}, Norbert and {Marsh}, Thomas R. and {McLaughlin}, Maura A. and {Pennucci}, Timothy T. and {Ransom}, Scott M. and {Stairs}, Ingrid H. and {van Leeuwen}, Joeri and {Verbiest}, Joris P.~W. and {Whelan}, David G.},
        title = "{A Massive Pulsar in a Compact Relativistic Binary}",
      journal = {Science},
         year = 2013,
        month = apr,
       volume = {340},
       number = {6131},
        pages = {448},
          doi = {10.1126/science.1233232},
archivePrefix = {arXiv},
       eprint = {1304.6875},
 primaryClass = {astro-ph.HE},
       adsurl = {https://ui.adsabs.harvard.edu/abs/2013Sci...340..448A}
}

@ARTICLE{Romani_etal_apj2022,
       author = {{Romani}, Roger W. and {Kandel}, D. and {Filippenko}, Alexei V. and {Brink}, Thomas G. and {Zheng}, WeiKang},
        title = "{PSR J0952-0607: The Fastest and Heaviest Known Galactic Neutron Star}",
      journal = {Astroph. Journal Letters},
         year = 2022,
        month = aug,
       volume = {934},
       number = {2},
          eid = {L17},
        pages = {L17},
          doi = {10.3847/2041-8213/ac8007},
archivePrefix = {arXiv},
       eprint = {2207.05124},
 primaryClass = {astro-ph.HE},
       adsurl = {https://ui.adsabs.harvard.edu/abs/2022ApJ...934L..17R}
}

@ARTICLE{Steiner_prc2006,
       author = {{Steiner}, Andrew W.},
        title = "{High-density symmetry energy and direct Urca process}",
      journal = {\prc},
         year = 2006,
        month = oct,
       volume = {74},
       number = {4},
          eid = {045808},
        pages = {045808},
          doi = {10.1103/PhysRevC.74.045808},
archivePrefix = {arXiv},
       eprint = {nucl-th/0607040},
 primaryClass = {nucl-th},
       adsurl = {https://ui.adsabs.harvard.edu/abs/2006PhRvC..74d5808S}
}

@ARTICLE{LattimerLim_apj2013,
       author = {{Lattimer}, James M. and {Lim}, Yeunhwan},
        title = "{Constraining the Symmetry Parameters of the Nuclear Interaction}",
      journal = {\apj},
         year = 2013,
        month = jul,
       volume = {771},
       number = {1},
          eid = {51},
        pages = {51},
          doi = {10.1088/0004-637X/771/1/51},
archivePrefix = {arXiv},
       eprint = {1203.4286},
 primaryClass = {nucl-th},
       adsurl = {https://ui.adsabs.harvard.edu/abs/2013ApJ...771...51L}
}

@ARTICLE{Haensel_etal_AA2000,
       author = {{Haensel}, P. and {Levenfish}, K.~P. and {Yakovlev}, D.~G.},
        title = "{Bulk viscosity in superfluid neutron star cores. I. Direct Urca processes in npemu matter}",
      journal = {A\&A},
         year = 2000,
        month = may,
       volume = {357},
        pages = {1157-1169},
          doi = {10.48550/arXiv.astro-ph/0004183},
archivePrefix = {arXiv},
       eprint = {astro-ph/0004183},
 primaryClass = {astro-ph},
       adsurl = {https://ui.adsabs.harvard.edu/abs/2000A&A...357.1157H}
}

@ARTICLE{Sawyer_prd1989,
       author = {{Sawyer}, Raymond F.},
        title = "{Bulk viscosity of hot neutron-star matter and the maximum rotation rates of neutron stars}",
      journal = {\prd},
         year = 1989,
        month = jun,
       volume = {39},
       number = {12},
        pages = {3804-3806},
          doi = {10.1103/PhysRevD.39.3804},
       adsurl = {https://ui.adsabs.harvard.edu/abs/1989PhRvD..39.3804S}
}

@ARTICLE{Jones_prd2001,
       author = {{Jones}, P.~B.},
        title = "{Bulk viscosity of neutron-star matter}",
      journal = {\prd},
         year = 2001,
        month = oct,
       volume = {64},
       number = {8},
          eid = {084003},
        pages = {084003},
          doi = {10.1103/PhysRevD.64.084003},
       adsurl = {https://ui.adsabs.harvard.edu/abs/2001PhRvD..64h4003J}
}

@ARTICLE{Baiko_aa2001,
       author = {{Baiko}, D.~A. and {Haensel}, P. and {Yakovlev}, D.~G.},
        title = "{Thermal conductivity of neutrons in neutron star cores}",
      journal = {A\&A},
         year = 2001,
        month = jul,
       volume = {374},
        pages = {151-163},
          doi = {10.1051/0004-6361:20010621},
archivePrefix = {arXiv},
       eprint = {astro-ph/0105105},
 primaryClass = {astro-ph},
       adsurl = {https://ui.adsabs.harvard.edu/abs/2001A&A...374..151B}
}

@ARTICLE{Metzger_lrr2020,
       author = {{Metzger}, Brian D.},
        title = "{Kilonovae}",
      journal = {Living Reviews in Relativity},
         year = 2020,
        month = dec,
       volume = {23},
       number = {1},
          eid = {1},
        pages = {1},
          doi = {10.1007/s41114-019-0024-0},
archivePrefix = {arXiv},
       eprint = {1910.01617},
 primaryClass = {astro-ph.HE},
       adsurl = {https://ui.adsabs.harvard.edu/abs/2020LRR....23....1M}
}

@ARTICLE{Siegel_NatRP2022,
       author = {{Siegel}, Daniel M.},
        title = "{r-Process nucleosynthesis in gravitational-wave and other explosive astrophysical events}",
      journal = {Nature Reviews Physics},
         year = 2022,
        month = apr,
       volume = {4},
       number = {5},
        pages = {306-318},
          doi = {10.1038/s42254-022-00439-1},
       adsurl = {https://ui.adsabs.harvard.edu/abs/2022NatRP...4..306S}
}

@ARTICLE{Alford_etal_univ2021,
       author = {{Alford}, Mark G. and {Haber}, Alexander and {Harris}, Steven P. and {Zhang}, Ziyuan},
        title = "{Beta Equilibrium Under Neutron Star Merger Conditions}",
      journal = {Universe},
         year = 2021,
        month = oct,
       volume = {7},
       number = {11},
          eid = {399},
        pages = {399},
          doi = {10.3390/universe7110399},
archivePrefix = {arXiv},
       eprint = {2108.03324},
 primaryClass = {nucl-th},
       adsurl = {https://ui.adsabs.harvard.edu/abs/2021Univ....7..399A}
}

@ARTICLE{ShterninYakovlev_prd2007,
       author = {{Shternin}, P.~S. and {Yakovlev}, D.~G.},
        title = "{Electron-muon heat conduction in neutron star cores via the exchange of transverse plasmons}",
      journal = {\prd},
         year = 2007,
        month = may,
       volume = {75},
       number = {10},
          eid = {103004},
        pages = {103004},
          doi = {10.1103/PhysRevD.75.103004},
archivePrefix = {arXiv},
       eprint = {0705.1963},
 primaryClass = {astro-ph},
       adsurl = {https://ui.adsabs.harvard.edu/abs/2007PhRvD..75j3004S}
}

@ARTICLE{ShterninYakovlev_prd2008,
       author = {{Shternin}, P.~S. and {Yakovlev}, D.~G.},
        title = "{Shear viscosity in neutron star cores}",
      journal = {\prd},
         year = 2008,
        month = sep,
       volume = {78},
       number = {6},
          eid = {063006},
        pages = {063006},
          doi = {10.1103/PhysRevD.78.063006},
archivePrefix = {arXiv},
       eprint = {0808.2018},
 primaryClass = {astro-ph},
       adsurl = {https://ui.adsabs.harvard.edu/abs/2008PhRvD..78f3006S}
}

@article{Lattimer_prl1991,
  title = {Direct URCA process in neutron stars},
  author = {Lattimer, James M. and Pethick, C. J. and Prakash, Madappa and Haensel, Pawel},
  journal = {Phys. Rev. Lett.},
  volume = {66},
  issue = {21},
  pages = {2701--2704},
  numpages = {0},
  year = {1991},
  month = {5},
  publisher = {American Physical Society},
  doi = {10.1103/PhysRevLett.66.2701},
  url = {https://link.aps.org/doi/10.1103/PhysRevLett.66.2701}
}

@article{BALDO_ppnp2016_203,
title = {The nuclear symmetry energy},
journal = {Progress in Particle and Nuclear Physics},
volume = {91},
pages = {203-258},
year = {2016},
issn = {0146-6410},
doi = {https://doi.org/10.1016/j.ppnp.2016.06.006},
url = {https://www.sciencedirect.com/science/article/pii/S0146641016300254},
author = {M. Baldo and G.F. Burgio}
}

@article{Lattimer_particles2023,
   title={Constraints on Nuclear Symmetry Energy Parameters},
   volume={6},
   ISSN={2571-712X},
   url={http://dx.doi.org/10.3390/particles6010003},
   DOI={10.3390/particles6010003},
   number={1},
   journal={Particles},
   publisher={MDPI AG},
   author={Lattimer, James M.},
   year={2023},
   month=Jan, pages={30–56} }

@ARTICLE{Raithel_ApJ2019,
       author = {{Raithel}, Carolyn A. and {{\"O}zel}, Feryal},
        title = "{Measurement of the Nuclear Symmetry Energy Parameters from Gravitational-wave Events}",
      journal = {\apj},
         year = 2019,
        month = nov,
       volume = {885},
       number = {2},
          eid = {121},
        pages = {121},
          doi = {10.3847/1538-4357/ab48e6},
archivePrefix = {arXiv},
       eprint = {1908.00018},
 primaryClass = {astro-ph.HE},
       adsurl = {https://ui.adsabs.harvard.edu/abs/2019ApJ...885..121R}
}

\section{Appendix}\label{appendix}

\subsection{Muon appearance}

\subsubsection{Self-consistent solution for the isospin asymmetry parameter}\label{app_solver}

For a fixed baryon number density $n_b$, the isospin asymmetry parameter $I$ is determined self-consistently by simultaneously imposing the conditions of charge neutrality and $\beta$-equilibrium for $npe\mu$ matter. The solution is obtained iteratively using Eq.~\ref{chemical-potential-order-corr} for truncated symmetry-energy expansions and Eq.~\ref{chemical-potential-exact} 
for the exact treatment. Numerical convergence is achieved when successive iterations satisfy:

\begin{equation}\label{eq:app_tol_condition}
    |I_{k+1}-I_k| < \mathrm{tol}(I)
\end{equation}

Under $\beta$-equilibrium, the difference between the neutron and proton chemical potentials, defined as $\Delta\mu_{np} \equiv \mu_n-\mu_p$, is constrained by the leptonic chemical potentials such that $\mu_e = \mu_\mu = \Delta\mu_{np}$. 
Consequently, the leptonic sector is highly sensitive to the self-consistent convergence of the asymmetry parameter $I$. Figure~\ref{fig:sly4_mu_diff} illustrates the dependence of $\Delta\mu_{np}$ on $I$ for various orders of corrections to  symmetry energy $E_{\mathrm{sym}}^{(2n)}$, contrasting the truncated expansions with the exact treatment described above. 

\subsubsection{Muon-threshold convergence test}\label{app_solver_tol}

\begin{table}[H]
    \centering
    \begin{tabular}{c|c|c|c|c}
        \toprule
        tol ($I$) & $exact$  & $n=1$    & $n=2$    & $n=3$ \\
        \midrule
        $10^{-2}$ & 0.119054 & 0.126554 & 0.122354 & 0.120704 \\
        $10^{-3}$ & 0.120704 & 0.128354 & 0.124154 & 0.122654 \\
        $10^{-4}$ & 0.121004 & 0.128504 & 0.124304 & 0.122804 \\
        $10^{-5}$ & 0.121004 & 0.128654 & 0.124304 & 0.122804 \\
        $10^{-6}$ & 0.121004 & 0.128654 & 0.124304 & 0.122804 \\
        \bottomrule
    \end{tabular}
    \caption{Baryon number density $n_b$ (in fm$^{-3}$) at which muon appears in $\beta$-equilibrated matter are quoted in this table for the Skyrme model interaction parameter corresponding to SLy4 EOS. The rows demonstrate convergence with the iterative solver with tolerance $\mathrm{tol}(I)$. Results are shown for the symmetry energy computed at various order corrections from $\mathcal{O}(I^{2n})$ expansions for symmetry energy $E_{sym}^{(2n)}$ corresponding to the $exact$ (Eq.~\ref{asymm-energy_i}), $n=1$ (Eq.~\ref{asymm-O(1)}), $n=2$ (Eq.~\ref{asymm-O(2)}), and $n=3$ (Eq.~\ref{asymm-O(3)}), see Subsection~\ref{sym_en_diff_order} for details).} 
    \label{tab:muon_threshold_sly4}
\end{table}

To evaluate the numerical stability and precision of the self-consistent solution for the isospin asymmetry parameter $I$, we analyzed the numerical convergence of the threshold baryon number density for the muon onset in $\beta$-equilibrated matter. The tolerance parameter $\mathrm{tol}(I)$ was systematically varied to verify the stability of this threshold density, 
providing a rigorous consistency check on the iterative solver across the density range.

Table~\ref{tab:muon_threshold_sly4} summarizes the baryon number densities ($n_b$) corresponding to the onset of muon formation for the interaction parameters of SLy4. The results are reported for the symmetry energy computed at various order corrections from $\mathcal{O}(I^{2n})$ expansions for symmetry energy $E_{sym}^{(2n)}$ corresponding to $exact$ (Eq.~\ref{asymm-energy_i}), $n=1$ (Eq.~\ref{asymm-O(1)}), $n=2$ (Eq.~\ref{asymm-O(2)}), and $n=3$ (Eq.~\ref{asymm-O(3)}).

\subsection{Additional leptonic-sector plots}\label{app_aadn_plots}

\begin{figure}[!h]
    \centering
    \includegraphics[width=0.5\textwidth]{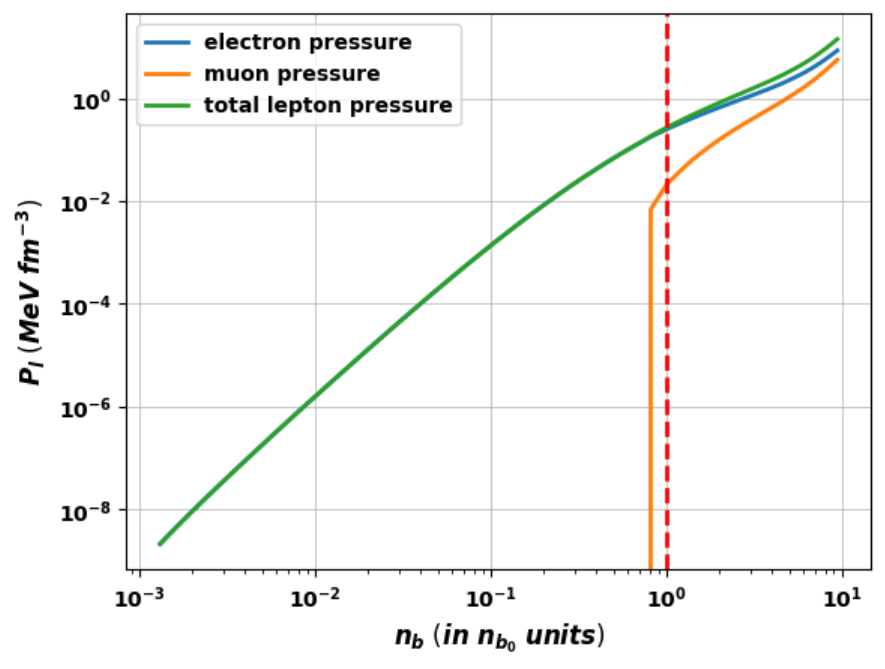}
    \caption{For $\beta$-equilibrium matter with the exact treatment of isospin asymmetry corrections for the case of SLy4 Skyrme parameters, the dependence of leptonic pressure (with individual contributions from electron and muon components) over the baryon number density $n_{b} \in [10^{-6}, 1.5] \text{ fm}^{-3}$ is presented here. The onset of muon appearance at $n_b = 0.121004~\text{fm}^{-3}$, and the saturation number density $n_{b0} \approx 0.1595 \text{ fm}^{-3}$ are also marked. The differences in baryon number density for muon onset are presented in Table~\ref{tab:muon_threshold_sly4}.}
    \label{fig:sly4_lepton_pressure}
\end{figure}

\begin{figure*}[htbp]
    \raggedright (a)\hspace*{\columnwidth}(b)\\[-0.5cm]
    \includegraphics[clip,width=\columnwidth]{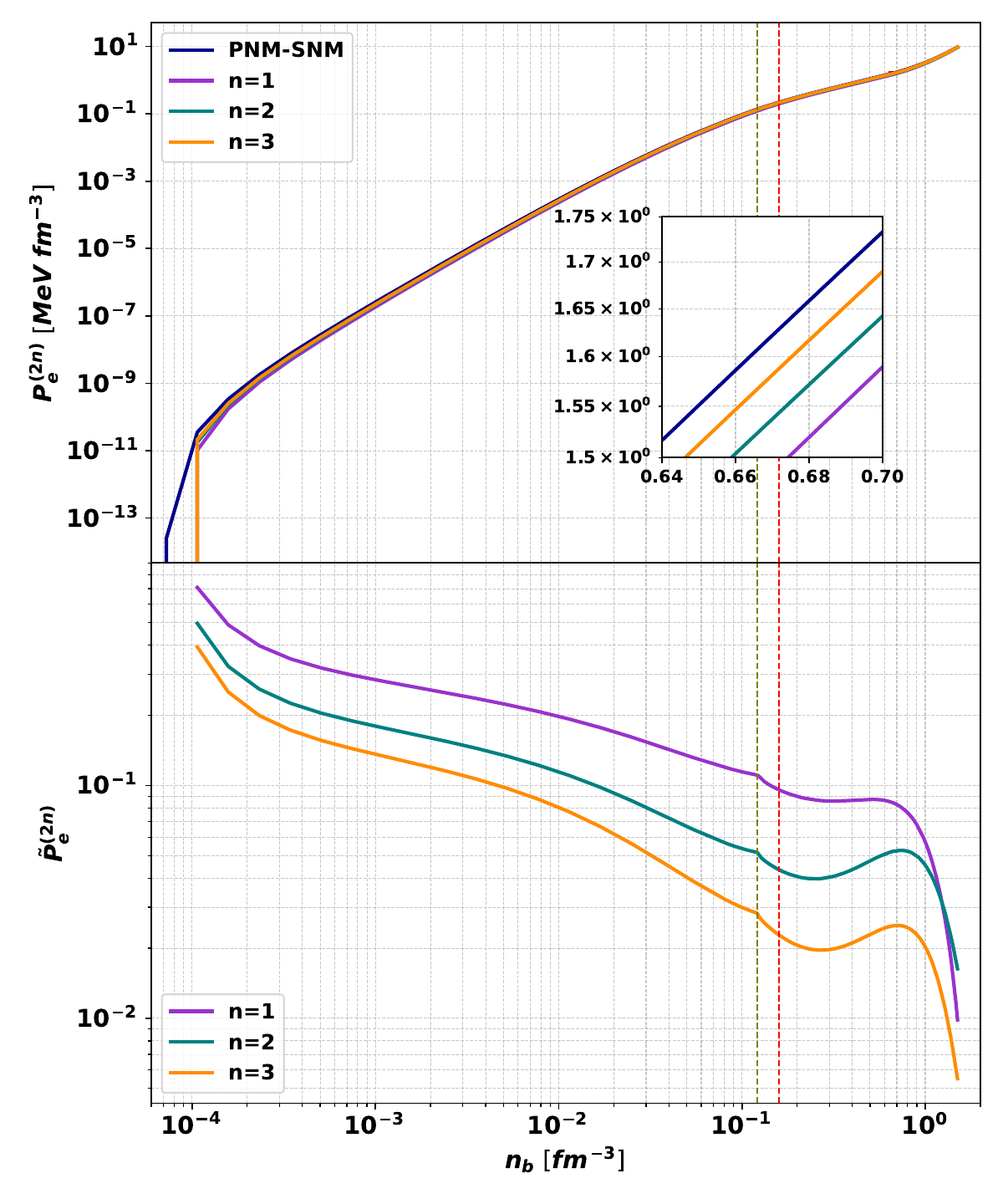}
    \includegraphics[clip,width=\columnwidth]{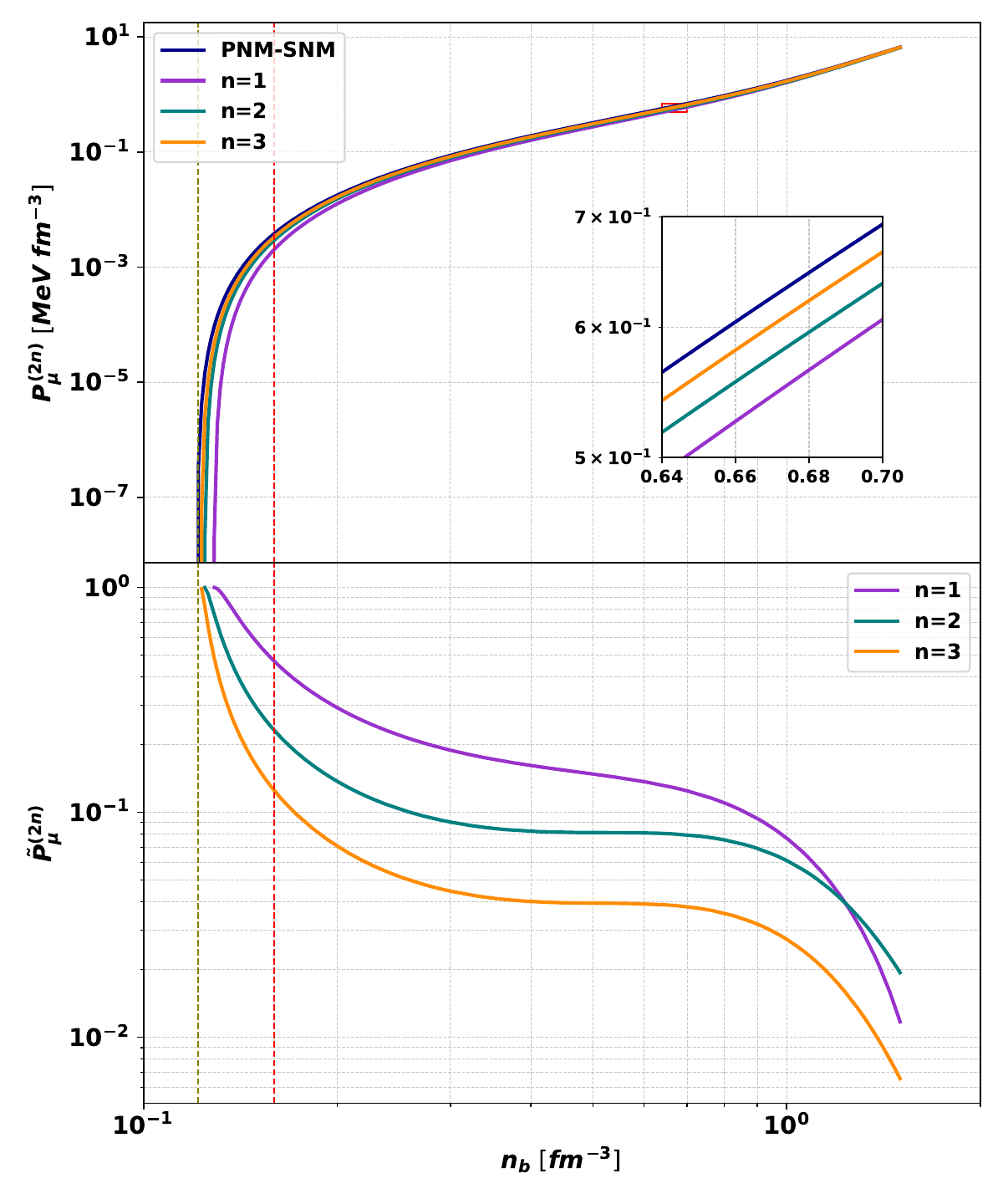} \\ 
    \caption{Individual contributions from electron pressure $P_{e}$ (\textit{in top-left panel}) and muon pressure $P_{\mu}$ (\textit{in top-right panel}) are shown separately over the same baryon density $n_b$ range for $\beta$-equilibrated neutron star matter obtained using the SLy4 Skyrme parametrization. The dependence corresponding to the exact symmetry energy and different truncation orders $\mathcal{O}(I^{2n})$ of the isospin-asymmetry expansion are shown. The corresponding lower panels display the corresponding fractional deviations with respect to the exact treatment, as defined in Eqn ~\ref{frac_diff_theta-tilde}. 
    }\label{fig:sly4_pres}
\end{figure*}

\begin{figure*}[htbp]
    \raggedright (a)\hspace*{\columnwidth}(b)\\[-0.5cm]
    \includegraphics[clip,width=\columnwidth]{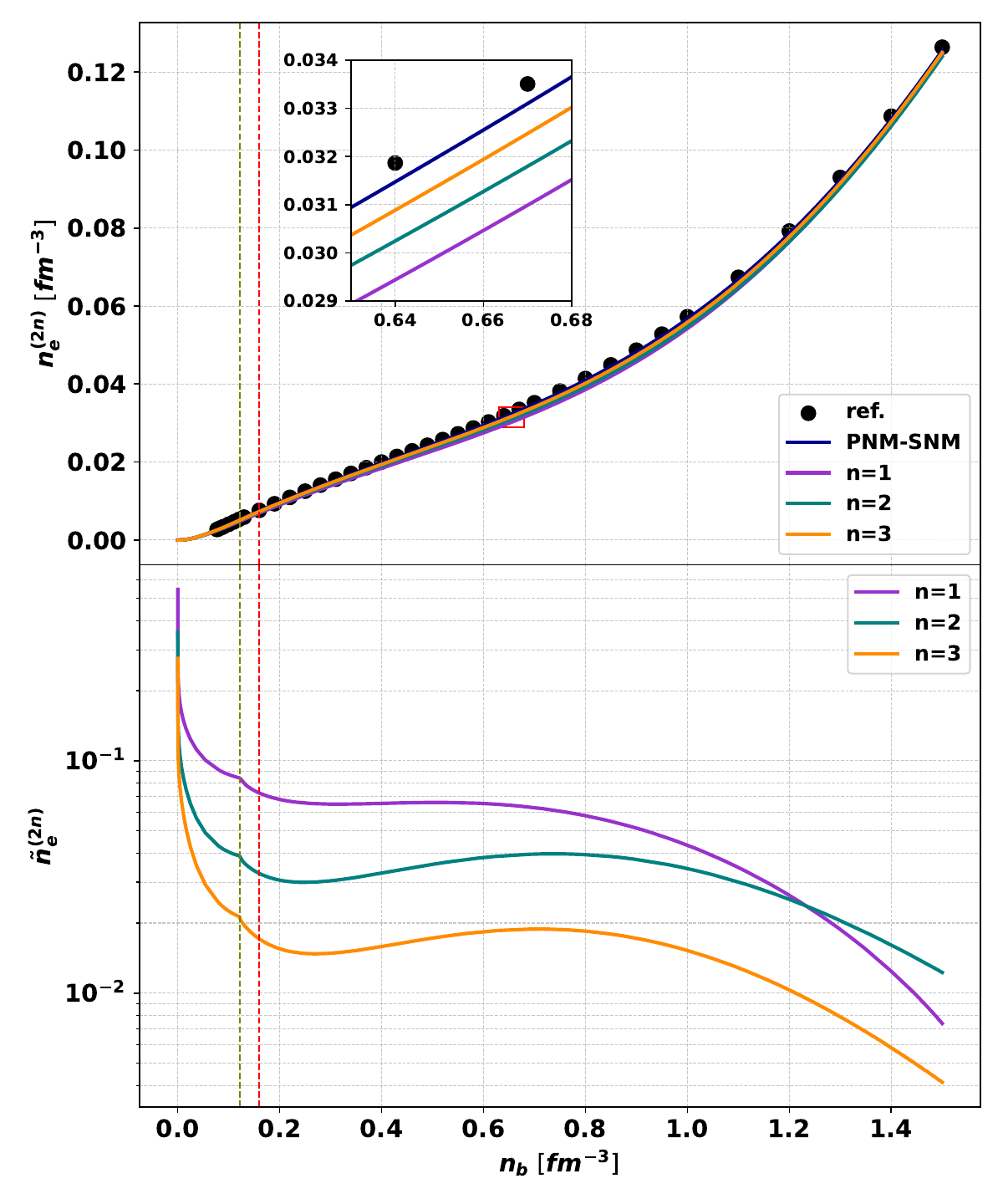}
    \includegraphics[clip,width=\columnwidth]{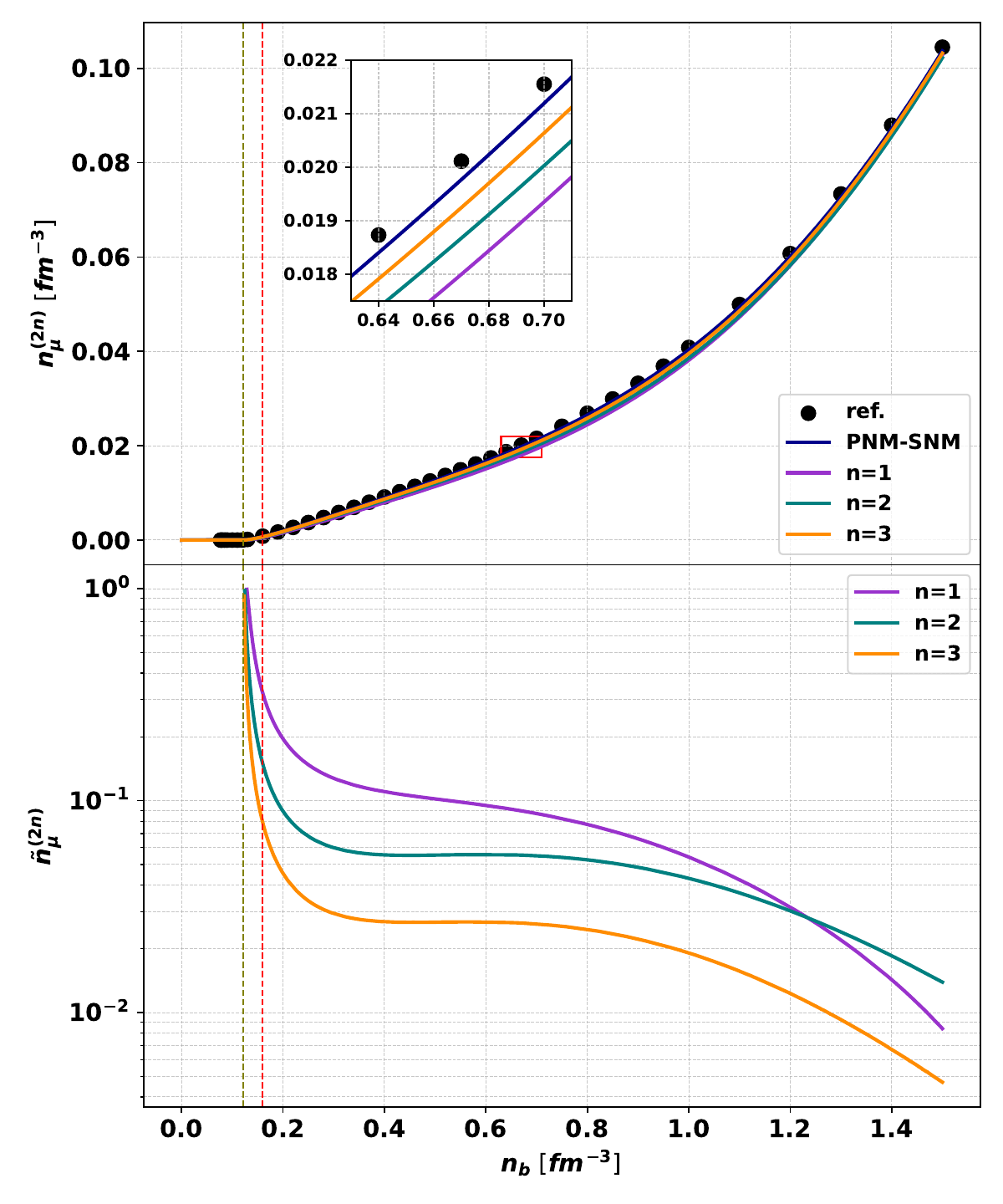} \\ 
    \caption{Similar to, Fig.~\ref{fig:sly4_pres}, the variation for Electron number density $n_{e}$ (\textit{in top-left panel}) and muon number density $n_{\mu}$ (\textit{in top-right panel}) over the same $n_b$ range for $\beta$-equilibrated neutron-star matter obtained using the SLy4 Skyrme parametrization are presented here. The dependence corresponding to the exact symmetry energy and different truncation orders $\mathcal{O}(I^{2n})$ of the isospin-asymmetry expansion are shown. The corresponding lower panels display the corresponding fractional deviations with respect to the exact treatment, as defined in Eqn ~\ref{frac_diff_theta-tilde}. An estimation of $n_{e}$ and $n_{\mu}$ have been reported in~\cite{DouchinHaensel_AA2001}, marked here as the ``ref''.
    }\label{fig:sly4_num_den}
\end{figure*}

In the main article, we demonstrated that higher-order isospin asymmetry corrections exert a more pronounced influence on composition-sensitive quantities than on the bulk thermodynamic properties of neutron-star matter. Notably, the leptonic sector exhibits significantly larger deviations from the conventional quadratic approximation of the symmetry energy than the baryonic sector. For completeness, we present here supplementary results detailing the individual electron and muon components in $\beta$-equilibrated matter using the SLy4 Skyrme parametrization. 

Figure~\ref{fig:sly4_lepton_pressure} displays the total lepton pressure alongside the individual contributions from electrons and muons, calculated using the exact treatment of the asymmetry corrections. The threshold baryon number density for muon onset is also indicated relative to the nuclear saturation density. 

The individual electron and muon pressures are shown in Fig.~\ref{fig:sly4_pres}, while their corresponding number densities and fractional differences are presented in Fig.~\ref{fig:sly4_num_den}. These quantities are evaluated over the baryon number density range $n_{b} \in [10^{-6}, 1.5] \ \mathrm{fm}^{-3}$ to contrast the effects of different truncation orders in the expansion of the symmetry energy.

\subsection{Contribution of beyond-quadratic terms to the symmetry energy for the Skyrme population}\label{app_eta_beyond_quadratic}

\begin{figure}[!h]
    \centering
    \includegraphics[width=\columnwidth]{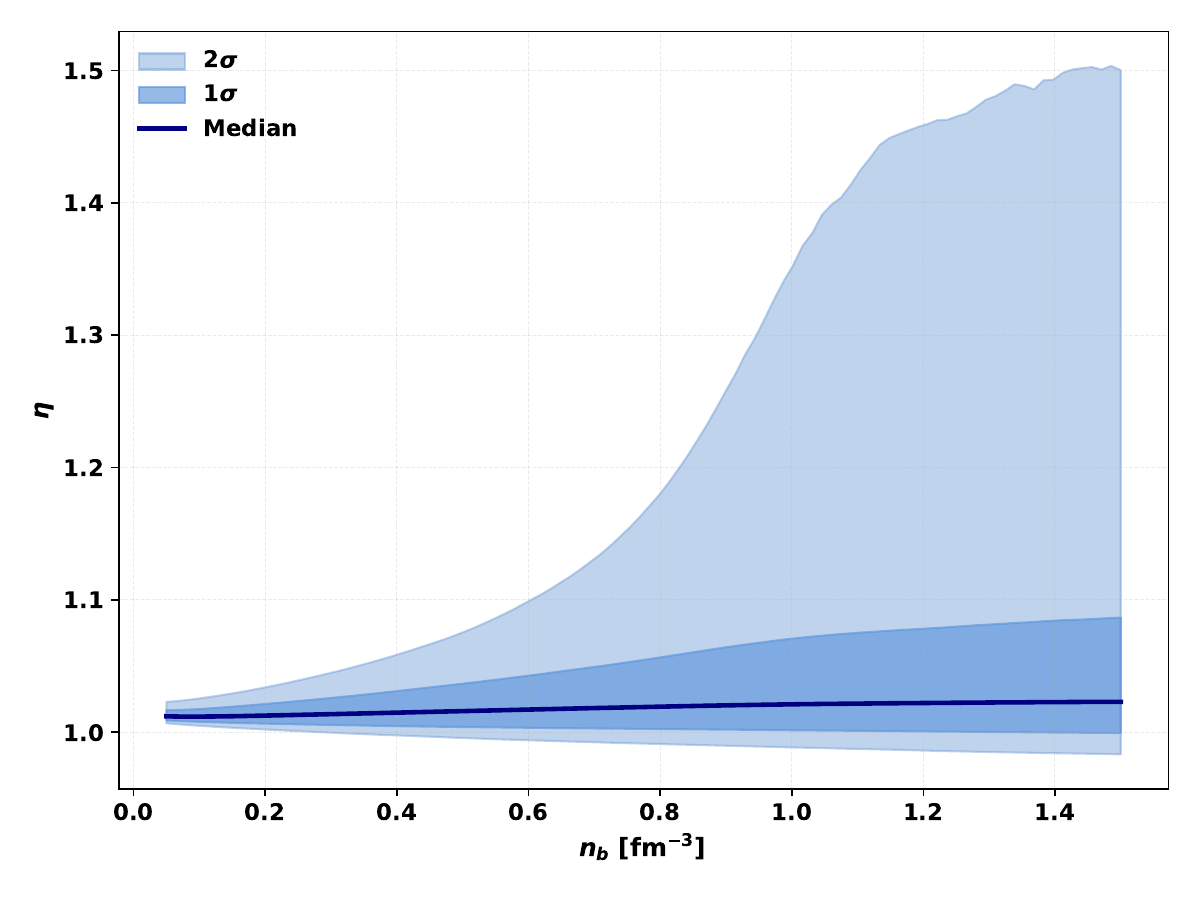}
    \caption{Difference in symmetry energy expressed through the parametrization $\eta$ introduced in~\citet{Steiner_prc2006}, shown at the $1 \sigma$ and $2 \sigma$ confidence levels for the population of physically viable Skyrme EOSs presented in Fig.~\ref{fig:pop_skyrme_physical_dist}.} 
    \label{fig:diff_symmE_eta}
\end{figure}

For completeness, we quantify the non-quadratic contributions to the symmetry energy using the parametrization $\eta$ reported by~\citet{Steiner_prc2006}. While the formulation in~\citet{Steiner_prc2006} specifically characterized the relative contribution of the quartic term, we generalize this parameter here to quantify the total higher-order isospin remainder beyond the conventional quadratic approximation. Figure~\ref{fig:diff_symmE_eta} illustrates the statistical distribution of $\eta$ across the population of physically viable Skyrme equations of state (EOSs) analyzed in this work.

The non-quadratic remainder, $Q(n)$, is defined as 
\begin{equation}\label{eq:app_q_definition}
    Q(n) = E_{\mathrm{sym}}^{(\mathrm{exact})}(n) - E_{\mathrm{sym}}^{(2)}(n),
\end{equation}
with the corresponding generalized parametrization given by 
\begin{equation}\label{eq:app_eta_definition}
    \eta = \frac{4E_{\mathrm{sym}}^{(2)}(n) + 5Q(n)}{4E_{\mathrm{sym}}^{(2)}(n) + Q(n)}.
\end{equation}

As detailed in the main text, deviations from the quadratic approximation increase systematically with baryon density. Furthermore, the variance among the different Skyrme EOSs widens progressively in the supra-saturation density regime. This behavior reflects both the growing importance of higher-order isospin terms and the substantial uncertainties inherent to the isovector sector of the nuclear interaction at high densities. 

\subsection{Deviation in dUrca onset}\label{app_dUrca_onset_deviation}

\begin{figure}[!h]
    \centering
    \includegraphics[width=\columnwidth]{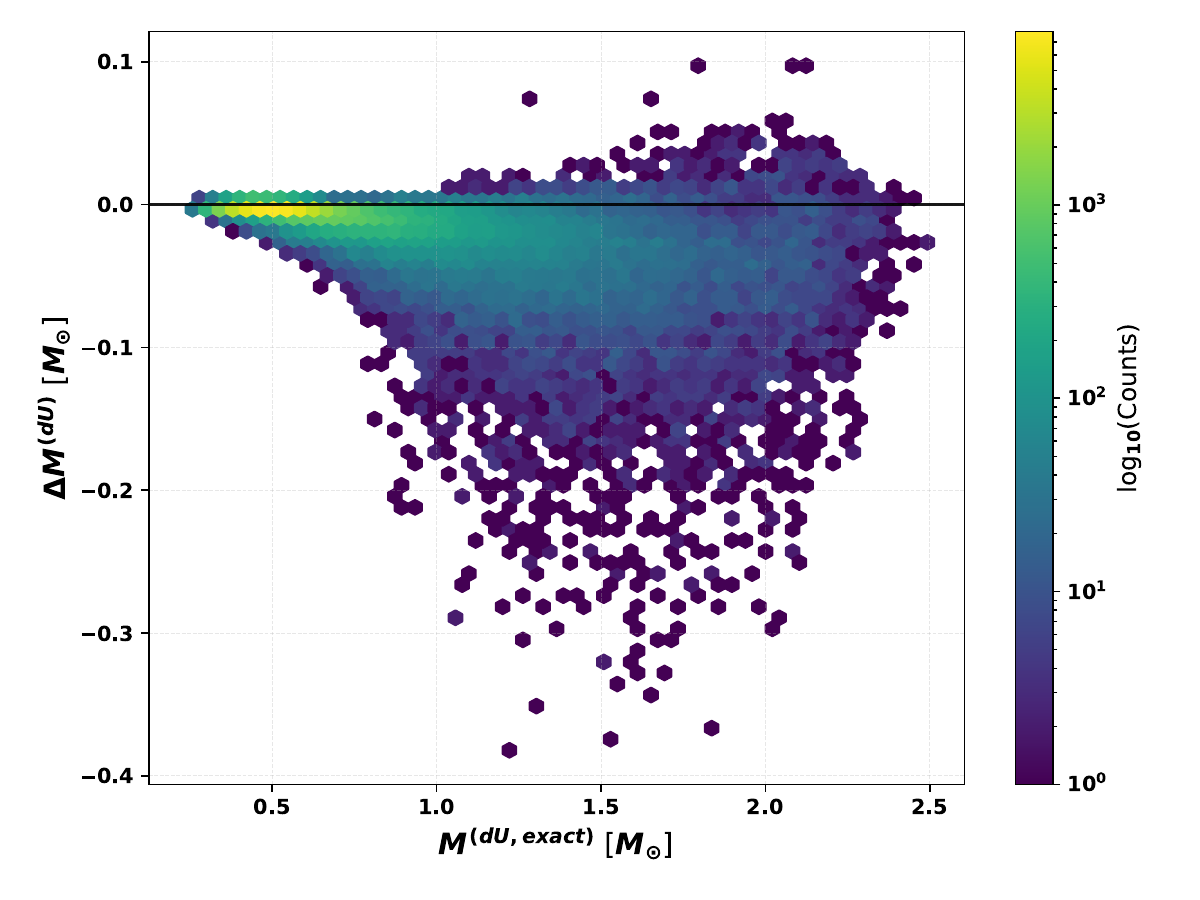}
    \caption{ Hexagonal density map showing the distribution of the direct-Urca onset mass deviation $\Delta M^{(dU)} = M^{(dU, exact)} - M^{(dU, 2)}$ as a function of the exact onset mass $M^{(dU, exact)}$ for the full ensemble of valid Skyrme EOSs. The color scale represents the logarithm of the number of EOS samples in each hexagonal bin. The distribution is strongly concentrated near $\Delta M^{(dU)} \simeq 0$, while a progressively broader negative tail develops toward larger onset masses, indicating that higher-order isospin corrections can lead to appreciable reductions in the predicted dUrca threshold mass for a small subset of EOSs. }
    \label{fig:diff_urca_threshold_mass_hexbin}
\end{figure}

To elucidate the statistical structure of the truncation-induced shift in the direct Urca threshold mass, Figure~\ref{fig:diff_urca_threshold_mass_hexbin} displays the joint distribution of the exact direct Urca onset mass, $M_{\mathrm{dU}}^{\mathrm{exact}}$, and the corresponding deviation 

\begin{equation}\label{eq:app_dUrca_mass_diff}
    \Delta M_{\mathrm{dU}} = M_{\mathrm{dU}}^{\mathrm{exact}} - M_{\mathrm{dU}}^{(2)}.
\end{equation}

The density of the equation of state (EOS) samples within this two-dimensional parameter space is resolved using a logarithmic hexagonal binning scheme. The distribution demonstrates that the majority of valid Skyrme EOSs cluster tightly around $\Delta M_{\mathrm{dU}} \simeq 0$, confirming that the conventional quadratic approximation reproduces the direct Urca onset mass with high global fidelity. 

Concurrently, the distribution exhibits a pronounced asymmetric broadening toward negative values of $\Delta M_{\mathrm{dU}}$, indicating a systematic overestimation of the threshold mass (i.e., a delayed onset) when employing the quadratic approximation relative to the exact treatment. Furthermore, the variance in $\Delta M_{\mathrm{dU}}$ increases toward intermediate and larger onset masses, revealing a heteroscedastic trend where EOS-to-EOS variability and sensitivity to higher-order isospin contributions become enhanced at higher threshold masses. 

\end{document}